\documentclass[vecphys]{svmult}

\usepackage{makeidx}         
\usepackage{graphicx}        
\usepackage{multicol}        
\usepackage[bottom]{footmisc}
\usepackage{amssymb}

\makeindex             

\newcommand{\setR}{\mathbb{R}}

\newcommand{\be}{\begin{equation}}
\newcommand{\en}{\end{equation}}
\newcommand{\bea}{\begin{eqnarray}}
\newcommand{\ena}{\end{eqnarray}}

\newcommand{\dd}{\mathrm{d}}

\newcommand{\mP}{m_{_{\mathrm{Pl}}}}

\newcommand{\ess}{{_\mathrm{S}}}
\newcommand{\ti}{{_\mathrm{T}}}
\newcommand{\muS}{\mu_\ess}
\newcommand{\muT}{\mu_\ti}
\newcommand{\muST}{\mu_{\ess,\ti}}
\newcommand{\omegaT}{\omega_\ti}
\newcommand{\omegaS}{\omega_\ess}
\newcommand{\omegaST}{\omega_{\ess, \ti}}

\newcommand{\calH}{\mathcal{H}}

\begin{document}

\title*{Inflationary Perturbations: the Cosmological
Schwinger Effect}
\titlerunning{The Cosmological Schwinger Effect} \author{J\'er\^ome
Martin\inst{1}}
\institute{Institut d'Astrophysique de Paris, UMR 7095 CNRS,
Universit\'e Pierre et Marie Curie, 98bis boulevard Arago, 75014
Paris, France \texttt{jmartin@iap.fr}}
%
%
\maketitle

\begin{abstract}

This pedagogical review aims at presenting the fundamental aspects of
the theory of inflationary cosmological perturbations of
quantum-mechanical origin. The analogy with the well-known Schwinger
effect is discussed in detail and a systematic comparison of the two
physical phenomena is carried out. In particular, it is demonstrated
that the two underlying formalisms differ only up to an irrelevant
canonical transformation. Hence, the basic physical mechanisms at play
are similar in both cases and can be reduced to the quantization of a
parametric oscillator leading to particle creation due to the
interaction with a classical source: pair production in vacuum is
therefore equivalent to the appearance of a growing mode for the
cosmological fluctuations. The only difference lies in the nature of the
source: an electric field in the case of the Schwinger effect and the
gravitational field in the case of inflationary perturbations. Although,
in the laboratory, it is notoriously difficult to produce an electric
field such that pairs extracted from the vacuum can be detected, the
gravitational field in the early universe can be strong enough to lead
to observable effects that ultimately reveal themselves as temperature
fluctuations in the Cosmic Microwave Background. Finally, the question
of how quantum cosmological perturbations can be considered as classical
is discussed at the end of the article.

\end{abstract}

\section{Introduction}
\label{sec:intro}

The scenario of inflation was invented in order to solve puzzling issues
associated with the standard hot Big Bang
theory~\cite{guth,inflation}. Soon after its advent, it was realized
that this scenario also contains a remarkable extra bonus: it gives a
well-motivated mechanism for structure formation that leads to a nearly
scale-invariant power spectrum~\cite{MuChi, pert}, namely exactly what
is needed in order to account for various astrophysical observations in
a satisfactory way~\cite{wmap}. However, this is not the only aspect
that deserves to be stressed. Indeed, even from a fundamental point of
view, this mechanism appears quite remarkable in the sense that it
combines general relativity with quantum mechanics. The main purpose of
this review article is to thoroughly discuss this aspect of the theory
of inflationary cosmological perturbations.

\par

This theory is in fact remarkable at two levels. Firstly, because it
relies on the phenomenon of particle creation which is a non-trivial
effect in quantum field theory. In this sense, it is equivalent to the
well-known Schwinger~\cite{schwinger} effect and this analogy will be
made explicit in this paper. The basic ingredient is a quantum scalar
field $\Phi $ (in practice this is rather a fermionic field $\Psi$ but,
for simplicity, we will restrict ourselves to the case of a scalar
field) interacting with a classical source, in the case of the Schwinger
effect, an electric field $E$. The Schwinger effect has not yet been
observed in the laboratory as it is difficult to produce an electric
field with the required strength but there are prospects to do so, in
particular at DESY with a Free Electron Laser (FEL) in the X-ray
band~\cite{Ring,AHRSV,xfel} but also at SLAC with the Linac Coherent
Light Source (LCLS)~\cite{linac}. Even if there is absolutely no reason
to doubt the reality of the Schwinger effect, observing pair creation in
the laboratory would clearly be a breakthrough and, in some sense, a
verification of the corresponding inflationary mechanism.

\par

Secondly, the theory of cosmological perturbations is also remarkable
for the following reason. In cosmology, what plays the role of the
constant electric field $E$ [originating from a time-dependent potential
vector $A_{\mu}(t)$] is the background gravitational field, i.e. the
Friedmann-Lema\^{\i}tre-Robertson-Walker (FLRW) scale factor $a(t)$, and
what plays the role of the quantum fermionic field $\Psi (t, \vec{x})$
is the quantum perturbed metric $\delta g_{\mu \nu}(t,\vec{x})$, that is
to say the small inhomogeneous fluctuations of the gravitational field
itself~\cite{MFB}. In the early Universe, the gravitational field is
quite strong, i.e. for instance $H/\mP\sim 10^{-5}$, where $H$ is the
Hubble parameter, and this is why the cosmological version of the
Schwinger effect can be efficient. From the previous considerations, it
is also clear that, in some sense, the inflationary mechanism relies on
quantum gravity which adds another interesting aspect to the problem. Of
course, we only deal with linearized quantum gravity and this is why we
do not have to face tricky questions associated with finiteness of
quantum gravity and/or renormalization. More precisely, in the case of
scalar perturbations, $\delta g_{\mu \nu}(t,\vec{x})$ is replaced by the
Mukhanov-Sasaki variable $v(t, \vec{x})$ which is a combination of the
Bardeen potential (the generalization of the Newtonian potential in
general relativity) and of the small fluctuations in the inflaton
field. For gravitational waves, the relevant quantity is
$h_{ij}(t,\vec{x})$, the transverse and traceless part of the perturbed
metric.
 
\par

Let us notice that the two above-mentioned aspects are features of the
theory of cosmological perturbations in general. The inflationary
aspect is in fact not necessary in order to have particles creation:
only a dynamical background is required. However, a quasi-exponential
expansion is mandatory if one wants to obtain a power spectrum which
is close to scale invariance as indicated by astrophysical
observations.

\par

The fact that the inflationary mechanism for structure formation relies
on general relativity and quantum mechanics also raises fundamental
interpretational questions. In particular, the question of how
classicality emerges is of special relevance in this
context~\cite{GP,PS}. Indeed, the perturbations are of
quantum-mechanical origin but no astrophysical observations suggest any
typically quantum-mechanical signature. Therefore, it is necessary to
understand how the perturbations have become classical (and in which
sense). This leads to very deep issues. For instance, if one invokes a
mechanism based on the phenomenon of decoherence~\cite{Zurek}, then one
has to discuss what plays the role of the environment. This question is
clearly non-trivial in the cosmological context. At the end of this
review article, we will address these questions using the Wigner
function~\cite{wigner} as a tool to understand when a system can be
considered as classical.

\par

This paper can be viewed as the third of a series on the inflationary
theory, the two first ones being Refs.~\cite{procbrazil,
procpoland}. The topics developed in those last two references will be
supposed to be known and we will often refer to them. The present paper
is organized as follows. In Sec.~\ref{sec:schwinger}, we briefly review
the Schwinger effect. In particular, we derive the rate of pair
production in the Schr\"odinger functional approach and stress the
importance of the Wentzel-Kramers-Brillouin (WKB) approximation as a
method to choose a well-defined initial state. In
Sec.~\ref{sec:quantizationsf}, we quantize a free scalar field in a FLRW
Universe and show that the basic physical phenomenon at play is
equivalent to that responsible for the Schwinger effect, namely particle
creation under the influence of a classical source. In particular, we
demonstrate that, up to a canonical transformation, the underlying
formalisms are the same. Roughly speaking, in both cases, one has to
deal with parametric oscillators. The only difference between the two
systems lies in the time dependence of the corresponding effective
frequencies. In Sec.~\ref{sec:pert}, we argue that the equations obeyed
by the cosmological perturbations (in particular during inflation) are
equivalent to the equations of motion of a free scalar field. We
emphasize that the relevant observable quantity is the two-point
correlation function since it is directly linked to the Cosmic Microwave
Background (CMB) temperature fluctuations. Finally, as mentioned above,
in Sec.~\ref{sec:classical}, we address the question of the classicality
of the cosmological perturbations.

\section{The Schwinger Effect}
\label{sec:schwinger}

\subsection{General Formalism}
\label{subsec:1-1}

The action of a complex (charged) scalar field interacting with an
electromagnetic field is given by
\begin{equation}
\label{eq:action}
S=-\int {\rm d}^4x\left(\frac12 \eta^{\alpha \beta }{\cal D}_{\alpha
}\Phi {\cal D}_{\beta }\Phi ^*+\frac12 m^2 \Phi \Phi ^*\right)\, ,
\end{equation}
where $\eta ^{\alpha \beta }$ is the flat (Minkowski) space-time metric
with signature $(-,+,+,+)$ and where the covariant derivative can be
expressed as
\begin{equation}
{\cal D}_{\alpha }\Phi \equiv \partial_{\alpha } \Phi +iqA_{\alpha
}\Phi\, ,
\end{equation}
$q$ being the charge of the field. The quantity $m$ represents the mass
of the scalar particle. Assuming the following configuration for the
vector potential $A_{\mu }=(0,0,0,-Et)$, where $E$ is the magnitude of
the static electric field aligned along the $z$ direction (by
convention), one obtains the following equation of motion
\begin{equation}
\label{eq:motionPhi}
\partial _t^2\Phi -\partial ^i\partial_i\Phi +2iqEt \partial _z\Phi 
+q^2E^2t^2\Phi +m^2\Phi =0\, .
\end{equation}
It turns out to be more convenient to Fourier transform the field since
this allows us to study the evolution of the system mode by mode.  For
this purpose, one decomposes the field according to
\begin{equation}
\label{eq:Ftransform}
\Phi (t, \vec{x})=\frac{1}{(2\pi )^{3/2}}\int {\rm d}^3 \vec{k}\, \Phi
_{\vec{k}}(t) {\rm e}^{i\vec{k}\cdot \vec{x}}\, .
\end{equation}
In the above expression, $\Phi _{\vec{k}}(t)$ is the time-dependent
Fourier amplitude of the mode characterized by the wave-vector
$\vec{k}$. Inserting the Fourier transform~(\ref{eq:Ftransform}) into
Eq.~(\ref{eq:action}), the action of the system takes the form
\begin{equation}
\label{eq:actionfourier}
S=-\int {\rm d}t\int _{\setR^3}{\rm d}\vec{k} \left[-\dot{\Phi
}_{\vec{k}}\dot{\Phi }_{\vec{k}}^*
+\left(k^2-2qEk_zt+q^2E^2t^2+m^2\right)\Phi _{\vec{k}}\Phi
_{\vec{k}}^*\right]\, ,
\end{equation}
where a dot denotes a derivative with respect to time. The variation of
this Lagrangian with respect to $\Phi _{\vec{k}}^*$ and $\dot{\Phi
}_{\vec{k}}^*$ leads to
\begin{equation}
\frac{\delta \bar{\cal L}}{\delta \Phi _{\vec{k}}^*}=
-\left(k^2-2qEk_zt+q^2E^2t^2+m^2\right)\Phi _{\vec{k}}\, ,\quad 
\frac{\delta \bar{\cal L}}{\delta \dot{\Phi }_{\vec{k}}^*}
\equiv p_k=\dot{\Phi }_{\vec{k}}\, ,
\end{equation}
where $p_k$ is the conjugate momentum of the Fourier component of the
field and $\bar{\cal L}$ denotes the Lagrangian density in Fourier
space. Using the two above formula, the Euler-Lagrange equation of
motion reads
\begin{equation}
\label{eqphi}
\ddot{\Phi}_{\vec{k}}
+\omega ^2(k,t)\Phi _{\vec{k}}=0 \, ,
\end{equation}
where the time dependent frequency $\omega (k,t)$ can be expressed as
\begin{equation}
\label{eq:defomega}
\omega ^2(k,t)\equiv k^2-2qEk_zt+q^2E^2t^2+m^2\, .
\end{equation}
Eq.~(\ref{eqphi}) is of course similar to the one one would have
obtained by directly substituting Eq.~(\ref{eq:Ftransform}) into
Eq.~(\ref{eq:motionPhi}). It is the equation of motion of a parametric
oscillator. Let us recall that a parametric oscillator is an harmonic
oscillator whose frequency depends on time. A typical example is a
pendulum with a varying length.

\par

Let us now pass to the Hamiltonian formalism. The Hamiltonian is
obtained from the Lagrangian by a standard Legendre transformation and
can be expressed as
\begin{equation}
\label{eq:Hcomplex}
H=\int _{\setR^3}{\rm d}\vec{k} \left(p_k\dot{\Phi }_{\vec{k}}^*
+p_k^*\dot{\Phi }_{\vec{k}}-\bar{\cal L}\right) =\int _{\setR^3}{\rm
d}\vec{k} \left[p_kp_k^*+\omega ^2\left(k,t\right)\Phi _{\vec{k}}\Phi
_{\vec{k}}^*\right]\, .
\end{equation}
For the following considerations, it turns out to be convenient to
also work with real variables instead of the complex $\Phi
_{\vec{k}}$. Therefore, we now introduce the definitions
\begin{equation}
\label{defphireal}
\Phi _{\vec{k}}\equiv \frac{1}{\sqrt{2}}\left(\Phi ^{_{\mathrm
R}}_{\vec{k}}+ i\Phi ^{_{\mathrm I}}_{\vec{k}}\right)\, , \quad
p_{\vec{k}}\equiv \frac{1}{\sqrt{2}}\left(p ^{_{\mathrm R}}_{\vec{k}}+
ip^{_{\mathrm I}}_{\vec{k}}\right)\, .
\end{equation}
Then, the Hamiltonian can be written as
\begin{equation}
\label{eq:Hreal}
H=\int _{\setR^3}{\rm
d}\vec{k} \left[\frac12 \left(p ^{_{\mathrm R}}_{\vec{k}}\right)^2+\frac12
\omega ^2\left(k,t\right)\left(\Phi ^{_{\mathrm
R}}_{\vec{k}}\right)^2
+\frac12 \left(p ^{_{\mathrm I}}_{\vec{k}}\right)^2+\frac12
\omega ^2\left(k,t\right)\left(\Phi ^{_{\mathrm
I}}_{\vec{k}}\right)^2\right]\, .
\end{equation}
One recognizes the Hamiltonian of a collection of parametric oscillators
with a time-dependent frequency given by Eq.~(\ref{eq:defomega}). Again,
one can check that the Hamilton equations deduced from
Eqs.~(\ref{eq:Hcomplex}) and~(\ref{eq:Hreal}) lead to an equation of
motion similar to the one already derived before, namely
Eq.~(\ref{eqphi}).

\subsection{Quantization}
\label{subsec:1-2}

Our next step is to describe the quantization of the system. More
precisely, the complex scalar field is quantized while the gauge field
remains classical. Therefore, we have to deal with the interaction of a
quantum field with a classical source. Quantization is achieved by
requiring the following commutation relations (a hat symbol is put on
letters denoting operators)
\begin{equation}
\label{commutreal}
\left[\hat{\Phi }^{_{\mathrm R}}_{\vec{k}}, \hat{p} ^{_{\mathrm
R}}_{\vec{p}}\right]=i\delta^{(3)}\left(\vec{k}-\vec{p}\right)\, ,
\quad \left[\hat{\Phi }^{_{\mathrm I}}_{\vec{k}}, \hat{p} ^{_{\mathrm
I}}_{\vec{p}}\right]=i\delta^{(3)}\left(\vec{k}-\vec{p}\right)\, .
\end{equation}
We choose to work in the Schr\"odinger picture where the states are
time-dependent and the operators constant. The above commutation
relations admit the following representation
\begin{equation}
\label{representation}
\hat{\Phi }^{_{\mathrm R}}_{\vec{k}}\Psi=\Phi ^{_{\mathrm R}}_{\vec{k}}
\Psi\, , \quad \hat{p}^{_{\mathrm R}}_{\vec{k}}\Psi=-i
\frac{\partial \Psi}{\partial \Phi ^{_{\mathrm R}}_{\vec{k}}}\, .
\end{equation}
The state of the system is described by a functional of the scalar
field, $\Psi[\Phi (t,\vec{x})]$ (in the present context, $\Psi$ is the
field functional and has clearly nothing to do with the fermionic field
mentioned before), which can also be viewed as a function of an infinite
number of variables, namely the values of $\Phi $ at each point in
space. Alternatively, one can also consider this functional as a
function of the infinite number of Fourier components of the field and
write
\begin{equation}
\Psi =\prod  _{\vec{k}}^n\Psi_{\vec{k}}\left(\Phi
^{_{\mathrm R}}_{\vec{k}}, \Phi
^{_{\mathrm I}}_{\vec{k}}\right)
=\prod  _{\vec{k}}^n\Psi_{\vec{k}}^{_{\mathrm R}}
\left(\Phi
^{_{\mathrm R}}_{\vec{k}}\right)
\Psi_{\vec{k}}^{_{\mathrm I}}
\left(\Phi
^{_{\mathrm I}}_{\vec{k}}\right)
\, .
\end{equation}
In the above equation, $n$ represents the number of modes that, in the
intermediate calculations, it is useful to keep finite (for instance, if
we imagine that the field lives in a finite box). However, at the end,
we will always consider the continuous case and take the limit
$n\rightarrow +\infty$.

\par

In the framework described before, the Schr\"odinger equation is a
functional differential equation. However, the Hamiltonian takes the
form of an infinite sum over $\vec{k}$, see for instance
Eqs.~(\ref{eq:Hcomplex}) and~(\ref{eq:Hreal}), without any interaction
between different modes. As a consequence, each mode evolves
independently and the corresponding Hamiltonian is represented by an
ordinary differential operator. Explicitly, one has
\begin{eqnarray}
\label{Hamilschwinger}
H _{\vec{k}}\Psi &=& \left(H _{\vec{k}}^{_{\mathrm R}}
+H _{\vec{k}}^{_{\mathrm I}}\right)\Psi
=-\frac{1}{2}\frac{\partial ^2\Psi}{\partial
  \left(\Phi ^{_{\mathrm
      R}}_{\vec{k}}\right)^2}+\frac12\omega ^2(k,t)\left(\Phi
^{_{\mathrm R}}_{\vec{k}}\right)^2\Psi
-\frac{1}{2}\frac{\partial ^2\Psi}{\partial
  \left(\Phi ^{_{\mathrm
      I}}_{\vec{k}}\right)^2}\nonumber \\ & & 
+\frac12 \omega ^2(k,t)\left(\Phi
^{_{\mathrm I}}_{\vec{k}}\right)^2\Psi
\, ,
\end{eqnarray}
where the frequency $\omega $ has been given by
Eq.~(\ref{eq:defomega}).

\par

Let us now consider the ground state of the system described before. We
will discuss the choice of the initial conditions and the meaning of the
vacuum state in the following but, as is well-known, it is given by a
Gaussian state
\begin{equation}
\label{wavefunction}
\Psi _{\vec{k}}^{_{\mathrm R}}\left(t, \Phi
^{_{\mathrm R}}_{\vec{k}}\right)
=N_{\vec{k}}\left(t \right){\rm e}^{-\Omega _{\vec{k}}\left(t \right)
\left(\Phi
^{_{\mathrm R}}_{\vec{k}}\right)^2}\, ,\quad 
\Psi _{\vec{k}}^{_{\mathrm I}}\left(t, \Phi
^{_{\mathrm I}}_{\vec{k}}\right)
=N_{\vec{k}}\left(t \right){\rm e}^{-\Omega _{\vec{k}}\left(t \right)
\left(\Phi
^{_{\mathrm I}}_{\vec{k}}\right)^2}\, ,
\end{equation}
where $N_{\vec{k}}(t)$ and $\Omega _{\vec{k}}(t)$ are functions that
can be determined using the Schr\"odinger equation $i\partial _{t}\Psi
=H_{\vec{k}}\Psi$. This leads to
\begin{equation}
\label{eq:NOm}
i\frac{\dot{N}_{\vec{k}}}{N_{\vec{k}}}=\Omega _{\vec{k}}\, , \quad
\dot{\Omega}_{\vec{k}} =-2i\Omega _{\vec{k}}^2+\frac{i}{2}\omega
^2\left(k,t\right)\, .
\end{equation}
The equation for the complex quantity $\Omega _{\vec{k}}$ is a
non-linear Ricatti equation. When a particular solution is known, the
general solution can be found by means of two successive
quadratures~\cite{Ince}. But this non-linear first order differential
equation can also be transformed into a linear second order differential
equation. It turns out that this last one is exactly the equation for
the Fourier mode function, Eq.~(\ref{eqphi}). Therefore, the solutions
to Eqs.~(\ref{eq:NOm}) read
\begin{eqnarray}
\label{eq:solNOm}
N_{\vec{k}}=\left(\frac{2\Re \Omega _{\vec{k}}}{\pi}\right)^{1/4} \,
,\quad \Omega
_{\vec{k}}=-\frac{i}{2}\frac{\dot{f}_{\vec{k}}}{f_{\vec{k}}}\, ,
\end{eqnarray}
where $f_{\vec{k}}$ obeys the equation $\ddot{f}_{\vec{k}}+\omega
^2(k,t)f_{\vec{k}} =0$, that is to say, as already mentioned, the same
equation as the Fourier component of the field, namely
Eq.~(\ref{eqphi}). The quantity $N_{\vec{k}}$ is obtained by normalizing
the wave-function. One can check that this leads to an equation
consistent with the first formula in Eqs.~(\ref{eq:NOm}). Therefore, one
obtains that the ground quantum state of the field is given by
\begin{equation}
\Psi =\prod _{\vec{k}}^n\left(\frac{2\Re \Omega
_{\vec{k}}}{\pi}\right)^{1/2} {\rm e}^{-\Omega _{\vec{k}}\left(t
\right)\left[ \left(\Phi ^{_{\mathrm R}}_{\vec{k}}\right)^2+\left(\Phi
^{_{\mathrm I}}_{\vec{k}}\right)^2\right]}\, .
\end{equation}
Once a particular solution for the mode function has been singled out,
the functions $N_{\vec{k}}$ and $\Omega _{\vec{k}}$, and hence the
wave-function of the field, are completely specified.

\par

One can now use the above-mentioned state in order to calculate the
amplitude associated with the transition between two states $\Psi _1$
and $\Psi _2$. It is defined by
\begin{eqnarray}
\left\langle \Psi _1\vert \Psi_2\right \rangle &=& \int \prod  _{\vec{k}}^n
{\rm d}\Phi^{_{\mathrm R}}_{\vec{k}}
{\rm d}\Phi^{_{\mathrm I}}_{\vec{k}}
\left(\frac{2\Re \Omega _{1, \vec{k}}}{\pi}\right)^{1/2}
\left(\frac{2\Re \Omega _{2, \vec{k}}}{\pi}\right)^{1/2} {\rm e}^{-\sum
_{\vec{p}}^n\left(\Omega _{1, \vec{p}}^*+\Omega _{2, \vec{p}}\right)
\left(\Phi^{_{\mathrm R}}_{\vec{p}}\right)^2}\nonumber \\
& &\times 
{\rm e}^{-\sum
_{\vec{p}}^n\left(\Omega _{1, \vec{p}}^*+\Omega _{2, \vec{p}}\right)
\left(\Phi^{_{\mathrm I}}_{\vec{p}}\right)^2}\, ,
\end{eqnarray}
from which, after having performed the Gaussian integration, one
deduces that
\begin{eqnarray}
\left\vert \left\langle \Psi _1\vert \Psi_2\right \rangle \right \vert ^2
&=& 
\det \left[\frac{4 \Re \Omega _{1, \vec{k}}\Re \Omega _{2, \vec{k}}}
{\left(\Omega _{1, \vec{k}}^*+\Omega _{2, \vec{k}}\right)
\left(\Omega _{1, \vec{k}}+\Omega _{2, \vec{k}}^*\right)}\right]\, .
\end{eqnarray}
At this point, one has to use the specific form of $\Omega _{\vec{k}}$,
in particular its expression given by Eq.~(\ref{eq:solNOm}) in terms of
the function $f_{\vec{k}}$. One obtains
\begin{eqnarray}
\left\vert \left\langle \Psi _1\vert \Psi_2\right \rangle \right \vert
^2 &=& \det \left[\frac{\left(\dot{f}_{1,\vec{k}}f_{1,\vec{k}}^*-
\dot{f}_{1,\vec{k}}^*f_{1,\vec{k}}\right)
\left(\dot{f}_{2,\vec{k}}f_{2,\vec{k}}^*-
\dot{f}_{2,\vec{k}}^*f_{2,\vec{k}}\right)}
{\left(\dot{f}_{1,\vec{k}}^*f_{2,\vec{k}}-
\dot{f}_{2,\vec{k}}f_{1,\vec{k}}^*\right)
\left(\dot{f}_{2,\vec{k}}^*f_{1,\vec{k}}-
\dot{f}_{1,\vec{k}}f_{2,\vec{k}}^*\right)} \right]\, .
\end{eqnarray}
In this formula $f_{1,\vec{k}}$ is the mode function for the initial
state while $f_{2,\vec{k}}$ is the same quantity but for the final
state. As usual, one can always expand $f_{2,\vec{k}}$ over the basis
$(f_{1,\vec{k}},f_{1,\vec{k}}^*)$ and write
\begin{equation}
f_{2,\vec{k}}=\alpha
_{\vec{k}}f_{1,\vec{k}}+\beta_{\vec{k}}f_{1,\vec{k}}^*\, .
\end{equation}
Then, using the fact that the Wronskian
$\dot{f}_{1,\vec{k}}f_{1,\vec{k}}^*- \dot{f}_{1,\vec{k}}^*f_{1,\vec{k}}$
is a conserved quantity (as can be easily checked by differentiating it
and using the equation satisfied by $f_{\vec{k}}$), one arrives
at~\cite{Kiefer}
\begin{eqnarray}
\label{eq:det}
\left\vert \left\langle \Psi _1\vert \Psi_2\right \rangle \right \vert
^2 &=& \det \left(\frac{1}{\vert \alpha _{\vec{k}}\vert ^2} \right)\,
.
\end{eqnarray}
Therefore, the determination of the transition amplitude amounts to
integrating the equation controlling the evolution of the mode
function. Once this is done, the coefficient $\alpha _{\vec{k}}$ is
known and the quantity $ \left\vert \left\langle \Psi _1\vert
\Psi_2\right \rangle \right\vert ^2$ can be determined. In the next
section, we discuss an explicit example.

\subsection{Particle creation}
\label{subsec:particlecreation}

We now use the formalism developed above in order to study the creation
of quantum scalar particles due to interaction with a classical
source. This is the well-known Schwinger effect~\cite{schwinger}. In the
following, we will demonstrate that the inflationary mechanism for
cosmological perturbations is exactly similar to the one discussed here,
see also Refs.~\cite{Kiefer,SP}.

\par

In order to determine the coefficient $\alpha _{\vec{k}}$, one can
proceed as follows. Let us use the dimensionless variable $\tau \equiv
\sqrt{qE}t-k_z/\sqrt{qE}$. Then, the equation of motion for the Fourier
component of the field, see Eq.~(\ref{eqphi}), takes the form
\begin{equation}
\label{eom}
\frac{{\rm d}^2\Phi _{\vec{k}}}{{\rm d}\tau ^2}+\left(\Upsilon+\tau
^2\right) \Phi _{\vec{k}}=0\, ,
\end{equation}
with $\Upsilon \equiv (k_{\perp}^2+m^2)/(qE)$. The quantity $k_{\perp}$
is defined by $k_{\perp}^2\equiv k^2-k_z^2=k_x^2+k_y^2$ (let us recall
that we have chosen an electrical field aligned along the
$z$-direction). Eq.~(\ref{eom}) can be integrated exactly, see
Eq.~(9.255.2) of Ref.~\cite{Grad} and the solution can be expressed as
\begin{equation}
\label{exactsol}
\Phi _{\vec{k}}\left(\tau
\right)=A_{\vec{k}}D_{-(1+i\Upsilon)/2}\left[\left(1+i\right)\tau
\right] +B_{\vec{k}}D_{-(1+i\Upsilon)/2}\left[-\left(1+i\right)\tau
\right]\, ,
\end{equation}
where $A_{\vec{k}}$ and $B_{\vec{k}}$ are two constants fixed by the
initial conditions and $D_p(z)$ is a parabolic cylinder function of
order $p$.

\par

Despite the previous change of variable, Eq.~(\ref{eom}) has retained
the form of an equation for a parametric oscillator but the frequency is
now given by
\begin{equation}
\omega (\tau )\equiv \sqrt{\Upsilon +\tau ^2}\, .
\end{equation}
This equation is well-suited to the WKB approximation. This
approximation is not only useful to get an approximate form of the
solution but can also be used in order to choose initial conditions that
are well-motivated. Here, since we do already know the exact solution,
it is clearly this last application we shall be concerned with.

\par

\begin{figure}[t]
\centering \includegraphics[height=6cm, width=10cm]{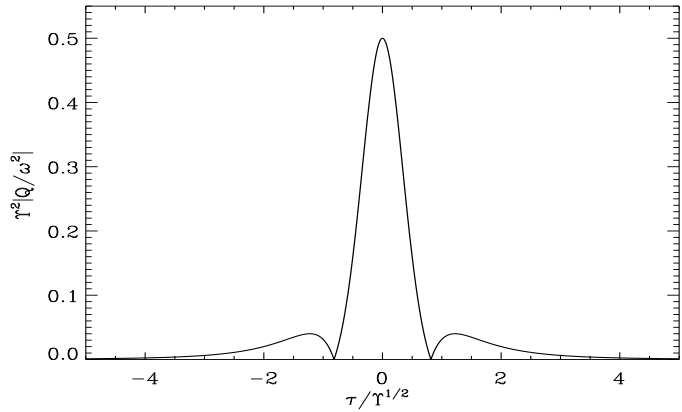}
\caption{Evolution of the quantity $\vert Q/\omega ^2\vert $ with time
$\tau $ in the case of the Schwinger effect. In the limit
$\tau/\sqrt{\Upsilon} \rightarrow \pm \infty$, $\vert Q/\omega ^2\vert$
vanishes and the notion of adiabatic vacuum is available.}
\label{fig:wkb}
\end{figure}

By definition, the WKB mode function $(2\omega )^{-1/2}{\rm e}^{\pm
i\int \omega {\rm d}\tau }$ obeys the following equation of motion
$\ddot{\Phi }_{\vec{k}}+\left(\omega ^2-Q\right)\Phi _{\vec{k}}=0$,
where the quantity $Q$ is defined by
\begin{equation}
Q\equiv \frac34 \frac{1}{\omega ^2}\left(\frac{{\rm d}\omega }{{\rm
d}\tau }\right)^2-\frac12 \frac{1}{\omega }\frac{{\rm d}^2\omega
}{{\rm d}\tau ^2}\, .
\end{equation}
Therefore, one sees that the WKB mode function is a good approximation
to the actual one as soon as $\vert Q/w^2\vert \ll 1$. This last
condition defines the regime where the WKB approximation is valid. Let
us compute this quantity in the case of Eq.~(\ref{eom}). Straightforward
calculations lead to the following expression
\begin{equation}
\left\vert \frac{Q}{\omega ^2}\right\vert = \frac{1}{\Upsilon
^2}\frac{1}{2\left(1+\tau ^2/\Upsilon\right)^2}\left\vert\frac{5 \tau
^2/\Upsilon}{2\left(1+\tau ^2/\Upsilon\right)}-1\right\vert \, ,
\end{equation}
The quantity $\vert Q/\omega^2\vert $ is represented in
Fig.~\ref{fig:wkb}. It is clear that, in the limits $\tau
/\sqrt{\Upsilon }\rightarrow \pm \infty$, the WKB approximation is
valid. This means that there exists a well-defined vacuum state (or
adiabatic vacuum) in the ``in'' region, $\vert 0^-\rangle $, and in the
``out'' region, $\vert 0^+\rangle $.

\par

When the WKB approximation is satisfied, an approximate solution of the
mode functions is available and, as already briefly mentioned above, is
given by
\begin{eqnarray}
\label{approxsol}
\Phi _{\vec{k}}\left(\tau \right)&\simeq &\frac{\alpha
_{\vec{k}}}{\sqrt{2\omega }}\exp\left[- i\int_{\tau _{\rm ini}} ^{\tau
}\omega (\theta ){\rm d}\theta \right] +\frac{\beta
_{\vec{k}}}{\sqrt{2\omega }}\exp\left[+ i\int_{\tau _{\rm ini}} ^{\tau
}\omega (\theta ){\rm d}\theta \right] \nonumber \\ &\equiv & \alpha
_{\vec{k}}\Phi _{\rm{wkb}, \vec{k}}\left(\tau \right) +\beta
_{\vec{k}}\Phi _{\rm{wkb}, \vec{k}}^*\left(\tau \right) \, ,
\end{eqnarray}
where $\tau _{\rm ini}<0$ is some arbitrary initial time. 

\par

One can now use the WKB approximation in order to choose an initial
state in the following way. We require that the system is in the
adiabatic vacuum in the ``in'' region, $\vert 0^-\rangle$, that is to
say when $\tau /\sqrt{\Upsilon }\rightarrow -\infty$. Technically, this
means that one has $\lim _{\tau/\sqrt{\Upsilon }\rightarrow -\infty
}\Phi _{\vec{k}} =\Phi _{\rm{wkb}, \vec{k}}$ or $\alpha _{\vec{k}}=1$
and $\beta _{\vec{k}}=0$ (hence satisfying $\left\vert \alpha
_{\vec{k}}\right\vert ^2-\left\vert \beta _{\vec{k}}\right\vert
^2=1$). This criterion completely specifies the coefficients
$A_{\vec{k}}$ and $B_{\vec{k}}$ in Eq.~(\ref{exactsol}) and, as a
consequence, also completely determines the coefficients $\alpha
_{\vec{k}}$ and $\beta _{\vec{k}}$ in the ``out'' region (when $\tau
/\sqrt{\Upsilon }\rightarrow +\infty$) that are needed in order to
compute the transition amplitude. Around $\tau \sim 0$, see
Fig.~\ref{fig:wkb}, the WKB approximation is violated and we have
particle creation. In the ``out'' region the vacuum is defined by $\vert
0^+\rangle $ and, therefore, the number of particles present in this
region is measured by evaluating the amplitude $\langle 0^-\vert
0^+\rangle $.

\par

We now briefly explain how this can be done at the technical level. The
phase can be computed exactly and reads
\begin{eqnarray}
\int_{\tau _{\rm ini}} ^{\tau }\omega (\theta ){\rm d}\theta
&=&\frac{\Upsilon }{2}\Biggl[\frac{\tau }{\sqrt{\Upsilon }}
\sqrt{1+\frac{\tau ^2}{\Upsilon }}-\frac{\tau _{\rm
ini}}{\sqrt{\Upsilon }} \sqrt{1+\frac{\tau ^2_{\rm ini}}{\Upsilon }}
\nonumber \\ & & +\ln \left(\frac{\tau }{\sqrt{\Upsilon
}}+\sqrt{1+\frac{\tau ^2}{\Upsilon }}\right) -\ln \left(\frac{\tau
_{\rm ini}}{\sqrt{\Upsilon }}+\sqrt{1+\frac{\tau _{\rm
ini}^2}{\Upsilon }}\right)\Biggr]\, .
\end{eqnarray}
The arguments of the logarithms are always positive even if $\tau $ is
negative (hence the corresponding expression with $\tau _{\rm ini}<0$ is
also meaningful). In the limit $\left \vert \tau \right \vert \Upsilon
^{-1/2}\gg 1$, the phase goes to
\begin{equation}
\int_{\tau _{\rm ini}} ^{\tau }\omega (\theta ){\rm
d}\theta\rightarrow \frac{1}{2}\left(\tau \left\vert \tau \right\vert
+\left \vert \tau _{\rm ini}\right\vert ^2\right) +\frac{\Upsilon}{2}
\left(\frac{\vert \tau \vert }{\tau }\ln \left \vert \tau \right \vert
+\ln \left \vert \tau _{\rm ini} \right \vert \right)
\end{equation}
and, therefore, the WKB mode function takes the form
\begin{equation}
\Phi _{\rm{wkb}, \vec{k}}=\frac{1}{\sqrt{2}} {\rm e}^{-i\left(\tau
\vert \tau \vert +\tau _{\rm ini}^2\right)/2} \left\vert \tau
\right\vert ^{-1/2-i\vert \tau \vert \Upsilon/(2\tau )}\left \vert
\tau _{\rm ini}\right \vert ^{-i\Upsilon /2}\, .
\end{equation}
Then, in the limit $\tau \rightarrow -\infty $, the exact solution
given by Eq.~(\ref{exactsol}) can be expressed as
\begin{eqnarray}
\Phi _{\vec{k}} &\simeq & A_{\vec{k}}\frac{\sqrt{2\pi
}}{\Gamma\left(\frac12+\frac{i\Upsilon }{2}\right)} {\rm
e}^{-i\pi\left(1-i\Upsilon
\right)/4}\left(1-i\right)^{-\left(1-i\Upsilon\right)/2} \sqrt{2}\,
{\rm e}^{i\left \vert \tau _{\rm ini}\right \vert ^2/2} \left\vert
\tau _{\rm ini} \right\vert ^{i\Upsilon /2}\Phi _{\rm{wkb},
\vec{k}}\nonumber \\
&+&
\left[A_{\vec{k}}{\rm
e}^{i\pi\left(1+i\Upsilon
\right)/2}+B_{\vec{k}}\right]\left(1+i\right)^{-\left(1+i\Upsilon\right)/2}
\sqrt{2}\, {\rm e}^{-i\left \vert \tau _{\rm ini}\right \vert
^2/2}\left\vert \tau _{\rm ini} \right\vert ^{-i\Upsilon /2}\Phi
_{\rm{wkb}, \vec{k}}^* \, .  \nonumber \\
\end{eqnarray}
Since, as explained above, we choose the initial state such that $\alpha
_{\vec{k}}=1$ and $\beta _{\vec{k}}=0$, this amounts to requiring
\begin{eqnarray}
\label{Aschwinger}
A_{\vec{k}} &=& \frac{\Gamma\left(\frac12+\frac{i\Upsilon
}{2}\right)}{\sqrt{2\pi }} {\rm e}^{i\pi\left(1-i\Upsilon
\right)/4}\left(1-i\right)^{\left(1-i\Upsilon\right)/2}\frac{1}{\sqrt{2}}
{\rm e}^{-i\left \vert \tau _{\rm ini}\right \vert ^2/2} \left\vert
\tau _{\rm ini} \right\vert ^{-i\Upsilon /2}\, ,\\
\label{Bschwinger}
B_{\vec{k}}&=&-A_{\vec{k}} {\rm e}^{i\pi \left(1+i\Upsilon\right)/2}\,
.
\end{eqnarray}
Finally, one considers the behavior of the mode function in the limit
$\tau \rightarrow +\infty $ and, using again the WKB mode function is
this regime, one can find the corresponding coefficients $\alpha
_{\vec{k}}$ and $\beta _{\vec{k}}$. One obtains
\begin{equation}
\left \vert \alpha _{\vec{k}}\right \vert ^2=1+{\rm e}^{-\pi \Upsilon
} \, ,\quad \left \vert \beta _{\vec{k}}\right \vert ^2={\rm e}^{-\pi
\Upsilon }\, .
\end{equation}
These expressions still satisfy $\left\vert \alpha _{\vec{k}}\right\vert
^2-\left\vert \beta _{\vec{k}}\right\vert ^2=1$ as required.

\par

We have now reached our final goal and can return to the calculation of
the determinant in Eq.~(\ref{eq:det}). Using the coefficient $\alpha
_{\vec{k}}$ obtained above, one has
\begin{eqnarray}
\left\vert \left\langle 0^-\vert 0^+\right \rangle \right \vert
^2 &=& \det \left(\frac{1}{1+{\rm e}^{-\pi \Upsilon
}}\right)=\exp\left[-{\rm Tr} \ln \left(1+{\rm e}^{-\pi \Upsilon
}\right)\right]\,
.
\end{eqnarray}
The evaluation of the trace is standard and leads to the well-known
result first obtained by Schwinger in the early fifties~\cite{schwinger}
\begin{eqnarray}
\label{schwingerresult}
\left\vert \left\langle 0^-\vert 0^+\right \rangle \right \vert
^2 &=& \exp\left[-\frac{VT}{(2\pi
)^3}\sum_{n=1}^{\infty}\frac{(-1)^{n+1}}{n^2} \left(qE\right)^2{\rm
e}^{-n\pi m^2/(qE)} \right]\, .
\end{eqnarray}
The physical interpretation of this formula is clear. The argument of
the exponential gives the number of pairs created in the space-time
volume $VT$ due to the interaction of the quantum scalar field with the
classical electric field. One can define a critical electric field (we
have restored the fundamental constants) by
\begin{equation}
E_{\rm cri}=\frac{m^2c^3}{q\hbar}\, ,
\end{equation}
which is such that the number of particles created is significant only
if $E\gg E_{\rm cri}$. This condition can be understood by noting that
this is just the requirement that the work performed by the force $qE$
over the Compton length $\lambda =\hbar /(mc)$ is larger than the rest
energy $2mc^2$. In the case of pairs $e^+e^-$, the critical electric
field is given by $E_{\rm cri}\sim 1.3\times 10^{18}\, \mbox{V}\times
\mbox{m}^{-1}$. It is also interesting to remark that the dependence of
$\left\vert \left\langle 0^-\vert 0^+\right \rangle \right \vert ^2$ in
$E$ is non-perturbative. This is one of the few example in quantum field
theory where an exact result can be obtained (of course, this is not
``full quantum theory'' but rather ``potential theory'' since the
radiative corrections to the Schwinger mechanism are not taken into
account).

\par

We will see that the inflationary mechanism of production of
cosmological perturbations is similar to the Schwinger mechanism.
Therefore, observing this latter effect in the laboratory could be seen
as an indication that we are on the right track as far as the
inflationary mechanism is concerned. For instance, at DESY, there are
plans to construct a Free Electron Laser (FEL) in the X-ray band which
would effectively produce a very strong electric field and, hence, to
observe the Schwinger mechanism~\cite{Ring,AHRSV}. Unfortunately, even
with a FEL, it is inconceivable to produce a static field with the
required strength given present day technology. However, the situation
is different for an oscillating electric field~\cite{IZ} (other
configurations have been studied in Refs.~\cite{FGKA,AFY,FY}) and, in
this case, it seems possible to extract pairs from the vacuum. This
would also be a validation of the Schwinger mechanism since only the
time-dependence of $\omega (k,t)$ is changed but not the other basic
ingredients. It is also interesting to notice that, in the context of
the inflationary theory, this case is in fact very similar to the
reheating~\cite{turner, preheating} stage where the effective frequency
of the perturbations is alternating due to the oscillations of the
inflaton field at the bottom of its potential.

\par

To conclude this section, let us recall that the basic ingredient at
play here is particle creation due to the interaction of a quantum field
with a classical source. When the WKB approximation is valid, a
well-defined notion of vacuum state exists, and when the WKB
approximation is violated, particle creation occurs. We will see that,
in the case of inflationary cosmological perturbations, exactly the same
mechanism is available.

\section{Quantization of a Free Scalar Field in Curved Space-time}
\label{sec:quantizationsf}

Before considering inflation itself, let us now discuss the case of a
free real scalar field in curved space-time since this is the simplest
example which allows us to capture all the essential features of the
theory of inflationary cosmological perturbations.

\subsection{General Formalism}
\label{subsec:sfgeneral}

We consider the question of quantizing a (massless) scalar field in
curved space-time. The starting point is the following action
\begin{equation}
S=-\int {\rm d}^4x \sqrt{-g}g^{\mu \nu}\frac12 \partial _{\mu }\Phi
\partial _{\nu }\Phi \, ,
\end{equation}
which, in a flat FLRW Universe whose metric is given by ${\rm
d}s^2=a^2(\eta )(-{\rm d}\eta ^2+\delta _{ij}{\rm d}x^i{\rm d}x^j)$,
$\eta $ being the conformal time, reads
\begin{equation}
\label{actionsf2}
S=\frac12 \int {\rm d}^4x a^2(\eta )\left(\Phi '^2 -\delta ^{ij}\partial
_i\Phi \partial _j\Phi \right)\, .
\end{equation} 
It follows immediately that the conjugate momentum to the scalar 
field can be expressed as
\begin{equation}
\Pi (\eta ,\vec{x})=a^2\Phi '(\eta ,\vec{x})\, ,
\end{equation}
where a prime denotes a derivative with respect to conformal time. As
before, it is convenient to Fourier expand the field $\Phi (\eta
,\vec{x})$ over the basis of plane waves (therefore, we make explicit
use of the fact that the space-like hyper-surfaces are flat). This gives
\begin{equation}
\label{decompgw}
\Phi (\eta ,\vec{x})=\frac{1}{a(\eta )} \frac{1}{(2\pi )^{3/2}}\int
{\rm d}\vec{k} \mu _{\vec{k}}(\eta ){\rm e}^{i\vec{k}\cdot \vec{x}}\,
.
\end{equation} 
We have chosen to re-scale the Fourier component $\mu _{\vec{k}}$ with a
factor $1/a(\eta )$ for future convenience. Since the scalar field is
real, one has $\mu _{\vec{k}}^*=\mu _{-\vec{k}}$. The next step consists
in inserting the expression of $\Phi (\eta ,\vec{x})$ into the
action~(\ref{actionsf2}). This leads to 
\begin{eqnarray}
S &=& \frac{1}{2} \int {\rm d}\eta \int _{\setR ^{3+}}{\rm
d}^3\vec{k} \biggl[\mu _{\vec{k}}'{}^*\mu _{\vec{k}}'+ \mu
_{\vec{k}}'\mu _{\vec{k}}'{}^* -2\frac{a'}{a}\left(\mu _{\vec{k}}'\mu
_{\vec{k}}^*+\mu _{\vec{k}}'{}^*\mu _{\vec{k}}\right) \nonumber \\ & &
+\biggl(\frac{a'{}^2}{a^2}-k^2\biggr) \left(\mu _{\vec{k}}\mu
_{\vec{k}}^*+ \mu _{\vec{k}}^*\mu _{\vec{k}}\right)\biggr]\, .
\end{eqnarray}
Notice that the integral over the wave-numbers is calculated over half
the space in order to sum over independent variables
only~\cite{cohen}. This formula is similar to
Eq.~(\ref{eq:actionfourier}) for the case of the Schwinger effect (of
course, in this last case, we do not have $\Phi _{\vec{k}}=\Phi
_{-\vec{k}}^*$ since the field is charged and, hence, the integral is
performed over all the momentum space).

\par 

Equipped with the Lagrangian in the momentum space (which, in the
following, as it was the case in the previous Section, we denote by
$\bar{\cal L}$), one can check that it leads to the correct equation of
motion. Since we have $\delta \bar{\cal L}/\delta \mu
_{\vec{k}}^*=1/2[-2{\cal H} \mu _{\vec{k}}'{}+2({\cal H}^2-k^2)\mu
_{\vec{k}}'{}]$, the Euler-Lagrange equation $ {\rm d}[\delta \bar{\cal
L}/\delta \mu _{\vec{k}}'{}^*]/{\rm d}\eta -\delta \bar{\cal L}/\delta
\mu _{\vec{k}}^*=0$ reproduces the correct equation of motion for the
variable $\mu _{\vec{k}}$, namely
\begin{equation}
\label{eq:motion}
\frac{{\rm d}^2\mu _{\vec{k}}}{{\rm d}\eta ^2}+
\omega ^2(k,\eta )\mu _{\vec{k}}
=0\, ,
\end{equation}
that is to say, again, the equation of a parametric oscillator, as in
Eq.~(\ref{eqphi}), but with a frequency now given by
\begin{equation}
\omega ^2(k,\eta )=k^2-\frac{a''}{a}\, .
\end{equation}
This last formula should be compared with Eq.~(\ref{eq:defomega}). In
the case of the Schwinger effect, the frequency was time-dependent
because of the interaction of the scalar field with the time-dependent
potential vector. Here, the frequency is time-dependent because the
scalar field lives in a time-dependent background, or, in some sense,
because the scalar field interacts with the classical gravitational
background. Therefore, we already see at this stage that we can have
particle creation due to the interaction with a classical gravitational
field (instead of a classical electric field in the previous
Section). Of course, the two cases are not exactly similar in the sense
that the time dependence of $\omega ^2$ is different. Indeed, in the
Schwinger case, $\omega ^2(k,t)$ typically contains terms proportional
to $t$ and $t^2$, see Eq.~(\ref{eq:defomega}), while, in the
inflationary case, the term $a''/a$ is typically proportional to $1/\eta
^2$. As a consequence, the solution to the mode equation and the
particle creation rate will be different even if, again, the basic
mechanism at play is exactly the same in both situations.

\par

The mode amplitude $\mu _{\vec{k}}$ is complex but one can also work
with real variables $\mu ^{_{\mathrm R}}_{\vec{k}}$ and $\mu
^{_{\mathrm I}}_{\vec{k}}$, as was done previously in
Eq.~(\ref{defphireal}), defined such that
\begin{equation}
\mu _{\vec{k}}\equiv \frac{1}{\sqrt{2}}\left(\mu ^{_{\mathrm
R}}_{\vec{k}}+ i\mu ^{_{\mathrm I}}_{\vec{k}}\right)\, .
\end{equation}
In terms of these variables, the relation $\mu _{\vec{k}}^*=\mu
_{-\vec{k}}$ reads $\mu ^{_{\mathrm R}}_{\vec{k}}=\mu ^{_{\mathrm
R}}_{-\vec{k}}$ and $\mu ^{_{\mathrm I}}_{\vec{k}}=-\mu ^{_{\mathrm
I}}_{-\vec{k}}$. Then, the action (or Lagrangian) of the system takes
the form
\begin{eqnarray}
\label{lagrangianreal}
S &=& \frac{1}{2} \int {\rm d}\eta \int _{\setR ^{3+}}{\rm d}^3\vec{k}
\biggl\{\left(\mu ^{_{\mathrm R}}_{\vec{k}}{}'\right)^2 +\left(\mu
^{_{\mathrm I}}_{\vec{k}}{}'\right)^2 -2\frac{a'}{a}\left(\mu
^{_{\mathrm R}}_{\vec{k}}\mu ^{_{\mathrm R}}_{\vec{k}}{}'+\mu
^{_{\mathrm I}}_{\vec{k}}\mu ^{_{\mathrm I}}_{\vec{k}}{}' \right)
\nonumber \\ & & +\biggl(\frac{a'{}^2}{a^2}-k^2\biggr) \left[
\left(\mu ^{_{\mathrm R}}_{\vec{k}}\right)^2 +\left(\mu ^{_{\mathrm
I}}_{\vec{k}}\right)^2 \right]\biggr\}\, .
\end{eqnarray}
One can check that it also leads to the correct equations of motion
for the two real variables $\mu ^{_{\mathrm R}}_{\vec{k}}$ and $\mu
^{_{\mathrm I}}_{\vec{k}}$.

\par

We can now pass to the Hamiltonian formalism. The conjugate momentum to
$\mu _{\vec{k}}$ is defined by the formula
\begin{equation}
\label{momentum}
p_{\vec{k}}\equiv \frac{\delta \bar{{\cal L}}}{\delta \mu
_{\vec{k}}'{}^*} =\mu _{\vec{k}}'-\frac{a'}{a} \mu _{\vec{k}}\, .
\end{equation}
One can check that the definitions of the conjugate momenta in the real
and Fourier spaces are consistent in the sense that they are related by
the (expected) expression
\begin{equation}
\Pi (\eta ,\vec{x})=\frac{a(\eta )}{(2\pi )^{3/2}}\int {\rm
d}{\vec{k}}\, p_{\vec{k}}{\rm e}^{i{\vec{k}}\cdot {\vec{x}}}\, .
\end{equation}
We see that the definition of the conjugate momentum $p_{\vec{k}}$ as
the derivative of the Lagrangian in Fourier space with respect to $\mu
_{\vec{k}}'{}^*$ and not to $\mu _{\vec{k}}'$ is consistent with the
expression of the momentum in real space. Otherwise the momentum $\Pi
(\eta ,\vec{x})$ in real space would have been expressed in terms of
$p_{\vec{k}}^*$ instead of $p_{\vec{k}}$. Moreover, one can also check
that
\begin{equation}
p^{_{\mathrm R}}_{\vec{k}}\equiv \frac{\delta \bar{{\cal L}}}{\delta
\mu ^{_{\mathrm R}}_{\vec{k}}{}'}= \mu ^{_{\mathrm R}}_{\vec{k}}{}'
-\frac{a'}{a} \mu ^{_{\mathrm R}}_{\vec{k}}\, , \quad p^{_{\mathrm
I}}_{\vec{k}}\equiv \frac{\delta \bar{{\cal L}}}{\delta \mu
^{_{\mathrm I}}_{\vec{k}}{}'}= \mu ^{_{\mathrm I}}_{\vec{k}}{}'
-\frac{a'}{a} \mu ^{_{\mathrm I}}_{\vec{k}} \, ,
\end{equation}
and, clearly, we have 
\begin{equation}
p_{\vec{k}}\equiv \frac{1}{\sqrt{2}}\left(p ^{_{\mathrm
R}}_{\vec{k}}+ ip^{_{\mathrm I}}_{\vec{k}}\right)\, ,
\end{equation}
as expected.

\par

We are now in a position where we can compute explicitly the Hamiltonian
in the momentum space. The Hamiltonian density, $\bar{\cal H}$, is
defined in terms of the Hamiltonian $H$ of the system through the
relation
\begin{equation}
H=\int _{\setR ^{3+}}{\rm d}^3\vec{k}\bar{\cal H}=\int _{\setR
^{3+}}{\rm d}^3\vec{k} \left(p_{\vec{k}}\mu
_{\vec{k}}'{}^*+p_{\vec{k}}^*\mu _{\vec{k}}' -\bar{\cal L}\right)\, ,
\end{equation}
and we obtain
\begin{equation}
\label{Hamil}
\bar{\cal H}=p_{\vec{k}}p_{\vec{k}}^*+k^2\mu _{\vec{k}} \mu
_{\vec{k}}^*+\frac{a'}{a}\left(p_{\vec{k}}\mu _{\vec{k}}^*
+p_{\vec{k}}^*\mu _{\vec{k}}\right)\, .
\end{equation}
Let us make some comments on this expression. If the background
gravitational field is not time-dependent, that is to say if the scalar
field lives in Minkowski space-time where $a'=0$, then the above
Hamiltonian reduces to a free Hamiltonian : there is simply no classical
``pump field''. From the Schwinger effect point of view, this would be
similar to a situation where there is no external classical electric
field. In these two cases, no particle creation would occur. Moreover,
one can also check that the Hamilton equations
\begin{equation}
\frac{{\rm d}\mu _{\vec{k}}^*}{{\rm d}\eta}
=\frac{\partial \bar{\cal H}}{\partial p_{\vec{k}}}
=cp_{\vec{k}}^*+\frac{a'}{a}\mu _{\vec{k}}^*\, ,
\quad
\frac{{\rm d}p_{\vec{k}}^*}{{\rm d}\eta}
=-\frac{\partial \bar{\cal H}}{\partial \mu _{\vec{k}}}
=-\frac{a'}{a}p_{\vec{k}}^*-\frac{k^2}{c}\mu _{\vec{k}}^*\, ,
\end{equation}
lead to the correct equation of motion given by Eq.~(\ref{eq:motion}).
Finally, in terms of the real variables, the Hamiltonian density reads
\begin{equation}
\label{Hamilreal}
\bar{\cal H}=\frac12\left[
\left(p ^{_{\mathrm
R}}_{\vec{k}}\right)^2+2\frac{a'}{a}
\mu ^{_{\mathrm
R}}_{\vec{k}}p ^{_{\mathrm
R}}_{\vec{k}}+k^2\left(\mu ^{_{\mathrm
R}}_{\vec{k}}\right)^2\right]+
\frac12\left[
\left(p ^{_{\mathrm
I}}_{\vec{k}}\right)^2+2\frac{a'}{a}
\mu ^{_{\mathrm
I}}_{\vec{k}}p ^{_{\mathrm
I}}_{\vec{k}}+k^2\left(\mu ^{_{\mathrm
I}}_{\vec{k}}\right)^2\right]\, .
\end{equation}
We notice that $\bar{\cal H}$ is simply the sum of two identical
Hamiltonians for parametric oscillator, one for $\mu ^{_{\mathrm
R}}_{\vec{k}}$ and the other for $\mu ^{_{\mathrm I}}_{\vec{k}}$.

\par

The expressions~(\ref{Hamil}) and~(\ref{Hamilreal}) should be compared
to Eqs.~(\ref{eq:Hcomplex}) and (\ref{eq:Hreal}). We see that, although
similar, the formulae are not identical. However, as we are now going to
show, this difference is only apparent. Indeed, let us now restart from
the Lagrangian given by Eq.~(\ref{lagrangianreal}). One can always add a
total derivative without modifying the underlying theory. If one adds
the following term
\begin{equation}
\frac12\frac{{\rm d}}{{\rm d}\eta }\left[\frac{a'}{a}\left(\mu
^{_{\mathrm R}}_{\vec{k}}\right)^2+\frac{a'}{a}\left(\mu ^{_{\mathrm
I}}_{\vec{k}}\right)^2\right]\, ,
\end{equation}
then the Lagrangian takes the form
\begin{eqnarray}
\label{lagrangianreal2}
S &=& \frac{1}{2} \int {\rm d}\eta \int _{\setR ^{3+}}{\rm
d}^3\vec{k} \biggl\{\left(\mu ^{_{\mathrm
R}}_{\vec{k}}{}'\right)^2
+\left(\mu ^{_{\mathrm
I}}_{\vec{k}}{}'\right)^2-\omega ^2\left(k,\eta \right)
\left[
\left(\mu ^{_{\mathrm
R}}_{\vec{k}}\right)^2
+\left(\mu ^{_{\mathrm
I}}_{\vec{k}}\right)^2
\right]\biggr\}\, ,
\end{eqnarray}
where $\omega ^2(k,\eta)=k^2-a''/a$. In this case, the conjugate momenta
are simply given by $p ^{_{\mathrm R}}_{\vec{k}}=\mu ^{_{\mathrm
R}}_{\vec{k}}{}'$ and $p ^{_{\mathrm I}}_{\vec{k}}=\mu ^{_{\mathrm
I}}_{\vec{k}}{}'$. As a consequence, the Hamiltonian now reads
\begin{eqnarray}
\label{Hamilreal2}
H &=&\int _{\setR ^{3+}}{\rm d}^3\vec{k} \left\{\frac12 \left(\hat{p}
^{_{\mathrm R}}_{\vec{k}}\right)^2 +\frac12 \left(\hat{p} ^{_{\mathrm
I}}_{\vec{k}}\right)^2 +\frac12 \omega ^2\left(k,\eta
\right)\left[\left(\hat{\mu }^{_{\mathrm R}}_{\vec{k}}\right)^2
+\left(\hat{\mu }^{_{\mathrm I}}_{\vec{k}}\right)^2\right]\right\}\, .
\end{eqnarray}
This time, the Hamiltonian is exactly similar to the Schwinger
Hamiltonian given by Eq.~(\ref{eq:Hreal}). This is another manifestation
of the fact that, except for the exact time dependence of the effective
frequency, the physical phenomenon, namely particle creation under the
influence of an external classical field, is the same in both cases.

\par

Let us now investigate in more detail the relation between the
Hamiltonian given by Eq.~(\ref{Hamilreal}) and the one of
Eq.~(\ref{Hamilreal2}). We have just seen that the two corresponding
theories differ by a total derivative and, hence, are physically
equivalent. Another way to discuss the same property is through a
canonical transformation. For this purpose, let us consider the
following Hamiltonian
\begin{equation}
\label{Hamiltoy1}
H_1\left(p_1,q_1\right)=\frac12 p_1^2+\frac{a'}{a}p_1q_1+\frac12
k^2q_1^2 \, ,
\end{equation}
where $a(\eta )$ is an arbitrary function of the time. Clearly, $H_1$
plays the role of the Hamiltonian in Eq.~(\ref{Hamilreal}) and $a(\eta
)$ is the scale factor. Then, let us consider a canonical
transformation of type II~\cite{goldstein} such that
$(q_1,p_1)\rightarrow (q_2,p_2)$, the generating function of which is
given by (a similar transformation has also been studied in
Refs.~\cite{CR, PPP})
\begin{equation}
\label{generating}
G_2\left(q_1,p_2,\eta \right)=q_1p_2-\frac12 \frac{a'}{a}q_1^2\, .
\end{equation}
From this function, it is easy to establish the relation between the
``old'' variables and the ``new'' ones. One obtains
\begin{equation}
\label{transclass}
p_1=\frac{\partial G_2}{\partial q_1}=p_2-\frac{a'}{a}q_1\, ,\quad 
q_2=\frac{\partial G_2}{\partial p_2}=q_1 \, .
\end{equation}
In particular, the first relation reproduces Eq.~(\ref{momentum}) with
the correct sign. Finally, the ``new'' Hamiltonian is given by
\begin{equation}
H_2\left(p_2,q_2\right)=H_1+\frac{\partial G_2}{\partial \eta
}=\frac12 p_2^2+\frac12\left(k^2-\frac{a''}{a}\right)q_2^2\, .
\end{equation}
Clearly this Hamiltonian is similar to the Hamiltonian of
Eq.~(\ref{Hamilreal2}).

\par

Therefore, the two versions of the theory, the one given by the
Hamiltonian~(\ref{Hamilreal}), which is what we naturally obtain in
the case of cosmological perturbations (see Sec.~\ref{sec:pert}), and
the one which leads to the Hamiltonian~(\ref{Hamilreal2}) ``\`a la
Schwinger'' are simply connected by a canonical transformation and,
thus, are physically identical. In the following, we will see that
this is also the case at the quantum level.

\subsection{Quantization and the squeezed states formalism}
\label{sub:squeezed}

So far, the discussion has been purely classical. We now study the
quantization of the system starting with the Heisenberg picture. The
quantization in the functional picture that we used for the Schwinger
effect will be investigated in the next sub-Section. At the quantum
level, $\mu _{\vec{k}}$ and $p_{\vec{k}}$ become operators satisfying
the commutation relation
\begin{equation}
\label{commutfourier}
\left[\hat{\mu }_{\vec{k}}, \hat{p}_{\vec{p}}^{\dagger}\right]=i\delta
^{(3)}\left(\vec{k}-\vec{p}\right)\, .
\end{equation}
Clearly, factor ordering is now important. The quantum Hamiltonian is
obtained from the classical one by properly symmetrizing the
expression~(\ref{Hamil}). This leads to
\begin{equation}
\label{Hamilsym}
\hat{H}=\int _{\setR ^{3+}}{\rm d}^3\vec{k}
\left[\hat{p}_{\vec{k}}\hat{p}_{\vec{k}}^{\dagger}+k^2\hat{\mu}
_{\vec{k}} \hat{\mu }_{\vec{k}}^{\dagger }+\frac{a'}{2a}
\left(\hat{p}_{\vec{k}}\hat{\mu}
_{\vec{k}}^{\dagger} +\hat{\mu}
_{\vec{k}}^{\dagger}\hat{p}_{\vec{k}}
+\hat{p}_{\vec{k}}^{\dagger}\hat{\mu }_{\vec{k}}
+\hat{\mu }_{\vec{k}}\hat{p}_{\vec{k}}^{\dagger}
\right)\right]\, .
\end{equation}
In addition, this guarantees the hermiticity of the Hamiltonian. The
next step consists in introducing the normal variable
$\hat{c}_{\vec{k}}$~\cite{cohen} (which becomes the annihilation
operator, $\hat{c}_{\vec{k}}^{\dagger}$ becoming the creation operator)
defined by
\begin{equation}
\label{defnormal}
\hat{c}_{\vec{k}}(\eta )\equiv \sqrt{\frac{k}{2}}\hat{\mu }_{\vec{k}}+
\frac{i}{\sqrt{2k}}\hat{p}_{\vec{k}}\, . 
\end{equation}
Equivalently, one can also express $\hat{\mu }_{\vec{k}}$ and
$\hat{p}_{\vec{k}}$ in terms of the normal variable and its hermitic
conjugate. This gives the following two relations
\begin{equation}
\hat{\mu }_{\vec{k}}=\frac{1}{\sqrt{2k}}\left(\hat{c}_{\vec{k}}
+\hat{c}_{-\vec{k}}^{\dagger}\right)\, ,\quad \hat{p}_{\vec{k}}=
-i\sqrt{\frac{k}{2}}\left(\hat{c}_{\vec{k}}
-\hat{c}_{-\vec{k}}^{\dagger}\right)\, .
\end{equation}
Then, from the commutation relation~(\ref{commutfourier}), or
equivalently from the relation in real space $[\hat{\Phi }(\eta
,\vec{x}),\hat{\Pi }(\eta ,\vec{y})] =i\delta
^{(3)}(\vec{x}-\vec{y})$, it follows that $[c_{\vec{k}}(\eta ),
c_{\vec{p}}^{\dagger}(\eta )]=\delta ^{(3)}({\vec{k}}- {\vec{p}})$. In
terms of the normal variables, the scalar field and its conjugate
momentum can be expressed as
\begin{eqnarray}
\hat{\Phi} (\eta ,\vec{k}) &=& \frac{1}{a(\eta )} \frac{1}{(2\pi
)^{3/2}} \int \frac{{\rm d}{\vec{k}}}{\sqrt{2k}}\, \left[\hat{c}
_{\vec{k}}(\eta ){\rm e}^{i{\vec{k}}\cdot {\vec{x}}} +\hat{c}
_{\vec{k}}^{\dagger}(\eta ){\rm e}^{-i{\vec{k}}\cdot {\vec{x}}}\right]
\, , \\ \hat{\Pi }(\eta ,\vec{x}) &=& -\frac{a(\eta )}{(2\pi
)^{3/2}}\int {\rm d}{\vec{k}}\, i\sqrt{\frac{k}{2}} \left[\hat{c}
_{\vec{k}}(\eta ){\rm e}^{i{\vec{k}}\cdot {\vec{x}}} -\hat{c}
_{\vec{k}}^{\dagger}(\eta ){\rm e}^{-i{\vec{k}}\cdot {\vec{x}}}\right]
\, .
\end{eqnarray}
Obviously, when $a'=0$, we recover the flat space-time limit and so we
expect the time dependence of the normal variables to be just $\hat{c}
_{\vec{k}}(\eta )\propto {\rm e}^{ik\eta }$.

\par

We can now calculate the Hamiltonian operator in terms of the creation
and annihilation operators. Using Eq.~(\ref{Hamilsym}) one obtains
\begin{equation}
\hat{H}=\frac{1}{2}\int _{\setR ^3}{\rm d}^3\vec{k} \left[
k\left(c_{\vec{k}}c_{\vec{k}}{}^{\dagger } +c_{-{\vec{k}}}{}^{\dagger
}c_{-{\vec{k}}}\right ) -i \frac{a'}{a}\left(c_{\vec{k}}c_{-{\vec{k}}}
-c_{-{\vec{k}}}{}^{\dagger }c_{{\vec{k}}}{}^{\dagger }\right
)\right]\, ,
\end{equation}
where it is important to notice that the integral is now calculated in
$\setR^3$ and not in $\setR^{3+}$. Let us analyze this Hamiltonian. The
first term is the standard one and represents a collection of harmonic
oscillators. The most interesting part is the second term. This term is
responsible for the quantum creation of particles in curved
space-time. It can be viewed as an interacting term between the scalar
field and the classical background. The coupling function $ia'/a$ is
proportional to the derivative of the scale factor and, therefore,
vanishes in flat space-time. From the structure of the interacting term,
i.e. in particular the product of two creation operators for the mode
${\vec k}$ and $-{\vec{k}}$, we can also see that we have creation of
pairs of quanta with opposite momenta during the cosmological expansion
(thus momentum is conserved as it should), exactly as we had particle
creation due to the interaction of the scalar field with a classical
electric field in the previous section.

\par

We can now calculate the time evolution of the quantum operators (here,
we are working in the Heisenberg picture). Everything is known if we can
determine the temporal behavior of the creation and annihilation
operators; this behavior is given by the Heisenberg equations which read
\begin{equation}
\frac{{\rm d}c_{{\vec{k}}}}{{\rm d}\eta
}=-i\left[c_{{\vec{k}}},\hat{H}\right]\, , \quad \frac{{\rm
d}c_{{\vec{k}}}{}^{\dagger }}{{\rm d}\eta }
=-i\left[c_{{\vec{k}}}{}^{\dagger },\hat{H}\right]\, .
\end{equation}
Inserting the expression for the Hamiltonian derived above, we arrive at
the equations
\begin{equation}
i\frac{{\rm d}c_{{\vec{k}}}}{{\rm d}\eta }=kc_{{\vec{k}}}
+i\frac{a'}{a}c_{-{\vec{k}}}{}^{\dagger }\, ,\quad i\frac{{\rm
d}c_{{\vec{k}}}{}^{\dagger }}{{\rm d}\eta } =-kc_{{\vec{k}}}{}^{\dagger
} +i\frac{a'}{a}c_{-{\vec{k}}}\, .
\end{equation}
This system of equations can be solved by means of a Bogoliubov
transformation and the solution can be written as
\begin{eqnarray}
\label{Bogtrans}
c_{\vec{k}}(\eta ) &=& u_k(\eta )c_{\vec{k}}(\eta _{\rm ini})
+v_k(\eta )c_{-{\vec{k}}}{}^{\dagger }(\eta _{\rm ini})\, ,
\\
\label{Bogtrans2}
c_{\vec{k}}{}^{\dagger }(\eta ) &=& u_k^*(\eta )
c_{\vec{k}}{}^{\dagger }(\eta _{\rm ini})
+v_k^*(\eta )c_{-{\vec{k}}}{}(\eta _{\rm ini})\, ,
\end{eqnarray}
where $\eta _{\rm ini}$ is a given initial time and where the
functions $u_k(\eta )$ and $v_k(\eta )$ satisfy the equations
\begin{equation}
\label{eqsuv}
i\frac{{\rm d}u_k(\eta )}{{\rm d}\eta }=ku_k(\eta )
+i\frac{a'}{a}v_k^*(\eta )\, ,\quad 
i\frac{{\rm d}v_k(\eta )}{{\rm d}\eta }=kv_k(\eta )
+i\frac{a'}{a}u_k^*(\eta )\, .
\end{equation}
In addition, these two functions must satisfy $\vert u_k\vert ^2 -\vert
v_k\vert ^2=1$ such that the commutation relation between the creation
and annihilation operators is preserved in time. A very important fact
is that the initial values of $u_k$ and $v_k$ are fixed and, from the
Bogoliubov transformation, read
\begin{equation}
\label{iniquant}
u_k(\eta _{\rm ini})=1\, , \quad v_k(\eta _{\rm ini})=0\, .
\end{equation}
Therefore, we remark that, in some sense, the initial conditions are
fixed by the procedure of quantization. In fact, Eqs.~(\ref{iniquant})
imply that the initial state has been chosen to be the vacuum $\vert
0\rangle$ at time $\eta =\eta _{\rm ini}$. {\it A priori}, it is not
obvious that this choice is well-motivated but it turns out to be the
case in an inflationary universe. This property constitutes one of the
most important aspect of the inflationary scenario. Here, we do not
discuss further this issue but we will come back to the problem of
fixing the initial conditions at the beginning of inflation in the
following.

\par

At this point, the next move is to establish the link between the
formalism exposed above and the classical picture. For this purpose,
it is interesting to establish the equation of motion obeyed by the
function $u_k+v_k^*$. Straightforward manipulations from
Eqs.~(\ref{eqsuv}) lead to
\begin{equation}
\label{eomquant}
\left(u_k+v_k^*\right)''
+\biggl(k^2-\frac{a''}{a}\biggr)\left(u_k+v_k^*\right)=0\, .
\end{equation}
We see that the function $u_k+v_k^*$ obeys the same equation as the
variable $\mu _{\vec{k}}$. This is to be expected since, using the
Bogoliubov transformation, the scalar field operator can be re-written
as
\begin{eqnarray}
\hat{\Phi }(\eta ,{\bf x}) &=& \frac{1}{a(\eta )}
\frac{1}{(2\pi )^{3/2}} \int \frac{{\rm d}{\vec{k}}}{\sqrt{2k}}
\biggl[\left(u_k+v_k^*\right)(\eta )c_{\vec{k}}(\eta _{\rm ini}) {\rm
e}^{i{\vec{k}}\cdot {\vec{x}}} \nonumber \\ & &
+\left(u_k^{*}+v_k\right)(\eta )c_{\vec{k}}^{\dag}(\eta _{\rm ini})
{\rm e}^{-i{\vec{k}}\cdot{\vec{x}}}\biggr]\, .
\end{eqnarray}
Therefore, if we are given a scale factor $a(\eta )$, we can now
calculate the complete time evolution of the quantum scalar field by
means of the formalism presented above.

\par
 
In fact, the Bogoliubov transformation~(\ref{Bogtrans}) and
(\ref{Bogtrans2}) can be expressed in a different manner which is useful
in order to introduce the squeezed states formalism. For this purpose,
let us come back to the functions $u_k$ and $v_k$. We have seen that, in
order for the commutator of the creation and annihilation operators to
be preserved in time, these two functions must satisfy $\vert u_k\vert
^2 -\vert v_k\vert ^2=1$. This means that we can always write
\begin{equation}
u_k={\rm e}^{i\theta _k}\cosh r_k\, ,\quad 
v_k={\rm e}^{-i(\theta _k-2\phi _k)}\sinh r_k \, ,
\end{equation}
where the quantities $r_k$, $\theta _k$ and $\phi _k$ are functions of
time. They are called the squeezing parameter, rotation angle and
squeezing angle respectively. These functions obey the equations
\begin{eqnarray}
\frac{{\rm d}r_k}{{\rm d}\eta }&=&\frac{a'}{a}\cos 2\phi _k\, ,
\quad 
\frac{{\rm d}\phi _k}{{\rm d}\eta }=-k-\frac{a'}{a}\sin 2\phi _k
\coth 2r_k\, ,
\\
\frac{{\rm d}\theta _k}{{\rm d}\eta }&=&-k-\frac{a'}{a}\sin 2\phi _k
\tanh r_k \, .
\end{eqnarray}
These expressions can be used for an explicit calculation of $r_k$,
$\theta _k$ and $\phi _k$ when a specific scale factor $a(\eta )$ is
given.  Now, the crucial property is that the Bogoliubov
transformation~(\ref{Bogtrans}), (\ref{Bogtrans2}) which solves the
perturbed Einstein equations can be cast into the following
form~\cite{GS1,GS2,GS3,AFJP}
\begin{eqnarray}
\label{Boggeo}
c_{\vec{k}}(\eta ) &=& R(\theta )S(r, \varphi )c_{\vec{k}}(\eta _{\rm i})
S^{\dagger }(r, \varphi )R^{\dagger} (\theta )\, ,
\\
\label{Boggeo2}
c_{\vec{k}}^{\dagger }(\eta )&=&
R(\theta )S(r, \varphi )c_{\vec{k}}^{\dagger }(\eta _{\rm i})
S^{\dagger }(r, \varphi )R^{\dagger} (\theta )\, ,
\end{eqnarray}
where the operators $R(\theta )$ and $S(r, \varphi )$ are given by
\begin{eqnarray}
R(\theta )&=&\exp \biggl\{-i\theta _k\biggl[ c_{\vec{k}}^{\dagger
}(\eta _{\rm i}) c_{\vec{k}}(\eta _{\rm i}) + c_{-{\vec{k}}}^{\dagger
}(\eta _{\rm i}) c_{-{\vec{k}}}(\eta _{\rm i})\biggr]\biggr\}\, , \\
S(r, \varphi )&=&\exp \biggl\{r_k\biggl[{\rm e}^{-2i\phi _k}
c_{\vec{k}}^{\dagger }(\eta _{\rm i}) c_{\vec{k}}(\eta _{\rm i}) -{\rm
e}^{2i\phi _k} c_{-{\vec{k}}}^{\dagger }(\eta _{\rm i})
c_{-{\vec{k}}}(\eta _{\rm i})\biggr]\biggr\}\, .
\end{eqnarray}
Eqs.~(\ref{Boggeo}) and~(\ref{Boggeo2}) allows us to interpret the
Bogoluibov transformation in a new manner: indeed we can also see the
time evolution of the creation and annihilation operators as rotations
in the Hilbert space.

\par

The previous considerations are valid in the Heisenberg picture. What
happens in the Schr\"odinger picture where the operators no longer
evolve but the states become time-dependent? For the sake of simplicity,
let us ignore $\theta _k$ and $\phi _k$ by setting $\theta _k=\phi
_k=0$. As mentioned above, let us also postulate that the system is
originally placed in the vacuum state $\vert 0\rangle $. Then, the
previous results imply that, after the cosmological evolution, the mode
characterized by the wave-vector $\vec{k}$ will evolve into the
following state~\cite{GS1,GS2,GS3,AFJP}
\begin{equation}
\label{2mode}
\exp \biggl\{r_k\biggl[ c_{\vec{k}}^{\dagger }(\eta _{\rm i})
c_{\vec{k}}(\eta _{\rm i}) - c_{-{\vec{k}}}{}^{\dagger }(\eta _{\rm
i}) c_{-{\vec{k}}}(\eta _{\rm i})\biggr]\biggr\}\vert 0\rangle \, ,
\end{equation}
which is, by definition, a two-mode vacuum squeezed state. This state is
a very peculiar state and is of particular relevance in other branches
of physics as well, most notably in quantum optics~\cite{Shum}.

\par

We now discuss the properties of such a quantum state. For this purpose,
it is interesting to recall that a state containing a fixed number of
particles, $\vert n\rangle $, can be obtained by successive action of
the creation operator on the vacuum. Explicitly, one has
\begin{equation}
\vert n\rangle =\frac{(c_{\vec{k}}^{\dagger })^n}{\sqrt{n!}}  \vert
0\rangle \, .
\end{equation}
Let us also introduce the coherent (Glauber) quantum
state~\cite{cohen}. It is defined by the following expression
\begin{equation}
\label{coherent}
\vert \alpha \rangle ={\rm e}^{-\vert \alpha \vert ^2/2}
\sum _{n=0}^{\infty }\frac{\alpha ^n}{\sqrt{n!}}\vert n \rangle \, ,
\end{equation}
where $\alpha $ is a complex number. The coherent state is especially
important in quantum optics since they represent, in a sense to be
specified, the most classical state. We will come back to this question
in the last Section. Finally we also define two new operators $B$ and
$T$ by
\begin{equation}
\label{defBT}
B_{\vec{k}}=\frac{r}{2}\biggl[(c_{\vec{k}})^2 -(c_{\vec{k}}{}^{\dagger
  })^2\biggr]\, , \quad T_{\vec{k}}={\rm e}^{B_{\vec{k}}}\, ,
\end{equation}
where $r$ is a real number (in fact, our squeezing parameter). The
operators $B_{\vec{k}}$ and $T_{\vec{k}}$ possess various interesting
properties, in particular $B_{\vec{k}}$ is anti-unitary,
$B_{\vec{k}}^{\dagger }=-B_{\vec{k}}$, and, as a consequence,
$T_{\vec{k}}$ is unitary, $T_{\vec{k}}T_{\vec{k}}^{\dagger }=1$. The
general definition of a squeezed state $\vert s\rangle $ is given by
\begin{equation}
\label{defs}
\vert s \rangle \equiv T_{\vec{k}}^{\dagger }\vert \alpha \rangle 
={\rm e}^{-B_{\vec{k}}}\vert \alpha \rangle \, .
\end{equation}
Let us notice that this is the expression for a one-mode squeezed state
while, in Eq.~(\ref{2mode}), we have to deal with a two-mode squeezed
state (hence the presence of operators $c_{\vec{k}}$ and $c_{-\vec{k}}$
that arises from the fact that we have pair creation, while in the above
definition we only have operators $c_{\vec{k}}^2$ and
$c_{\vec{k}}^{\dagger }{}^2$). The properties of one and two-mode
squeezed states are similar and, here, for simplicity, we focus on the
one-mode state only. Moreover, in our case, $\vert \alpha \rangle =\vert
0\rangle $ which means that, in the cosmological case, we have a
two-mode vacuum state.

\par

Why is this state called a squeezed state? To answer this question, we
introduce two new operators that are linear combinations of the creation
and annihilation operators, namely
\begin{equation}
(c_{\vec{k}})_P\equiv \frac{1}{2}\left(c_{\vec{k}}+c_{\vec{k}}^{\dagger
}\right) \, \quad (c_{\vec{k}})_Q\equiv
\frac{1}{2i}\left(c_{\vec{k}}-c_{\vec{k}}^{\dagger }\right) \, .
\end{equation}
These new operators are annihilation and creation operators of standing
waves since, in a Fourier expansion of the field, they would stand in
front of $\cos k\eta $ and $\sin k\eta $ rather than ${\rm e}^{ik\eta }$
and ${\rm e}^{-ik\eta }$ in the case of the standard creation and
annihilation operators. Then, it is straightforward to demonstrate that
\begin{equation}
\langle s\vert (c_{\vec{k}})_P \vert s\rangle =\frac{\alpha +\alpha
^*}{2}{\rm e}^r \, , \quad \langle s\vert (c_{\vec{k}})_Q \vert
s\rangle = \frac{\alpha -\alpha ^*}{2i}{\rm e}^{-r}\, .
\end{equation}
Let us now calculate the mean value of the squares of these
operators. We have
\begin{equation}
\langle s\vert (c_{\vec{k}})_P^2 \vert s\rangle 
=\frac{{\rm e}^{2r}}{4}(\alpha ^2+\alpha ^*{}^2+2\alpha \alpha ^*+1)\, ,
\end{equation}
and a similar expression for $\langle s\vert (c_{\vec{k}})_Q^2 \vert
s\rangle $ (but with ${\rm e}^{-2r}$ instead of ${\rm e}^{2r}$). We are
now in a position where the dispersion in the squeezed state of the
operators $(c_{\vec{k}})_P$ and $(c_{\vec{k}})_Q$ can be calculated. One
finds
\begin{equation}
\label{squeeze}
\Delta (c_{\vec{k}})_P =\sqrt{\langle s\vert (c_{\vec{k}})_P^2 \vert
s\rangle - \langle s\vert (c_{\vec{k}})_P \vert s\rangle
^2}=\frac{{\rm e}^r}{2}\, ,\quad \Delta (c_{\vec{k}})_Q=\frac{{\rm
e}^{-r}}{2}\, ,
\end{equation}
and, therefore, from these equations one deduces that
\begin{equation}
\Delta (c_{\vec{k}})_P \Delta (c_{\vec{k}})_Q=\frac{1}{4}\, .
\end{equation}
We see that the lower bound of the Heisenberg uncertainty relations is
reached but, contrary to a coherent state, the dispersion is not equal
for the two operators. On the contrary, the dispersion can be very small
on one component and very large on the other hence the name ``squeezed
state''. In the cosmological situation, this is actually the
case. Indeed, Refs.~\cite{GS1,GS2,GS3} have shown that, for modes whose
wavelengths are of the order of the Hubble length today, that is to say
the modes that contribute the most to the ``large angle'' CMB multipoles
$C_{\ell }$ (corresponding to a frequency of $\omega \sim
10^{-17}\mbox{Hz}$), one has $r\sim 120$. From Eqs.~(\ref{squeeze}), we
see that this corresponds to a very strong squeezing, in fact much
larger than what can be achieved in the laboratory~\cite{GS3}.

\par

It is also clear that a strongly squeezed state is not a classical
state in the sense that it is very far from the coherent state for
which $\Delta (c_{\vec{k}})_P=\Delta (c_{\vec{k}})_Q$. On the other
hand, since the mean value of $N_{\vec{k}}=c_{\vec{k}}^{\dagger
}c_{\vec{k}}$ is given by
\begin{equation}
\left \langle s\left \vert N_{\vec{k}} \right \vert s\right \rangle 
=\sinh ^2 r\, ,
\end{equation} 
a strongly vacuum squeezed state contains a very large number of
particles and this criterion is often taken as a criterion of
classicality. Therefore, we see that the meaning of classicality for a
strongly squeezed state is a subtle
issue~\cite{GP,PS,GS1,GS2,GS3,SWW,KP} since different criterions seem to
give different answers. We will come back to this point in the last
Section of this review article.

\subsection{Quantization in the functional approach}
\label{subsec:functionalapproach}

Let us now discuss the quantization in the functional approach where
each Fourier mode is described by a wave-function. For this purpose, we
use the description in terms of real variables. This will allow us to
emphasize again the complete analogy that exists between the Schwinger
effect and the theory of inflationary cosmological perturbations of
quantum-mechanical origin. We restart from Eq.~(\ref{Hamilreal}) and,
since we deal with quantum operators, we symmetrize the corresponding
expressions. In this case, the quantum Hamiltonian reads
\begin{eqnarray}
\hat{H}&=&\int _{\setR ^{3+}}{\rm d}^3\vec{k} \Biggl[\frac12
\left(\hat{p} ^{_{\mathrm R}}_{\vec{k}}\right)^2+\frac{a'}{2a} \left(
\hat{\mu }^{_{\mathrm R}}_{\vec{k}}\hat{p} ^{_{\mathrm R}}_{\vec{k}}+
\hat{p} ^{_{\mathrm R}}_{\vec{k}}\hat{\mu }^{_{\mathrm R}}_{\vec{k}}
\right) +\frac{k^2}{2}\left(\hat{\mu }^{_{\mathrm
R}}_{\vec{k}}\right)^2 +\frac12 \left(\hat{p} ^{_{\mathrm
I}}_{\vec{k}}\right)^2\nonumber \\ & &+\frac{a'}{2a} \left( \hat{\mu
}^{_{\mathrm I}}_{\vec{k}}\hat{p} ^{_{\mathrm I}}_{\vec{k}}+ \hat{p}
^{_{\mathrm I}}_{\vec{k}}\hat{\mu }^{_{\mathrm I}}_{\vec{k}} \right)
+\frac{k^2}{2}\left(\hat{\mu }^{_{\mathrm
I}}_{\vec{k}}\right)^2\Biggr] \equiv H^{_{\mathrm R}}+H^{_{\mathrm
I}}\, .
\end{eqnarray}
We also have the following commutation relations that are compatible
with Eq.~(\ref{commutfourier})
\begin{equation}
\label{commutrealcosmo}
\left[\hat{\mu }^{_{\mathrm R}}_{\vec{k}}, \hat{p} ^{_{\mathrm
R}}_{\vec{p}}\right]=i\delta^{(3)}\left(\vec{k}-\vec{p}\right)\, ,
\quad \left[\hat{\mu }^{_{\mathrm I}}_{\vec{k}}, \hat{p} ^{_{\mathrm
I}}_{\vec{p}}\right]=i\delta^{(3)}\left(\vec{k}-\vec{p}\right)\, .
\end{equation}
In the Schr\"odinger picture, similarly to Eqs.~(\ref{representation}),
the above-mentioned operators admit the following representation
\begin{equation}
\hat{\mu }^{_{\mathrm R}}_{\vec{k}}\Psi=\mu ^{_{\mathrm R}}_{\vec{k}}
\Psi\, , \quad \hat{p}^{_{\mathrm R}}_{\vec{k}}\Psi=-i
\frac{\partial \Psi}{\partial \mu ^{_{\mathrm R}}_{\vec{k}}}\, .
\end{equation}
Therefore, one deduces that the Hamiltonian (here, the Hamiltonian for
the real part of $\mu _{\vec{k}}$, hence for a fixed Fourier mode) can
be written as
\begin{equation}
\label{Hamilcosmo}
H^{_{\mathrm R}}_{\vec{k}}\Psi=-\frac{1}{2}\frac{\partial
  ^2\Psi}{\partial \left(\mu ^{_{\mathrm
  R}}_{\vec{k}}\right)^2}-\frac{i}{2}\frac{a'}{a}\Psi
  -i\frac{a'}{a}\mu ^{_{\mathrm R}}_{\vec{k}}\frac{\partial
  \Psi}{\partial \mu ^{_{\mathrm R}}_{\vec{k}}}+\frac{k^2}{2}\left(\mu
  ^{_{\mathrm R}}_{\vec{k}}\right)^2\Psi\, .
\end{equation}
Again, if $a'=0$, we recover the Hamiltonian of an harmonic oscillator
(instead of the Hamiltonian of a parametric oscillator when $a'\neq
0$).

\par

Let us now study the ground state of the theory. As done in
Eq.~(\ref{wavefunction}), we have the following Gaussian state,
\begin{equation}
\label{wavefunctioncosmo}
\Psi _{\vec{k}}^{_{\mathrm R}}(\eta, \mu ^{_{\mathrm
R}}_{\vec{k}})=N_{\vec{k}}\left(\eta \right){\rm e}^{-\Omega
_{\vec{k}}\left(\eta \right) \left(\mu ^{_{\mathrm
R}}_{\vec{k}}\right)^2}\, ,
\end{equation}
where $N_{\vec{k}}$ and $\Omega _{\vec{k}}$ are two functions to be
determined. They are found by means of the Schr\"odinger equation
$i\partial _{\eta }\Psi _{\vec{k}}^{_{\rm R}}=H^{_{\mathrm
R}}_{\vec{k}}\Psi_{\vec{k}}^{_{\rm R}}$ that leads to
\begin{eqnarray}
\label{eqnorm}
i\frac{N_{\vec{k}}'}{N_{\vec{k}}} &=& \Omega
_{\vec{k}}-\frac{i}{2}\frac{a'}{a}\, , \quad
\label{eqphase}
\Omega _{\vec{k}}'= -2i\Omega ^2_{\vec{k}}-2\frac{a'}{a}\Omega
_{\vec{k}}+i\frac{k^2}{2}\, .
\end{eqnarray}
The analogy with Eqs.~(\ref{eq:NOm}) is obvious. We notice, however,
that the structure of the equations is not exactly similar. This is due
to the presence of the terms proportional to $a'/a$ in the
Hamiltonian~(\ref{Hamilcosmo}) that have no equivalent in the
Hamiltonian~(\ref{Hamilschwinger}). Below, we briefly come back to this
point. These equations can be integrated and the solutions read
\begin{eqnarray}
\label{Nom}
N_{\vec{k}}=\left(\frac{2\Re \Omega _{\vec{k}}}{\pi}\right)^{1/4} \,
,\quad \Omega
_{\vec{k}}=-\frac{i}{2}\frac{\left(f_{\vec{k}}/a\right)'}{\left(f
_{\vec{k}} /a\right)}\, ,
\end{eqnarray}
where $f_{\vec{k}}$ obeys the equation $f
_{\vec{k}}''+\left(k^2-a''/a\right)f_{\vec{k}}=0$. Therefore, the
integration of the equation controlling the time evolution of the mode
function leads to a complete determination of the quantum state of the
system in full agreement with what was discussed before in the case of
the Schwinger effect. Again, the fact that the solution for $\Omega
_{\vec{k}}$ is given in terms of the function $f_{\vec{k}}/a$ and not
only in terms of $f_{\vec{k}}$, as one could have guessed from
Eq.~(\ref{eq:solNOm}), is due to the presence of the terms proportional
to $a'/a$ in Eq.~(\ref{Hamilcosmo}).

\par

Let us briefly come back to the derivation of the above
solution. Eq.~(\ref{eqphase}) is a Ricatti equation and, therefore, can
be solved in the usual way, namely by transforming this non-linear first
order differential equation into a linear second order differential
equation. In order to find $N_{\vec{k}}$, one requires that the wave
function is normalized, that is to say
\begin{equation}
\int \Psi _{\vec{k}}^{_{\mathrm R}}\Psi_{\vec{k}}^{_{\mathrm R}}{}^*
{\rm d}\mu ^{_{\mathrm R}}_{\vec{k}}=1\, ,
\end{equation}
which leads to the previous expression of $N_{\vec{k}}$. Moreover, there
is also the following consistency check. The real part of the second of
Eqs.~(\ref{eqphase}) reads
\begin{equation}
\label{realeqphase}
\left(\Re \Omega _{\vec{k}}\right)'=4\Re \Omega _{\vec{k}}\times \Im
\Omega _{\vec{k}}-2\frac{a'}{a} \Re \Omega _{\vec{k}}\, ,
\end{equation} 
and the imaginary part of the first of Eqs.~(\ref{eqnorm}) can be
written as $N'_{\vec{k}}/N_{\vec{k}}=\Im \Omega _{\vec{k}}-a'/(2a)$. It
is straightforward to check that, inserting the above solution for
$N_{\vec{k}}$ into the last equation, precisely leads to
Eq.~(\ref{realeqphase}).

\par

Let us now come back to the remark made before that the structure of
Eqs.~(\ref{eqnorm}) is not exactly similar to what we have in the
Schwinger case due to the presence of the terms proportional to
$a'/a$. The reason is clearly that we have used the Hamiltonian given by
Eq.~(\ref{Hamilreal}) which contains such terms. But, obviously, one can
also use the Hamiltonian given by Eq.~(\ref{Hamilreal2}). Then, assuming
again the Gaussian form~(\ref{wavefunctioncosmo}) for the wave-function,
the Schr\"odinger equation reduces to
\begin{equation}
\label{schrodingersf}
i\frac{N'_{\vec{k}}}{N_{\vec{k}}}=\Omega _{\vec{k}}\, ,\quad \Omega
_{\vec{k}}'=-2i\Omega ^2_{\vec{k}}+\frac{i}{2}\omega ^2(k,\eta )\, .
\end{equation}
which are now exactly similar to Eqs.~(\ref{eq:NOm}). As a
consequence, the solutions are also the same and read
\begin{eqnarray}
\label{solomegacosmo}
N_{\vec{k}}=\left(\frac{2\Re \Omega _{\vec{k}}}{\pi}\right)^{1/4} \,
,\quad \Omega _{\vec{k}}=-\frac{i}{2}\frac{f'_{\vec{k}}}{f
_{\vec{k}}}\, ,
\end{eqnarray}
where $f_{\vec{k}}$ obeys the mode function equation $f
_{\vec{k}}''+\omega ^2f_{\vec{k}}=0$.

\par

In Sec.~\ref{subsec:sfgeneral}, we have established, at the classical
level, the equivalence between the two formulations discussed above,
that is to say the one based on the Hamiltonian~(\ref{Hamilreal}), which
leads to a Gaussian wave-function with $N_{\vec{k}}$ and $\Omega
_{\vec{k}}$ given by Eqs.~(\ref{Nom}), and the one based on the
Hamiltonian~(\ref{Hamilreal2}), which also leads to a Gaussian
wave-function but with $N_{\vec{k}}$ and $\Omega _{\vec{k}}$ now given by
Eqs.~(\ref{solomegacosmo}). We now study this link at the quantum level
and, for this purpose, we reconsider the simple model introduced after
Eq.~(\ref{Hamiltoy1}). In Sec.~\ref{subsec:sfgeneral}, we showed that
the two formulations are connected by a canonical transformation and the
question is now to implement this canonical transformation at the
quantum level~\cite{CR,PPP,AM,LY,KW,OS}. For this purpose, one must find
a unitary operator $\hat{{\cal U}}$ such that the relations
\begin{equation}
\label{transquantique}
\hat{q}_2=\hat{{\cal U}}\hat{q}_1\hat{{\cal U}}^{\dagger }\, ,\quad
\hat{p}_2=\hat{{\cal U}}\hat{p}_1\hat{{\cal U}}^{\dagger }\, ,
\end{equation}
exactly reproduce the classical analogues~(\ref{transclass}). A
natural candidate would be the following operator
\begin{equation}
\hat{\cal U}={\rm e}^{i\hat{G}_2}
=\exp\left[-\frac{i}{2}\frac{a'}{a}\hat{q}_1^2+
\frac{i}{2}\left(\hat{q}_1\hat{p}_1+\hat{p}_1\hat{q}_1\right)\right]\, ,
\end{equation}
where $G_2$ is generating function introduced in
Eq.~(\ref{generating}). However, as already remarked in Ref.~\cite{CR},
this choice is too naive and does not work. In order to understand what
is going on, let us introduce a generalized version of
Eq.~(\ref{Hamiltoy1}), following Eq.~(2.21) of Ref.~\cite{CR}, which at
the classical level reads
\begin{equation}
\label{Hamiltoy1_v2}
H_1\left(p_1,q_1\right)=\frac12 \beta _3p_1^2+\beta _2(\eta
)p_1q_1+\frac12 \beta _1k^2q_1^2 \, ,
\end{equation}
where for simplicity we consider that $\beta _3 $ and $\beta _1$ are
constant while $\beta _2$ is a time-dependent function (in
Ref.~\cite{CR}, all the $\beta _i$'s are time-dependent
functions). Clearly, our case corresponds to $\beta _1=\beta _3=1$ and
$\beta _2=a'/a$. Then, as before, we consider a canonical transformation
of type II such that $(q_1,p_1)\rightarrow (q_2,p_2)$ with the following
generating function
\begin{equation}
\label{generating2}
G_2\left(q_1,p_2,\eta \right)=\beta _3^{-1/2}q_1p_2-\frac{\beta
_2}{2\beta _3}q_1^2\, .
\end{equation}
Setting $\beta _1=\beta _3=1$ and $\beta _2=a'/a$ in the above
expression reproduces Eq.~(\ref{generating}) as expected. Performing
standard calculations, one finds that the relation between the ``old''
variables and the ``new'' ones reads
\begin{equation}
\label{transclass2}
p_1=\frac{\partial G_2}{\partial q_1}=\beta _3^{-1/2}p_2-\frac{\beta
_2}{\beta _3}q_1\, ,\quad q_2=\frac{\partial G_2}{\partial p_2}=\beta
_3^{-1/2}q_1 \, ,
\end{equation}
and that the ``new'' Hamiltonian can now be expressed as
\begin{equation}
H_2\left(p_2,q_2\right)=H_1+\frac{\partial G_2}{\partial \eta
}=\frac12 p_2^2+\frac12\left(\beta _1\beta _3k^2-\beta _2'-\beta
_2^2\right)q_2^2\, .
\end{equation}
Notice, in particular, that the coefficient $\beta_3$ is no longer
present in the term $p_2^2/2$. Then, in agreement with Eqs.~(2.22) and
(2.23) of Ref.~\cite{CR}, let us consider the following operator
\begin{equation}
\label{defU} \hat{{\cal U}}(\hat{q}_1,\hat{p}_1,\eta
)=\exp\left(-\frac{i}{2}\beta
_2\hat{q}_1^2\right)\exp\left[-\frac{i}{4}(\ln \beta
_3)\left(\hat{q}_1\hat{p}_1+\hat{p}_1\hat{q}_1\right)\right]\, .
\end{equation}
Inserting this operator in Eqs.~(\ref{transquantique}) and using the
Baker-Campbell-Hausdorff formula, ${\rm e}^{\hat{A}}\hat{B}{\rm
e}^{-\hat{A}}=\hat{B}+[\hat{A},\hat{B}]+\cdots $, leads to the
transformation
\begin{equation}
\label{trans}
\hat{p}_1=\beta _3^{-1/2}\hat{p}_2-\frac{\beta
_2}{\beta _3}\hat{q}_1\, ,\quad \hat{q}_2=\beta
_3^{-1/2}\hat{q}_1 \, ,
\end{equation}
namely exactly Eqs.~(\ref{transclass2}), but now at the quantum
level. Therefore, we conclude that $\hat{\cal U}$ in Eq.~(\ref{defU}) is
the operator generating the correct quantum canonical transformation. In
addition, as one can check with the help of Eq.~(\ref{generating2}),
this operator is different from ${\rm e}^{iG_2}$, in particular due to
the presence of the factor $\ln \beta _3$. Let us also notice that a
similar operator has been considered recently in Refs.~\cite{PPP, AM},
which carries out an investigation very relevant for what is discussed
here, and that a factor akin to $\ln \beta _3$ was also present in the
operator $\hat{\cal U}$ of that paper [see Eq.~(2.46) where this factor
is written as ``$\ln \sqrt{12}/a$'']. Moreover, and this is the main
reason why we have considered a generalized version of
Eq.~(\ref{Hamiltoy1}), we notice that our case is in fact very special
since it corresponds to $\beta _3=1$ or $\ln \beta _3=0$ (or
``$\epsilon=0$'' in the language of Ref.~\cite{AM}). This means that, in
the operator~(\ref{defU}), the second exponential totally ``disappears''
while, of course, the term proportional to $q_1p_1$ remains present in
the classical generating function. Therefore, in our case, the quantum
generating operator is just given by
\begin{equation}
\label{defU2} \hat{{\cal U}}(\hat{q}_1,\hat{p}_1,\eta
)=\exp\left(-\frac{i}{2}\frac{a'}{a} \hat{q}_1^2\right)\, .
\end{equation}
Clearly, one can repeat the above calculations using this operator and
show that this leads to Eqs.~(\ref{transclass}) but at the quantum
level.

\par

Let us now turn to the transformation of the wave-function itself. It is
given by
\begin{equation}
\Psi \left(q_2\right)=N_2{\rm e}^{-\Omega _2q_2^2}= \hat{{\cal
U}}^{\dagger}(\hat{q}_1,\hat{p}_1,\eta )\Psi
\left(q_1\right)=\hat{{\cal U}}^{\dagger}(\hat{q}_1,\hat{p}_1,\eta
)N_1{\rm e}^{-\Omega _1q_1^2} \, ,
\end{equation}
from which, using Eq.~(\ref{defU2}), one deduces that 
\begin{equation}
\label{transwf}
N_2=N_1 \, ,\quad \Omega _2=\Omega _1 -i\frac{a'}{a}\, .
\end{equation}
The relation $N_2=N_1$ comes from the fact that the quantity $N_{1,2}$
is given by the real part of the function $\Omega _{1,2}$ and that
$\Omega _2$ and $\Omega _1$ differ by a complex factor only. The above
relation exactly reproduces what was observed in Eqs.~(\ref{Nom}) and
Eqs.~(\ref{solomegacosmo}). From these formulae, we see that $\Omega
_1=-i/2(f/a)'/(f/a)$, see Eqs.~(\ref{Nom}), while $\Omega _2=-if'/(2f)$,
see Eqs.~(\ref{solomegacosmo}), and they indeed satisfy
Eq.~(\ref{transwf}). Of course, the wave-functions after the quantum
canonical transformation is normalized because $\Psi _2^*\Psi_2=
\hat{\cal U}\hat{\cal U}^{\dagger }\Psi_1^*\Psi _1=\Psi_1^*\Psi _1$, the
operator $\hat{\cal U}$ being unitary.

\subsection{The Power Spectrum}
\label{subsec:psscalarfield}

Let us now calculate the two-point correlation function in the quantum
state where the scalar field is put by the cosmological evolution. As
will be discussed in the following, this quantity is relevant in
astrophysics because, in the case of cosmological perturbations, it is
directly observable; in particular it is directly linked to CMB
fluctuations. Its definition reads
\begin{equation}
\left\langle \hat{\Phi} (\eta ,\vec{x})\hat{\Phi }(\eta
,\vec{x}+\vec{r})\right \rangle = \int \prod _{\vec{k}}^n {\rm d}\mu
_{\vec{k}}^{_{\mathrm R}} {\rm d}\mu _{\vec{k}}^{_{\mathrm I}} \Psi^*
\left(\mu _{\vec{k}}^{_{\mathrm R}},\mu _{\vec{k}}^{_{\mathrm
I}}\right) \Phi (\eta ,\vec{x})\Phi (\eta ,\vec{x}+\vec{r}) \Psi
\left(\mu _{\vec{k}}^{_{\mathrm R}},\mu _{\vec{k}}^{_{\mathrm
I}}\right)\, .
\end{equation}
In this formula, the brackets mean the quantum average according to the
standard definition, i.e. $\langle\hat{A}\rangle\equiv \int{\rm d}x\,
\Psi^*\hat{A}(x)\, \Psi$. Then, using the Fourier expansion of the
scalar field and permuting the integrals, one obtains
\begin{eqnarray}
\left\langle \hat{\Phi} (\eta ,\vec{x})\hat{\Phi }(\eta
,\vec{x}+\vec{r})\right \rangle &=&\frac{1}{a^2}\frac{1}{(2\pi )^3} 
\int \int {\rm d}\vec{p}{\rm d}\vec{q}{\rm e}^{i\vec{p}\cdot \vec{x}}
{\rm e}^{i\vec{q}\cdot \left(\vec{x}+\vec{r}\right)}
\prod _{\vec{k}}^n\left(\frac{2 \Re \Omega _{\vec{k}}}{\pi }\right)
\nonumber \\
& & 
\int \left(\prod _{\vec{k}}^n {\rm d}\mu
_{\vec{k}}^{_{\mathrm R}} {\rm d}\mu _{\vec{k}}^{_{\mathrm I}}\right)
{\rm e}^{-2\sum _{\vec{k}}^n \Re \Omega _{\vec{k}}
\left[\left(\mu
_{\vec{k}}^{_{\mathrm R}}\right)^2+\left(\mu
_{\vec{k}}^{_{\mathrm I}}\right)^2\right]}\mu _{\vec{p}}\mu _{\vec{q}}
\, .
\nonumber \\
\end{eqnarray}
The above expression vanishes unless $\vec{p}=-\vec{q}$. Indeed, if
$\vert \vec{p}\vert \neq \vert \vec{q}\vert $ then the quantity $\mu
_{\vec{p}}\mu _{\vec{q}}$ is ``linear'' in $\mu _{\vec{p}}^{_{\mathrm R,
I}}$ and $\mu _{\vec{q}}^{_{\mathrm R, I}}$ and, consequently, the
Gaussian integral is zero. If $\vec{p}=\vec{q}$, then $\mu _{\vec{p}}\mu
_{\vec{q}}=\left(\mu _{\vec{p}}^{_{\mathrm R}}\right)^2-\left(\mu
_{\vec{p}}^{_{\mathrm I}}\right)^2$ and each term is indeed
non-vanishing but the sum is zero because of the minus sign. Therefore,
one obtains
\begin{eqnarray}
\left\langle \hat{\Phi} (\eta ,\vec{x})\hat{\Phi }(\eta
,\vec{x}+\vec{r})\right \rangle &=&\frac{2}{a^2}\frac{1}{(2\pi )^3} 
\int {\rm d}\vec{p}{\rm e}^{i\vec{p}\cdot \vec{r}}
\prod _{\vec{k}}^n\left(\frac{2 \Re \Omega _{\vec{k}}}{\pi }\right)
\nonumber \\
& & 
\int \prod _{\vec{k}}^n {\rm d}\mu
_{\vec{k}}^{_{\mathrm R}} {\rm d}\mu _{\vec{k}}^{_{\mathrm I}}
{\rm e}^{-2\sum _{\vec{k}}^n \Re \Omega _{\vec{k}}
\left[\left(\mu
_{\vec{k}}^{_{\mathrm R}}\right)^2+\left(\mu
_{\vec{k}}^{_{\mathrm I}}\right)^2\right]}\left(\mu
_{\vec{p}}^{_{\mathrm R}}\right)^2
\, .
\end{eqnarray}
The overall factor of $2$ originates from the fact that the integral
over $\mu _{\vec{k}}^{_{\mathrm R}}$ is equal to the integral over $\mu
_{\vec{k}}^{_{\mathrm I}}$. The next step is to perform the path
integral. In the above infinite product of integrals, all of them are of
the form ``$\int {\rm d}x {\rm e}^{-\alpha x^2}$'' except the one over
$\mu _{\vec{p}}^{_{\mathrm R}}$ which is of the form ``$\int {\rm d}x
x^2 {\rm e}^{-\alpha x^2}$''. Using standard results for Gaussian
integrals, one gets
\begin{eqnarray}
\left\langle \hat{\Phi} (\eta ,\vec{x})\hat{\Phi }(\eta
,\vec{x}+\vec{r})\right \rangle &=&\frac{2}{a^2}\frac{1}{(2\pi )^3} \int
{\rm d}\vec{p}{\rm e}^{i\vec{p}\cdot \vec{r}} \prod
_{\vec{k}}^n\left(\frac{2 \Re \Omega _{\vec{k}}}{\pi }\right) \prod
_{\vec{k}}^n\left(\frac{\sqrt{\pi}}{\sqrt{2 \Re \Omega
_{\vec{k}}}}\right) \nonumber \\ & & \times \prod
_{\vec{k}}^{n-1}\left(\frac{\sqrt{\pi}}{\sqrt{2 \Re \Omega
_{\vec{k}}}}\right) \frac{1}{2} \left[\frac{\sqrt{\pi}}{\sqrt{\left(2
\Re \Omega _{\vec{p}}\right)^3}}\right] \, .
\end{eqnarray}
The infinite product ``$\prod _{\vec{k}}^{n-1}$'' means a product over
all the wave-vectors but $\vec{p}$. Clearly, one can complete this
product by inserting an extra factor $\sqrt{\pi}/\sqrt{2 \Re \Omega
_{\vec{p}}}$ coming from the last term in the integral. Then, the last
two products exactly cancel the first one. Finally, one obtains the
simple expression
\begin{eqnarray}
\label{interps}
\left\langle \hat{\Phi} (\eta ,\vec{x})\hat{\Phi }(\eta
,\vec{x}+\vec{r})\right \rangle &=&\frac{1}{a^2}\frac{1}{(2\pi )^3}
\int {\rm d}\vec{p}\, {\rm e}^{i\vec{p}\cdot \vec{r}} \frac{1}{2\Re
\Omega _{\vec{p}}}\, .
\end{eqnarray}
Using the form of $\Omega _{\vec{p}}$ in the ground state wave-function,
see Eq.~(\ref{solomegacosmo}), one obtains
\begin{equation}
2 \Re \Omega _{\vec{p}}=-\frac{i}{2} \frac{\mu _{\vec{p}}'\mu
_{\vec{p}}^*-\mu _{\vec{p}}\mu _{\vec{p}}'^*}{ \mu _{\vec{p}}\mu
_{\vec{p}}^*} =\frac{1}{2 \mu _{\vec{p}}\mu _{\vec{p}}^*}\, ,
\end{equation}
where we have used the fact that, with the initial condition (we will
return to this point in the following) $\mu _{\vec{p}}(\eta )\rightarrow
(2p)^{-1/2}{\rm e}^{ip\eta }$ when $p\eta \rightarrow -\infty$, the
Wronskian is equal to $i$.

\par

At this point, one can also make the following remark about the
canonical transformation discussed in the previous subsection. It is
clear that the power spectrum must be the same before and after the
canonical transformation. Above, we used the form of $\Omega _{\vec{p}}$
given by Eq.~(\ref{solomegacosmo}). But one could have used the form
given by Eq.~(\ref{Nom}) in the same manner and without affecting the
final result. Technically, this can be seen in Eq.~(\ref{interps}) where
it is clear that the power spectrum only depends on $\Re \Omega
_{\vec{p}}$. Since we demonstrated before that the canonical
transformation only modifies the imaginary part of $\Omega _{\vec{p}}$,
the power spectrum remains indeed the same.

\par

Therefore, the final expression reads
\begin{eqnarray}
\label{powersf} \left\langle \hat{\Phi} (\eta ,\vec{x})\hat{\Phi }(\eta
,\vec{x}+\vec{r})\right \rangle &=&\frac{2}{a^2}\frac{1}{(2\pi )^3} \int
{\rm d}\vec{p}\, {\rm e}^{i\vec{p}\cdot \vec{r}} \mu _{\vec{p}}\mu
_{\vec{p}}^* \\ \label{powersf2} &=&\frac{1}{4\pi ^2}\int _0^{+\infty
}\frac{{\rm d}p}{p}\frac{\sin pr}{pr} p^2\left \vert \frac{\mu
_{\vec{p}}}{a}\right \vert ^2\, .
\end{eqnarray}
This expression is the standard one, usually derived in the Heisenberg
picture~\cite{procbrazil,procpoland}. Knowledge of the mode function
(including the initial conditions) is sufficient to estimate the power
spectrum. In the following, we consider the case of inflationary
cosmological perturbations and investigate which quantity plays the role
of $\mu _{\vec{k}}$ in that framework. This will allow us to discuss the
inflationary predictions.

\section{Inflationary Cosmological Perturbations of 
Quantum-Mechanical Origin}
\label{sec:pert}

\subsection{General Formalism}
\label{subsec:generalpert}

In this section, we finally consider our main subject, namely the theory
of inflationary cosmological perturbations of quantum-mechanical
origin~\cite{MFB,procbrazil,procpoland,Bardeen}. Our goal is to go
beyond the isotropic and homogeneous FLRW Universe, the metric of which
can be written as
\begin{equation}
{\rm d}s^2=a^2(\eta )\left[-{\rm d}\eta ^2+
\delta ^{(3)}_{ij}{\rm d}x^i{\rm d}x^j\right]\, ,
\end{equation}
and to study how small quantum perturbations around the above-mentioned
solution behave during inflation. As we will see, the basic physical
phenomenon and, hence, the corresponding formalism are similar to what
was discussed before. As already emphasized, we are mainly concerned
with inflation, that is to say a phase of accelerated expansion that
took place in the early universe. In general relativity, such a phase
can be obtained if the matter content is dominated by a fluid whose
pressure is negative. Since, at very high energies, quantum field theory
is the natural candidate to describe matter, it is natural and simple to
postulate that a scalar field (the ``inflaton'') was responsible for the
evolution of the universe in this regime. Therefore, the total action of
the system is given by
\begin{equation}
\label{actioninf}
S=-\frac{\mP^2}{16\pi }\int {\rm d}^4x\sqrt{-g}R -\int {\rm
d}^4x\sqrt{-g} \biggl[\frac{1}{2}g^{\mu \nu}\partial _{\mu} \varphi
\partial _{\nu}\varphi +V(\varphi )\biggr]\, ,
\end{equation}
where $\varphi $ is the inflaton field. Our discussion will be (almost)
independent of the detailed shape of the potential $V(\varphi) $ but,
clearly, deriving from high energy physics (for instance string theory)
what this shape could be (in particular explaining the required flatness
of the potential) is a major issue~\cite{LR,Kallosh}.

\par

Beyond homogeneity and isotropy, the most general form of the perturbed
line element can be expressed as \cite{MFB}:
\begin{eqnarray}
\label{metricgi}
{\rm d}s^2 &=& a^2(\eta )\{-(1-2\phi ){\rm d}\eta ^2+2({\rm
\partial}_iB){\rm d}x^i {\rm d}\eta +\bigl[(1-2\psi )\delta
_{ij}^{(3)} \nonumber \\ & & +2{\rm \partial }_i{\rm \partial
}_jE+h_{ij}\bigr]{\rm d}x^i{\rm d}x^j\}\ .
\end{eqnarray}
In the above expression, the functions $\phi $, $B$, $\psi $ and $E$
represent the scalar sector whereas the tensor $h_{ij}$, satisfying
$h_i{}^i=0=h_{ij}{}^{,j}$, represents the gravitational waves. These
functions must be small in comparison to one in order for the
perturbative treatment to be valid. There are no vector perturbations
because a single scalar field cannot seed rotational perturbations. At
the linear level, the two types of perturbations decouple and,
therefore, can be treated separately.

\par

In the case of scalar perturbations of the geometry evoked above, the
four functions are in fact redundant (thanks to our freedom to choose
the coordinate system) and, in fact, the scalar fluctuations of the
geometry can be characterized by a single quantity, namely the
gauge-invariant Bardeen potential $\Phi _{_{\rm B}}$~\cite{Bardeen} (not
to be confused with the scalar field $\Phi $ considered before) defined
by
\begin{eqnarray}
\Phi _{_{\rm B}}\left(\eta ,\vec{x}\right)&=& \phi
+\frac{1}{a}\left[a\left(B-E'\right)\right]'\, .
\end{eqnarray}
On the other hand, the fluctuations in the inflaton scalar field are
characterized by the following gauge-invariant quantity $\delta\phi
^{\rm (gi)}$
\begin{eqnarray}
\delta \varphi ^{\rm (gi)} \left(\eta ,\vec{x}\right)&=& \delta
\varphi +\varphi '\left(B-E'\right)\, .
\end{eqnarray}
We have therefore two gauge-invariant quantities but only one degree of
freedom since $\Phi _{_{\rm B}}$ and $\delta \varphi ^{\rm (gi)}$ are
coupled through the perturbed Einstein equations. As a consequence, in
the scalar sector of the theory, everything can be reduced to the study
of a single gauge-invariant variable (the so-called Mukhanov-Sasaki
variable) defined by~\cite{MuChi}
\begin{equation}
v\left(\eta ,\vec{x}\right)\equiv a\left[\delta \varphi^{\rm (gi)}+
\varphi'\frac{\Phi _{_{\rm B}}}{\calH} \right]\, .
\end{equation}
Let us notice that we will also work with the rescaled variable $\muS$
defined by $\mu _{_{\rm S}}\left(\eta ,\vec{x}\right)\equiv
-\sqrt{2\kappa }v$. Finally, density perturbations are also often
characterized by the so-called conserved quantity $\zeta \left(\eta ,
\vec{x}\right)$~\cite{Lyth1,MS1} defined by $\muS=-2a\sqrt{\gamma }\zeta
$, where $\gamma =1-{\calH}'/{\calH}^2$.

\par

In the tensor sector (which is automatically gauge invariant), the
quantity which plays the role of $\muS \left(\eta ,\vec{x}\right)$ is
$\muT\left(\eta ,\vec{x}\right)$, defined according to
$h_{ij}=(\muT/a)Q_{ij}$, where $Q_{ij}$ are the (transverse and
traceless) eigentensors of the Laplace operator on the space-like
sections~\cite{Bardeen}.

\par

As usual, it is more convenient to study the perturbations mode by mode
and, for this purpose, we will follow the evolution of the perturbations
in Fourier space. Therefore, the study of cosmological perturbations
during inflation reduces to investigating the behaviors of only two
variables: $\muS{}_{\vec{k}}\left(\eta \right)$ and
$\muT{}_{\vec{k}}\left(\eta \right)$.

\par

Let us now establish the equations of motion for our two basic
quantities. Since we want the variation of the action~(\ref{actioninf})
to give the first order equations of motion for
$\muS{}_{\vec{k}}\left(\eta \right)$ and $\muT{}_{\vec{k}}\left(\eta
\right)$, we have to expand the action pertubatively up to second order
in the metric perturbations and in the scalar field fluctuations. After
a lengthy and tedious calculation, one obtains~\cite{MFB}
\begin{eqnarray}
{}^{(2)}\delta S &=& {1 \over 2} \int {\rm d}^4x \biggl[(v')^2-
\delta^{i j} \partial _iv\partial _jv+{\left(a\sqrt{\gamma }\right)''
\over a\sqrt{\gamma }} v^2 \biggr] \nonumber \\ & & +\frac{m_{_{\rm
Pl}}^2}{64 \pi}\int {\rm d}^4x a^2(\eta )\left[(h^i{}_j)'(h^j{}_i)'
-\partial _k(h^i{}_j) \partial ^k(h^j{}_i)\right]\, ,
\end{eqnarray}
Notice that the constant $\mP$ does not appear explicitly in the scalar
part of the action because it has been absorbed via the background
Einstein equations (however, see also Ref.~\cite{PPP}). It is also
important to stress again that the previous expression is valid for any
potential $V(\varphi)$.

\par

Variation of the action leads to the following equation of motion for
the two quantities $\muS{}_{\vec{k}}\left(\eta \right)$ and
$\muT{}_{\vec{k}}\left(\eta \right)$
\begin{equation}
\label{eq:evolinf}
\frac{{\rm d}^2\muST{}_{\vec{k}}}{{\rm d}\eta ^2}+\omegaST^2(k,\eta )
\muST{}_{\vec{k}}=0 ,
\end{equation}
with 
\begin{equation}
\label{frequencycosmo}
\omegaS^2\left(k,\eta \right)=k^2 -\frac{(a\sqrt{\gamma
})''}{a\sqrt{\gamma }}\, , \quad \omegaT^2\left(k,\eta \right)=k^2
-\frac{a''}{a} \, .
\end{equation}
We have thus reached our goal and demonstrated that cosmological
perturbations obey exactly the same type of equation as a scalar field
interacting with a classical electric field (Schwinger effect), namely
the equation of a parametric oscillator as can be checked by comparing
Eq.~(\ref{eq:evolinf}) with Eq.~(\ref{eqphi}). The only difference lies
in the physical nature of the classical source. In the case of
cosmological perturbations, the (background) gravitational field is the
classical source. The time dependence of the frequencies $\omegaS$ and
$\omegaT$ is also different (recall that, in the case of the Schwinger
effect, $\omega^2$ contains terms proportional to $t$ and $t^2$). Here,
the dependence is fixed as soon as the behavior of the scale factor
$a(\eta )$ is known. It is also interesting to notice that, a priori,
the time dependence of $\omegaS$ is not the same as the one of
$\omegaT$. Indeed, $\omegaT$ depends on $a$ and its derivatives up to
second order while $\omegaS$ depends on the scale factor and its
derivatives up to the fourth order (since it contains a term $\gamma
''$, the quantity $\gamma $ containing itself a term $a''$). Finally,
the quantization of the theory proceeds as before and, as a consequence
of the interaction between the quantum cosmological perturbations and
the classical background, this results in the phenomenon of particle
creation, here graviton creation. Classically, this corresponds to the
amplification (``growing mode'') of the fluctuations.

\par

In the next section, we describe this phenomenon for an inflationary
scale factor.

\subsection{The Inflationary Effective Frequencies}
\label{subsec:frequencies}

So far, we have never specified $a(\eta )$ and, {\it a priori}, the
mechanism of graviton creation is valid for any scale factor provided it
is time-dependent. However, clearly, the detailed properties of the
transition amplitude $\left \langle \Psi _1\vert \Psi _2\right \rangle $
depend on the time behavior of the effective frequency $\omega ^2(k,\eta
)$ and, hence, on the form of $a(\eta )$. Obviously, in the case of the
Schwinger effect, a frequency different from the one given by
Eq.~(\ref{eq:defomega}) would have led to a number of created pairs
different from Eq.~(\ref{schwingerresult}).

\par

In order to evaluate $\omegaS^2(k, \eta )$ and $\omegaT^2(k,\eta )$ for
a typical inflationary model, one can use the slow-roll
approximation~\cite{SL,MS2,LLMS,flow}. Indeed, during inflation and by
definition, the kinetic energy to potential energy ratio and the scalar
field acceleration to the scalar field velocity ratio are small and this
suggests to view these quantities as parameters in which a systematic
expansion can be performed. Therefore, one introduces the two parameters
$\epsilon _1$ and $\epsilon _2$~\cite{flow} according to
\begin{equation}
\label{intersr}
\epsilon _1=3\frac{\dot{\varphi }^2/2}{\dot{\varphi
}^2/2+V(\varphi)}\, , \qquad \frac{\dd}{\dd t}\left(\frac{\dot{\varphi
}^2}{2}\right) = H\dot{\varphi }^2\left(\frac{\epsilon _2}{2}
-\epsilon _1\right) .
\end{equation}
{}From the above expressions, one sees that $\epsilon _1/3$ measures the
ratio of the kinetic energy to the total energy while $\epsilon _2>0$
(respectively $\epsilon _2<0$) represents a model where the kinetic
energy itself increases (respectively decreases) with respect to the
total energy. It is also interesting to notice that $\epsilon
_2=2\epsilon _1$ marks the frontier between models where the kinetic
energy increases ($\epsilon_2>2\epsilon _1$) and the models where it
decreases ($\epsilon_2<2\epsilon _1$). Provided the slow-roll conditions
are valid, that is to say $\epsilon _{1,2}\ll 1$, one can also invert
the previous expressions and express the slow-roll parameters only in
terms of the inflaton potential. This leads to
\begin{equation}
\epsilon _1\simeq \frac{\mP ^2}{16 \pi} \left( \frac{V'}{V}\right )^2,
\qquad \epsilon _2\simeq \frac{\mP ^2}{4\pi}\left[\left(
\frac{V'}{V}\right )^2 -\frac{V''}{V}\right]\, ,
\end{equation}
where, in the present context, a prime denotes a derivative with respect
to the scalar field $\varphi $. Concrete calculations of slow-roll
parameters for specific models can be found in Ref.~\cite{procbrazil}.

\par

Then, one can show that the two effective frequencies, to first order in
the slow-roll parameters, can be expressed as~\cite{MFB,Gpara}
\begin{equation}
\omegaS^2(k, \eta )\simeq k^2-\frac{2+3\epsilon _1-3\epsilon _2/2 }{\eta
^2}\, , \quad \omegaT^2(k, \eta )\simeq k^2-\frac{2+3\epsilon_1}{\eta ^2}\, .
\end{equation}
Several remarks are in order at this point. Firstly, and as already
mentioned previously, the time dependence in the inflationary case is
different from the Schwinger case: the effective frequency contains
terms proportional to $1/\eta ^2$. Therefore, although the basic
physical phenomenon is the same, one can expect the detailed predictions
to differ. Secondly, different inflationary models correspond to
different inflaton potentials (or to different time variations of the
scale factor) and, hence, to different values for the slow-roll
parameters. One notices that the effective frequencies are sensitive to
the details of the inflationary models since $\omegaS^2(k, \eta )$ and
$\omegaT^2(k, \eta )$ depend on $\epsilon _1$ and $\epsilon _2$.

\subsection{The WKB Approximation}
\label{subsec:wkb}

\begin{figure}[t]
\centering \includegraphics[height=6cm, width=10cm]{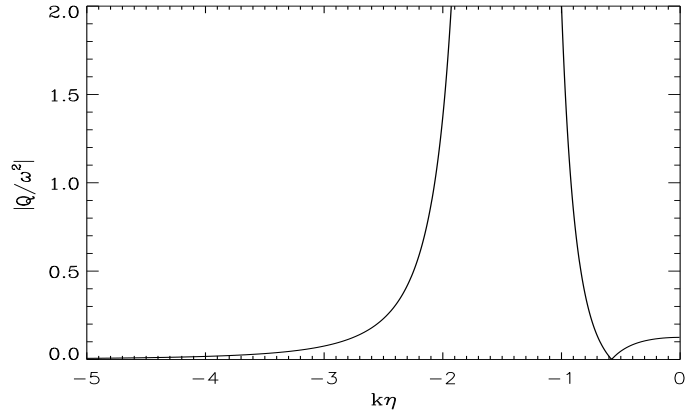}
\caption{Evolution of the quantity $\vert Q/\omega ^2\vert $ with the
quantity $k\eta $ for a typical model of inflation according to
Eq.~(\ref{wkbinftest}) (we have neglected the corrections proportional
to the slow-roll parameters). In the limit $k\eta \rightarrow -\infty$,
which corresponds to a wavelength much smaller than the Hubble radius,
$\vert Q/\omega ^2\vert$ vanishes and the notion of an adiabatic vacuum
is available.}  \label{fig:wkbinf}
\end{figure}

We have established the form of the effective frequencies in the case of
inflation. One must now solve the mode equations~(\ref{eq:evolinf}). For
this purpose, we now reiterate the analysis of
Sec.~\ref{subsec:particlecreation} using the WKB
approximation~\cite{MS3}. As was the case for the Schwinger effect, the
mode function can be found exactly. It is given in terms of Bessel
functions [instead of parabolic cylinder functions, see
Eq.~(\ref{exactsol})]
\begin{equation}
\label{exactcosmo}
\mu _{\vec{k}}(\eta )=\sqrt{k\eta }\left[A_{\vec{k}}J_{\nu }
\left(k\eta \right)+B_{\vec{k}}J_{-\nu }\left(k\eta \right)\right]\, ,
\end{equation}
where the orders are now functions of the slow-roll parameters, $\nu
_{_{\rm S}}=-3/2-\epsilon_1-\epsilon _2/2$ and $\nu _{_{\rm
T}}=-3/2-\epsilon_1 $. Then, one must choose the initial conditions. As
discussed in the case of the Schwinger effect, we use the WKB
approximation to discuss this question. The first step is to calculate
the quantity $Q$ in order to identify the regime where an adiabatic
vacuum is available. Straightforward calculations lead to
\begin{equation}
\label{wkbinftest}
\left\vert \frac{Q_{_{\rm S,T}}}{\omega _{_{\rm S,T}}^2}\right\vert
=\frac{1}{8}\left\vert \frac{1-3k^2\eta ^2}{\left(1-k^2\eta
^2/2\right)^3} \right\vert +{\cal O}\left(\epsilon _1,\epsilon
_2\right)\, .
\end{equation}
This quantity is represented in Fig.~\ref{fig:wkbinf}.
 
Let us now discuss this plot in more detail. The problem is
characterized by two scales: the wavelength of the corresponding Fourier
mode given by
\begin{equation}
\lambda \left(\eta \right)=\frac{2\pi }{k}a(\eta )\, ,
\end{equation}
where $k$ is the co-moving wavenumber, and the Hubble radius which can
be expressed as
\begin{equation}
\ell _{_{\rm H}}\left(\eta \right)=\frac{a^2}{a'}\, .
\end{equation}
We notice that the quantity $\left \vert Q/\omega ^2 \right \vert $
vanishes in the limit $k\eta \rightarrow -\infty$. This limit
corresponds to a regime where $\lambda \ll \ell _{_{\rm H}}$. In this
case, the wavelength is so small in comparison with the scale $\ell
_{_{\rm H}}$ characterizing the curvature of space-time that the Fourier
mode does not feel the FLRW Universe but behaves as if it were in flat
(Minkowski) space-time. Clearly, in this regime, an adiabatic vacuum
state is available since we recover the standard quantum field theory
description. In the limit $k\eta \rightarrow 0$, the quantity $\left
\vert Q/\omega ^2\right \vert $ goes to $1/8={\cal O}(1)$ as can be
checked in Fig.~\ref{fig:wkbinf}. This regime corresponds to the case
where $\lambda \gg \ell _{_{\rm H}}$, that is to say when the wavelength
of the Fourier mode is outside the Hubble radius. In this case, the
curvature of space-time is felt and, as a consequence, the WKB
approximation is violated and there is no unique vacuum state in this
limit.

\par

We have just seen that when a mode is sub-Hubble, that is to say
$\lambda \ll \ell _{_{\rm H}}$, the WKB approximation is valid. Let us
notice that, without a phase of inflation, all the Fourier modes of
astrophysical interest today would have been outside the Hubble radius
in the early Universe. It is only because, during inflation, the Hubble
radius is constant that, initially, the Fourier modes are inside the
Hubble radius. Therefore, although it was not designed for this purpose,
a phase of inflation automatically implies that the WKB approximation is
valid in the early Universe and, as a consequence, ensures that we can
choose a well-defined initial state. This is clearly an ``extra bonus''
of utmost importance. In the adiabatic regime, the solution for the mode
function can be written as
\begin{equation}
\label{wkbcosmo}
\mu _{\vec{k}}(\eta )=\alpha _{\vec{k}}\mu
_{{\rm wkb},\vec{k}}(\eta ) +\beta _{\vec{k}}\mu
_{{\rm wkb},\vec{k}}^*(\eta )\, ,
\end{equation}
where
\begin{equation}
\mu _{{\rm wkb},\vec{k}}(\eta ) \equiv \frac{1}{\sqrt{2\omega(k,\eta)}}
{\rm e}^{-i\int^{\eta }_{\eta _{\rm ini}}\omega(k,\tau){\rm d}\tau}\, .
\end{equation}
As done for the Schwinger effect, see Eq.~(\ref{exactsol}), we now
choose the initial conditions such that $\alpha _{\vec{k}}=1$, $\beta
_{\vec{k}}=0$, corresponding to only one WKB branch in
Eq.~(\ref{wkbcosmo}). This completely fixes the coefficients
$A_{\vec{k}}$ and $B_{\vec{k}}$ in Eq.~(\ref{exactcosmo}). One obtains
[compare with Eqs.~(\ref{Aschwinger}) and~(\ref{Bschwinger})]
\begin{eqnarray}
\frac{A_{\vec{k}}}{B_{\vec{k}}}=-{\rm e}^{i\pi \nu }\, ,\quad
B_{\vec{k}}=-\frac{2i\pi}{\mP}\frac{{\rm e}^{-i\nu (\pi /2)-i(\pi /4)
+ik\eta _{\rm ini}}}{\sqrt{k}\sin (\pi \nu )}\, .
\end{eqnarray}
Equipped with the above exact solution for the mode function, the
inflationary predictions can be determined.

\par

Before turning to this calculation, let us quickly come back to the fact
that the WKB approximation breaks down on super-Hubble scales. In fact,
this problem bears a close resemblance with a situation discussed by
atomic physicists at the time quantum mechanics was born. The subject
debated was the application of the WKB approximation to the motion in a
central field and, more specifically, how the Balmer formula for the
energy levels of hydrogenic atoms, can be recovered within the WKB
approximation. The effective frequency for hydrogenic atoms is given by
(obviously, in the atomic physics context, the wave equation is not a
differential equation with respect to time but to the radial coordinate
$r$)
\begin{equation}
\label{omeH}
\omega ^2(E,r)=\frac{2m}{\hbar^2}\biggl(E+\frac{Ze^2}{r}\biggr)
-\frac{\ell (\ell +1)}{r^2}\, ,
\end{equation}
where $Ze$ is the (attractive) central charge and $\ell$ the quantum
number of angular momentum. The symbol $E$ denotes the energy of the
particle and is negative in the case of a bound state. Apart from the
term $Ze^2/r$ and up to the identification $r \leftrightarrow \eta $,
the effective frequency has exactly the same form as $\omega _{_{\rm
S,T}}(k,\eta )$ during inflation, see
Eqs.~(\ref{frequencycosmo}). Therefore, calculating the evolution of
cosmological perturbations on super-Hubble scales, $\vert k\eta \vert
\rightarrow 0$, is similar to determining the behavior of the hydrogen
atom wave function in the vicinity of the nucleus, namely $r\rightarrow
0$. The calculation of the energy levels by means of the WKB
approximation was first addressed by Kramers \cite{Kramers} and by Young
and Uhlenbeck \cite{YU}. They noticed that the Balmer formula was not
properly recovered but did not realize that this was due to a misuse of
the WKB approximation. In $1937$ the problem was considered again by
Langer~\cite{Langer}. In a remarkable article, he showed that the WKB
approximation breaks down at small $r$, for an effective frequency given
by Eq.~(\ref{omeH}) and, in addition, he suggested a method to
circumvent this difficulty. Recently, this method has been applied to
the calculation of the cosmological perturbations in
Refs.~\cite{MS3,CFLG}. This gives rise to a new method of approximation,
different from the more traditional slow-roll approximation.

\subsection{The Inflationary Power Spectra}
\label{subsec:infps}

In this sub-Section we turn to the calculation of the inflationary
observables. The first step is to quantize the system. Obviously, this
proceeds exactly as for the Schwinger effect or for a scalar field in
curved space-time, the two cases that we have discussed before. We do
not repeat the formalism here. As before, in the functional
Schr\"odinger picture, the wave-function of the perturbations is given
by
\begin{equation}
\Psi =\prod _{\vec{k}}^n\Psi_{\vec{k}}\left(\mu ^{_{\mathrm
R}}_{\vec{k}}, \mu ^{_{\mathrm I}}_{\vec{k}}\right) =\prod
_{\vec{k}}^n\Psi_{\vec{k}}^{_{\mathrm R}} \left(\mu ^{_{\mathrm
R}}_{\vec{k}}\right) \Psi_{\vec{k}}^{_{\mathrm I}} \left(\mu
^{_{\mathrm I}}_{\vec{k}}\right) \, ,
\end{equation}
with 
\begin{equation}
\label{wavefunctionpert}
\Psi _{\vec{k}}^{_{\mathrm R}}\left(\eta, \mu
^{_{\mathrm R}}_{\vec{k}}\right)
=N_{\vec{k}}\left(\eta \right){\rm e}^{-\Omega _{\vec{k}}\left(\eta \right)
\left(\mu
^{_{\mathrm R}}_{\vec{k}}\right)^2}\, ,\quad 
\Psi _{\vec{k}}^{_{\mathrm I}}\left(\eta , \mu
^{_{\mathrm I}}_{\vec{k}}\right)
=N_{\vec{k}}\left(\eta \right){\rm e}^{-\Omega _{\vec{k}}\left(\eta \right)
\left(\mu
^{_{\mathrm I}}_{\vec{k}}\right)^2}\, ,
\end{equation}
where the functions $N_{\vec{k}}(\eta )$ and $\Omega _{\vec{k}}(\eta )$
are functions that can be determined using the Schr\"odinger
equation. This leads to expressions similar to Eqs.~(\ref{eq:NOm})
and~(\ref{schrodingersf}) where now $\omega ^2(k,t)$ should be replaced
by $\omega _{_{\rm S,T}}^2$ according to whether one considers the
scalar perturbations or the gravitational waves. In particular, the
function $\Omega _{\vec{k}}(\eta )$ is still given by $-i\mu
_{\vec{k}}'/\mu _{\vec{k}}$, see Eq.~(\ref{solomegacosmo}), where, in
the present context, $\mu _{\vec{k}}$ is given by the Bessel function of
Eq.~(\ref{exactcosmo}).

\par

At this stage, one could compute the amplitude $\langle 0^-\vert
0^+\rangle $ as one did in the case of the Schwinger effect. However, in
the context of inflation, this is not the observable one is interested
in. Indeed, we want to evaluate the amplitude of the fluctuations at the
end of inflation and on super-Hubble scales. In this regime, as
discussed before, there is no adiabatic state. So, in the context of
inflation, there exists a ``in'' vacuum state $\vert 0^-\rangle $ when
$k\eta \rightarrow -\infty $ but there is no ``out'' region and,
consequently, no $\vert 0^+\rangle$ state. Of course, if one follows the
evolution of the mode after inflation, then the unicity of the choice of
the vacuum state is restored when the mode re-enters the Hubble radius
either during the radiation or matter dominated eras. But our goal is to
compute the spectrum at the end of inflation. In other words, and
contrary to the Schwinger effect, the quantity $\langle 0^-\vert
0^+\rangle $ is not really relevant for the inflationary cosmological
perturbations.

\par

In fact, our goal is to calculate the anisotropies in the CMB (and/or to
understand the distribution of galaxies). The key point is that the
presence of cosmological perturbations causes anisotropies in the CMB:
this is the Sachs-Wolfe effect~\cite{sw,panek}. More precisely, on large
scales, one has
\begin{equation}
\label{tfluctuations} \hat{\frac{\delta T}{T}}(\vec{e}) \propto
\hat{\zeta} =-\frac{\hat{\muS}}{2a\sqrt{\gamma }}\, ,
\end{equation}
where $\vec{e}$ represents a direction in the sky. The exact link is
more complicated and has been discussed in details for instance in
Refs.~\cite{procpoland,panek}. In fact, it is convenient to expand this
operator on the celestial sphere, i.e. on the basis of spherical
harmonics
\begin{equation}
\label{dt/t}
\hat{\frac{\delta T}{T}}(\vec{e}) =\sum _{\ell =2}^{+\infty }\sum
_{m=-\ell }^{m=\ell } \hat{a}_{\ell m}Y_{\ell m}(\theta ,\varphi )\, .
\end{equation}
This allows us to calculate the vacuum two-point correlation function of
temperature fluctuations. One gets
\begin{equation}
\label{corrt}
\left \langle \hat{\frac{\delta T}{T}}(\vec{e}_1) \hat{\frac{\delta
T}{T}}(\vec{e}_2)\right \rangle =\sum _{\ell =2}^{+\infty
}\frac{(2\ell +1)}{4\pi }C_{\ell }P_{\ell } \left(\cos \gamma  
\right)\, ,
\end{equation}
where $P_{\ell }$ is a Legendre polynomial and $\gamma $ is the angle
between the two vectors $\vec{e}_1$ and $\vec{e}_2$. In the above
expression, the brackets mean the standard quantum average. In practice,
the observable two-point correlation function is rather defined by a
spatial average over the celestial sphere. These two averages are of
course not identical and the difference between them is at the origin of
the concept of ``cosmic variance''; see Ref.~\cite{GriJM} for a detailed
explanation. The $C_{\ell }$ 's are the multipole moments and have been
measured with great accuracy by the WMAP
experiment~\cite{wmap}. Clearly, as can be seen in
Eq.~(\ref{tfluctuations}), the above correlation function is related to
the two-point correlation function of the cosmological
fluctuations. Therefore, the relevant quantities to characterize the
inflationary perturbations of quantum-mechanical origin are
\begin{eqnarray}
\left \langle \hat{\zeta }(\eta ,{\vec{x}}) \hat{\zeta
}(\eta ,{\vec{x}}+{\vec{r}}) \right \rangle &=&
\int _0^{+\infty }
\frac{{\rm d}k}{k}\frac{\sin kr}{kr}k^3P_{\zeta} \, ,
\end{eqnarray} 
for scalar perturbations and, for tensor perturbations
\begin{eqnarray}
& &\left\langle \hat{h}_{ij}(\eta ,{\vec{x}}) \hat{h}^{ij}(\eta
,{\vec{x}}+{\vec{r}}) \right \rangle =\int _0^{+\infty }\frac{{\rm
d}k}{k}\frac{\sin kr}{kr} k^3P_h\, ,
\end{eqnarray}
One can then repeat the calculation done in
Sec.~\ref{subsec:psscalarfield} in order to evaluate the above
quantities. Indeed, the calculation proceeds exactly in the same way
since the wave-functional is still a Gaussian. This gives
\begin{equation}
\label{spec}
k^3P_{\zeta }(k)=\frac{k^3}{8\pi ^2}\biggl\vert \frac{\mu
_{_{\rm S}}}{a\sqrt{\gamma }}\biggr\vert ^2 , \quad 
k^3P_h(k)=\frac{2k^3}{\pi ^2}\biggl \vert \frac{\mu _{_{\rm T}}}{a}
\biggr \vert ^2 .
\end{equation}
These expressions should be compared with Eq.~(\ref{powersf2}).

\par

The two power spectra can be easily computed using the exact solution
for the mode function, see Eq.~(\ref{exactcosmo}). At first order in the
slow-roll parameters, one arrives at~\cite{SL,MS2,LLMS,flow}
\begin{eqnarray}
\label{pssr}
k^3P_{\zeta } &=& {H^2\over \pi \epsilon _1m_{_{\rm Pl}}^2}
\left[1-2\left(C+1\right)\epsilon_1 -C\epsilon _2 
-\left(2\epsilon _1+\epsilon _2\right)\ln {k\over
k_*}\right]\, , 
\\ 
\label{pssrgw}
k^3P_{h} &=& {16 H^2\over \pi m_{_{\rm Pl}}^2}
\left[1-2\left(C+1\right)\epsilon_1 -2\epsilon _1\ln {k\over
k_*}\right]\, ,
\end{eqnarray}
where $C$ is a numerical constant, $C\simeq -0.73$, and $k_*$ an
arbitrary scale called the ``pivot scale''. We see that the amplitude of
the scalar power spectrum is given by a scale-invariant piece (that is
to say which does not depend on $k$), $H^2/(\pi \epsilon_1 m_{_{\rm
Pl}}^2)$, plus logarithmic corrections the amplitude of which is
controlled by the slow-roll parameters, namely by the micro-physics of
inflation. The above remarks are also valid for tensor
perturbations. The ratio of tensor over scalar is just given by
$k^3P_{h}/k^3P_{\zeta }=16\epsilon _1$. This means that the
gravitational waves are always sub-dominant and that, when we measure
the CMB anisotropies, we essentially see the scalar modes. This is
rather unfortunate because this implies that one cannot measure the
energy scale of inflation since the amplitude of the scalar power
spectrum also depends on the slow-roll parameter $\epsilon _1$. Only an
independent measure of the gravitational waves contribution could allow
us to break this degeneracy. On the other hand, the spectral indices are
given by
\begin{equation}
n_{_{\rm S}}-1=\frac{\ln k^3P_{\zeta }}{{\rm d}\ln k}\biggl\vert
_{k=k_*}=- 2\epsilon _1-\epsilon _2\, , \quad n_{_{\rm T}}=\frac{\ln
k^3P_h}{{\rm d}\ln k}\biggl\vert _{k=k_*}=-2\epsilon_1 \, .
\end{equation}
As expected, the power spectra are always close to scale invariance
($n_{_{\rm S}}=1$ and $n_{_{\rm T}}=0$) and the deviation from it is
controlled by the magnitude of the two slow-roll parameters.

\par

\begin{figure}[t]
\centering
\includegraphics[height=5cm, width=10cm]{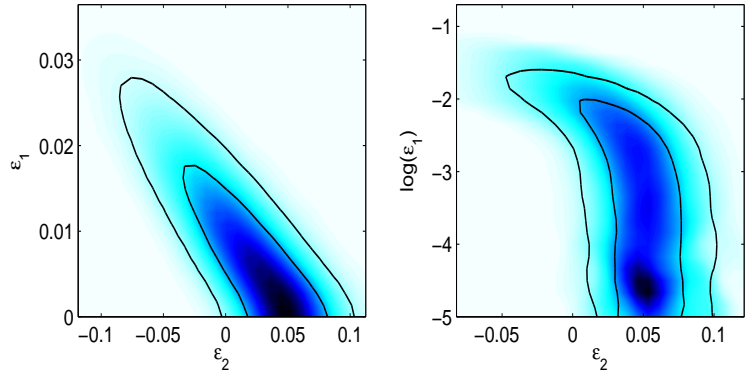}
\caption{$68\%$ and $95\%$ confidence intervals of the two-dimensional
marginalized posteriors in the slow-roll parameters plane, obtained at
leading order in the slow-roll expansion~\cite{MR}. The shading is the
mean likelihood and the left plot is derived under an uniform prior on
$\epsilon _1$ while the right panel corresponds to an uniform prior on
$\log \epsilon _1$.}
\label{wmap}       
\end{figure}

To conclude this section, let us signal that the slow-roll parameters
$\epsilon _1$ and $\epsilon _2$ are already constrained by the
astrophysical data, see Fig.~\ref{wmap} for the constraints coming from
the WMAP data. A complete analysis can be found in
Refs.~\cite{MR,ringeval}.

\section{The Classical Limit of Quantum Perturbations}
\label{sec:classical}

As discussed at length previously, the inflationary cosmological
perturbations are of quantum-mechanical origin. However, from the
observational point of view, it seems that we deal with a physical
phenomenon where quantum mechanics does not play a crucial role (even
does not a play a role at all). Therefore, from the conceptual point of
view, it is important to understand how the system can become
classical~\cite{GP,PS,PSS} (see also Ref.~\cite{CP}). We now turn to
this question.

\subsection{Coherent States}
\label{subsec:coherent}

It seems natural to postulate that a quantum system behaves classically
when it is placed in a state such that it follows (exactly or, at least,
approximatively) the classical trajectory. For the sake of illustration,
let us consider a simple one-dimensional system characterized by the
Hamiltonian
\begin{equation}
H(p,q)=\frac12 p^2+V(q)\, ,
\end{equation}
where, for the moment, the potential $V(q)$ is arbitrary. Solving the
classical Hamilton's equations (given some initial conditions)
\begin{equation}
\label{hamilharmo}
\frac{{\rm d}p}{{\rm d}t}=-\frac{\partial V(q)}{\partial q}\, ,\quad
\frac{{\rm d}q}{{\rm d}t}=p\, ,
\end{equation}
provides the classical solution $p_{\rm cl}$ and $q_{\rm cl}$. At the
technical level, the above-mentioned criterion of classicality amounts
to choosing a state $\vert \Psi \rangle $ such that
\begin{equation}
\label{classicHamil}
p_{\rm cl}(t)=\langle \Psi \vert \hat{p}(t)\vert \Psi \rangle\, ,\quad
q_{\rm cl}(t)=\langle \Psi \vert \hat{q}(t)\vert \Psi \rangle\, .
\end{equation}
This is clearly a non-trivial requirement as can be understood from the
Ehrenfest theorem. Indeed, this theorem shows that, for any state $\vert
\Psi \rangle$, one has
\begin{eqnarray}
\frac{{\rm d}}{{\rm d}t}\langle \Psi \vert \hat{p}(t)\vert \Psi \rangle
=-\left \langle \Psi \left \vert\frac{\partial \hat{V}(q)}{\partial
q}\right \vert \Psi \right\rangle\, , \quad \frac{{\rm d}}{{\rm
d}t}\langle \Psi \vert \hat{q}(t)\vert \Psi \rangle =\left \langle
\Psi \left\vert \hat{p}\right \vert \Psi \right\rangle\, ,
\end{eqnarray}
These equations resemble the Hamilton's equations~(\ref{hamilharmo}) but
are of course not identical. This implies that, placed in an arbitrary
state, the quantum system does not behave classically (i.e. the means of
the position and of the momentum do not obey the classical
equations). It would be the case only for a state $\vert \Psi \rangle$
such that
\begin{eqnarray}
\left \langle \Psi \left \vert\frac{\partial \hat{V}(q)}{\partial
q}\right \vert \Psi \right\rangle=\frac{\partial }{\partial q}
V\left(\langle \Psi \vert \hat{q}\vert \Psi \rangle\right)\, ,
\end{eqnarray}
which is not true in general since $\langle \Psi \vert \hat{q}^n\vert
\Psi \rangle \neq \langle \Psi \vert \hat{q}\vert \Psi \rangle^n$, but
obviously satisfied if the potential assumes the particular shape
$V(q)\propto q^2$, i.e. for the harmonic oscillator. In this case, the
means of the position and of the momentum do follow the classical
trajectory whatever the state $\vert \Psi \rangle$ is. This means that
Eqs.~(\ref{classicHamil}) are in fact not sufficient to define
classicality and that one needs to provide extra conditions. It seems
natural to require that the wave packet is equally localized in
coordinate and momentun $\Delta \hat{q}\equiv \sqrt{\langle
\hat{q}^2\rangle -\langle \hat{q}\rangle ^2}=\Delta\hat{p}$ to the
minimum allowed by the Heisenberg bound, namely $\Delta \hat{q}\Delta
\hat{p}=1/2$. This is another way to define a coherent state, see
Eq.~(\ref{coherent}) which, therefore, represents the ``most classical''
state of a quantum harmonic oscillator.

\par

We now demonstrate that the state~(\ref{coherent}) indeed satisfies the
above-mentioned properties. If the potential is given by
$V(q)=k^2q^2/2$, then the Hamilton's equations can be expressed as
$\dot{p}=-k^2q$ and $\dot{q}=p$ and the ``normal variable'' variable,
see also Eq.~(\ref{defnormal}),
\begin{equation}
\label{defalpha} \alpha \equiv
\sqrt{\frac{k}{2}}\left(q+\frac{i}{k}p\right)\, ,
\end{equation}
obeys the equation $\dot{\alpha }=-ik\alpha$ which allows us to write
the classical trajectory in phase space as
\begin{eqnarray}
\label{harmoclass}
p_{\rm cl}(t) &=& -i\sqrt{\frac{k}{2}}\left(\alpha _0{\rm e}^{-ikt}-
\alpha _0^*{\rm e}^{ikt}\right)\, ,\\ q_{\rm cl}(t) &=&
\frac{1}{\sqrt{2k}}\left(\alpha _0{\rm e}^{-ikt}+\alpha _0^*{\rm
e}^{ikt}\right)\, ,
\end{eqnarray}
with 
\begin{equation}
\alpha _0\equiv
\sqrt{\frac{k}{2}}\left[q(t=0)+\frac{i}{k}p(t=0)\right]\, ,
\end{equation}
Let us consider that, at time $t=0$, the system is placed in the state
$\vert \alpha _0\rangle$. At time $t>0$, the integration of the
Schr\"odinger equation leads to
\begin{equation}
\label{defpsi}
\vert \Psi(t)\rangle ={\rm e}^{-ikt/2}\left\vert \alpha(t) =\alpha
_0{\rm e}^{-ikt}\right \rangle\, .
\end{equation}
This result should be understood as follows. In the
expression~(\ref{coherent}) which defines a coherent state, the factor
$\alpha $ should be replaced with $\alpha _0{\rm e}^{-ikt}$ to get the
formula expressing the above state $\vert \Psi \rangle$. As already
mentioned, at the quantum level, the normal variable becomes the
annihilation operator [obtained from Eq.~(\ref{defalpha}) by simply
replacing $q$ and $p$ with their quantum counter-parts]. This implies
\begin{eqnarray}
\hat{q}&=& \frac{1}{\sqrt{2k}}\left(\hat{a}+\hat{a}^{\dagger }\right)\,
\quad \hat{p}=-i\sqrt{\frac{k}{2}}\left(\hat{a}-\hat{a}^{\dagger
}\right)\, .
\end{eqnarray}
Then, the crucial step is that any coherent state $\vert \alpha \rangle
$ is the eigenvector of $\hat{a}$ with the eigenvalue $\alpha$,
$\hat{a}\vert \alpha \rangle =\alpha \vert \alpha \rangle$. Using this
property, it is easy to show that, for the state $\vert \Psi \rangle$
defined by Eq.~(\ref{defpsi}), one has
\begin{eqnarray}
\left\langle \Psi \vert \hat{p}\vert \Psi \right\rangle &=&
-i\sqrt{\frac{k}{2}}\left(\alpha _0{\rm e}^{-ikt}- \alpha _0^*{\rm
e}^{ikt}\right)\, ,\\ 
\left\langle \Psi \vert \hat{q}\vert \Psi \right\rangle
&=& \frac{1}{\sqrt{2k}}\left(\alpha
_0{\rm e}^{-ikt}+\alpha _0^*{\rm e}^{ikt}\right)\, . 
\end{eqnarray}
In the same way, straightforward manipulations lead to 
\begin{eqnarray}
\left\langle \Psi \vert \hat{p}^2\vert \Psi \right\rangle &=&
\frac{k}{2}\left\{1+4\Im ^2\left[\alpha (t)\right]\right\}\, ,\quad
\left\langle \Psi \vert \hat{q}^2\vert \Psi \right\rangle =
\frac{1}{2k}\left\{1+4\Re ^2\left[\alpha (t)\right]\right\}\, ,
\end{eqnarray}
from which one deduces
\begin{equation}
\Delta \hat{q}=\sqrt{\frac{1}{2k}}\, \quad \Delta
\hat{p}=\sqrt{\frac{k}{2}}\, .
\end{equation}
We have thus reached our goal, i.e. we have shown that the
state~(\ref{defpsi}) follows the classical trajectory and that the
quantum dispersion around this trajectory is the same in position and
momentum and is minimal (that is to say the Heisenberg inequality is
saturated). Therefore, as announced, the coherent state is indeed the
``most classical'' state. It is also interesting to give the explicit
form of the wave-function. It reads
\begin{equation}
\label{wfcoherent}
\Psi_{\alpha }(q,t)={\rm e}^{i\theta
_{\alpha}}\left(\frac{k}{\pi}\right)^{1/4} {\rm e}^{-ikt/2}{\rm
e}^{iqp_{\rm cl}(t)}{\rm e}^{-k\left[q-q_{\rm cl}\left
(t\right)\right]^2/2}\, ,
\end{equation}
where the phase factor is defined by ${\rm e}^{i\theta _{\alpha}}\equiv
{\rm e}^{(\alpha ^*{}^2-\alpha ^2)/4}$.

\par

\begin{figure}[t]
\centering \includegraphics[height=5cm, width=5cm]{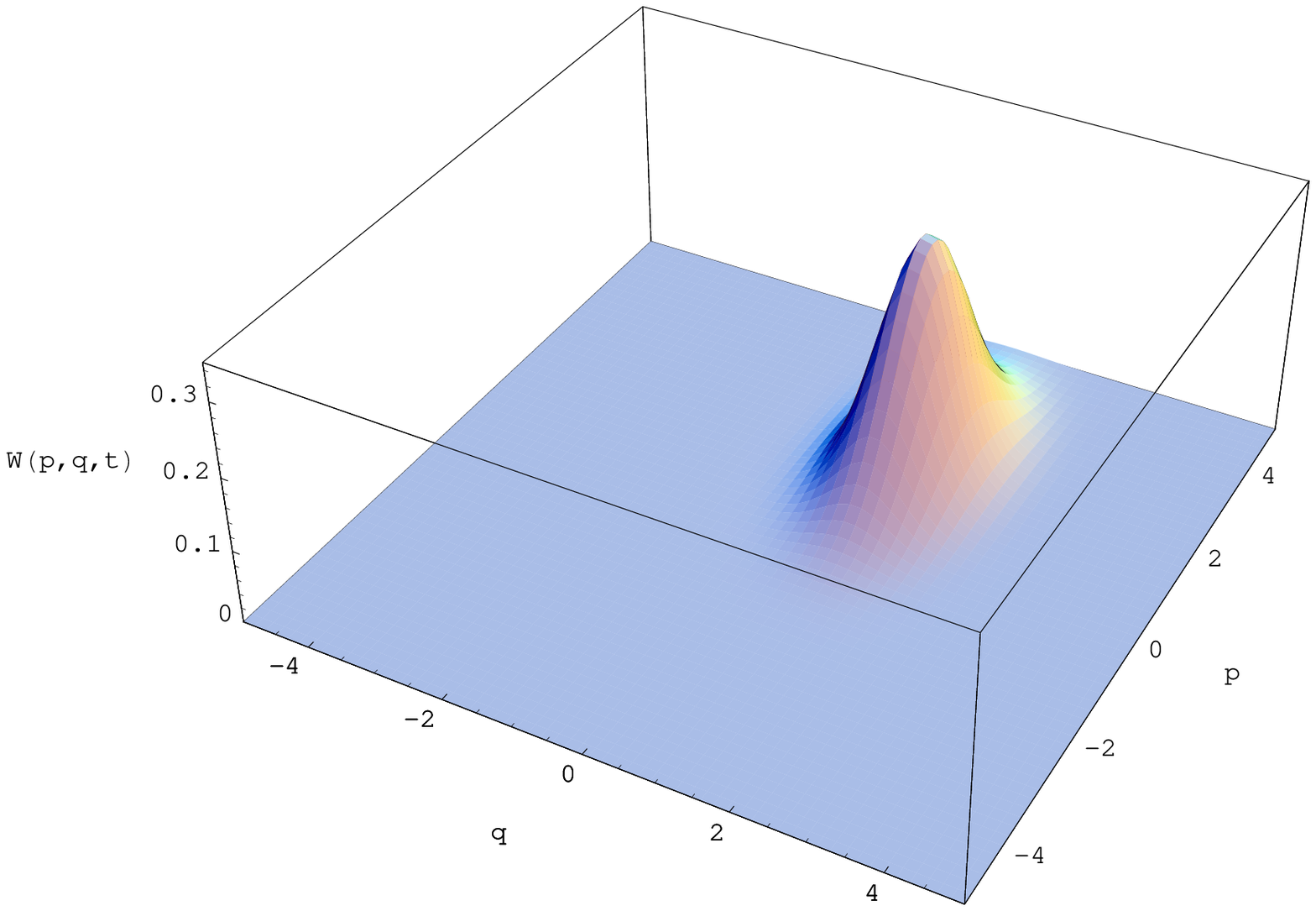}
\centering \includegraphics[height=5cm, width=5cm]{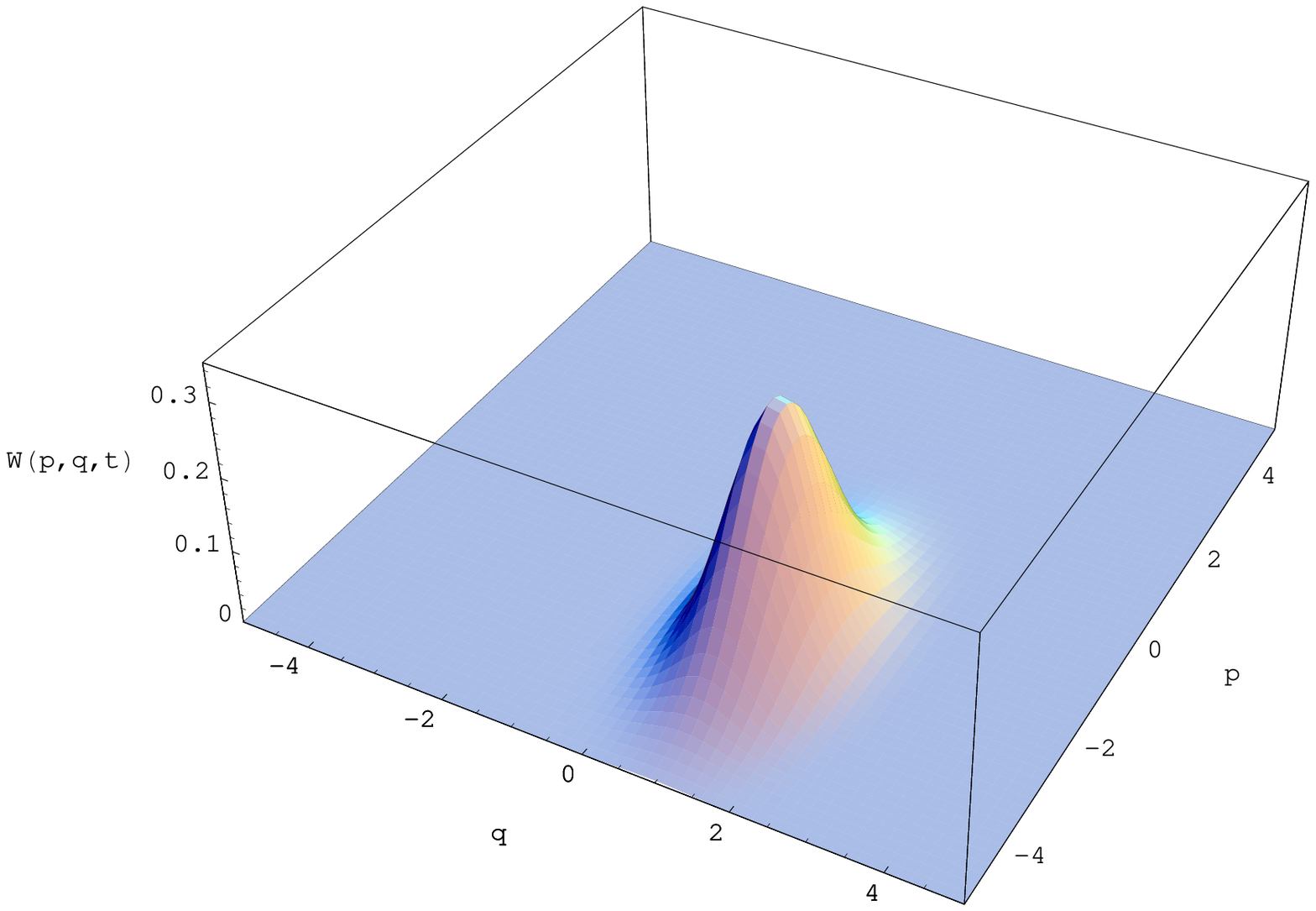}
\centering \includegraphics[height=5cm, width=5cm]{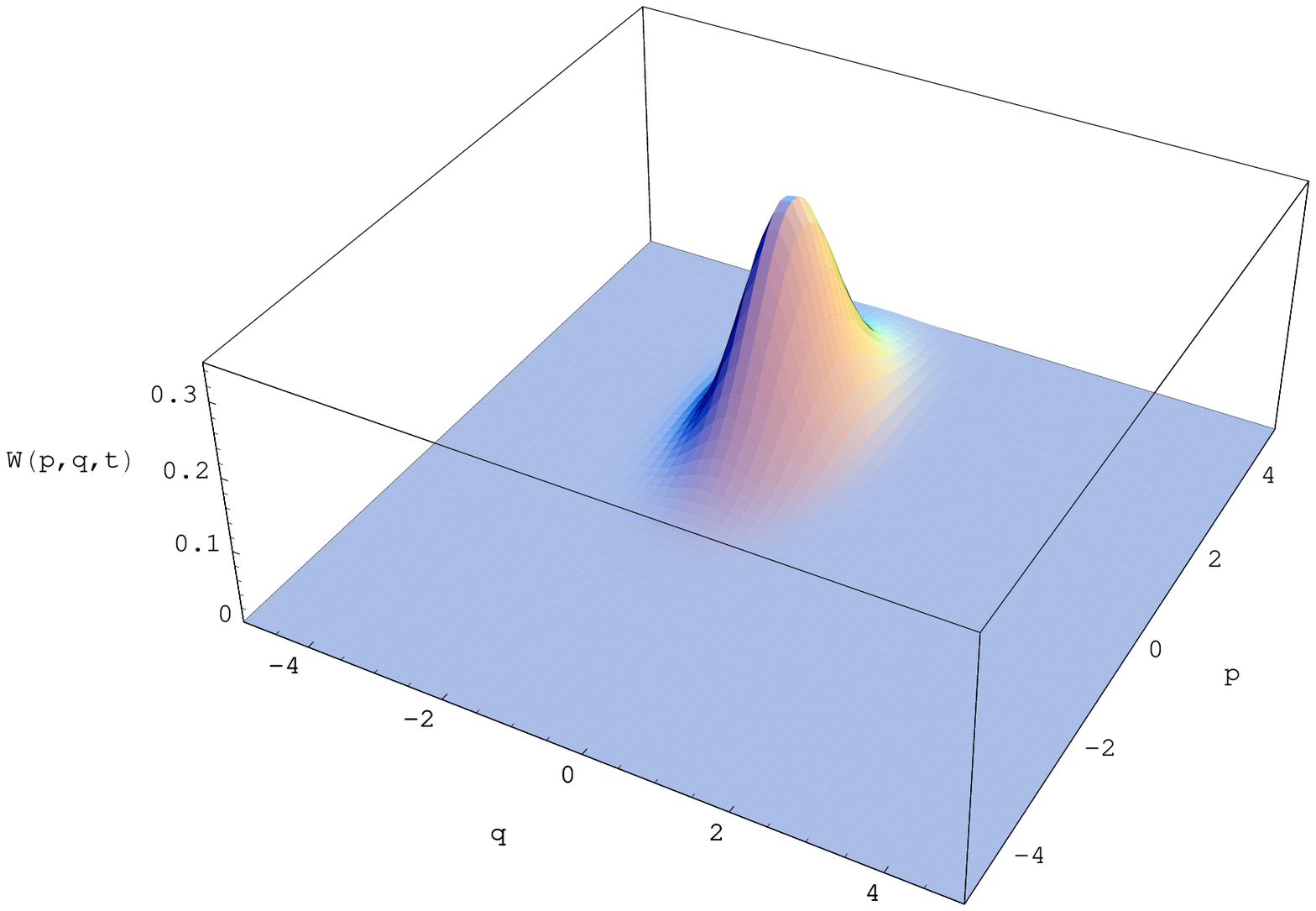}
\centering \includegraphics[height=5cm, width=5cm]{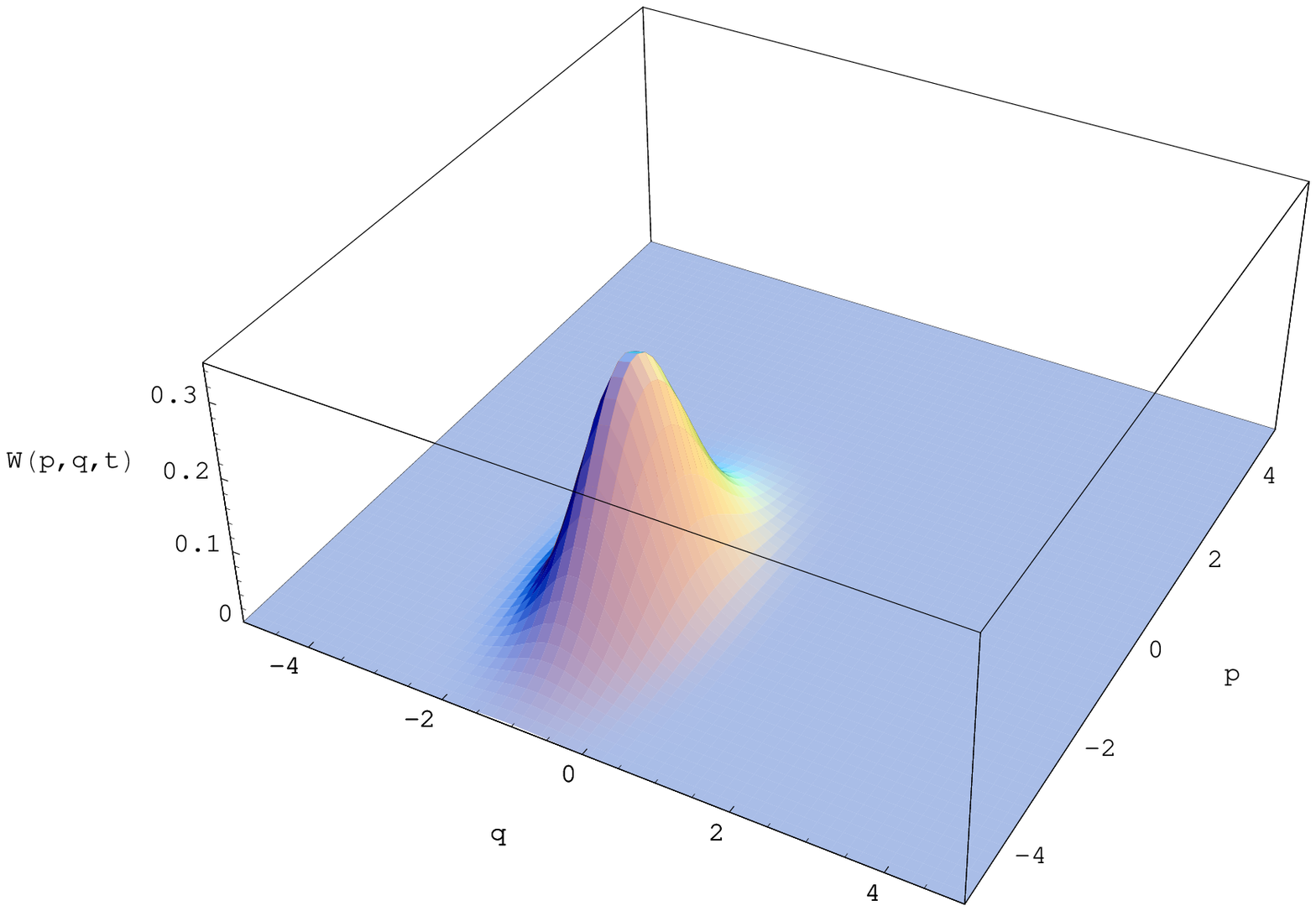}
\caption{Wigner function~(\ref{wignercoh}) for the coherent state $\vert
\alpha \rangle $ at different times. The (arbitrary) values $q_0=1$,
$p_0=1$ and $k=2$ have been used for this figure. This implies $\alpha
_0=\sqrt{2}{\rm e}^{i\pi/4}$ and, see Eqs.~(\ref{harmoclass}), $p_{\rm
cl}=2\sqrt{2}\sin \left(\pi/4-t\right)$ and $q_{\rm
cl}=\sqrt{2}\cos\left(\pi/4-t\right)$. The upper left panel represents
the Wigner function~(\ref{wignercoh}) at time $t=0$ while the upper
right, lower right and lower left panels correspond to $W(p,q,t)$ at
time $t=\pi/2$, $t=\pi$ and $t=3\pi/2$ respectively. The wave packet
follows the periodic (ellipsoidal) classical trajectory in phase space
and its shape remains unchanged during the motion.} 
\label{wignercoherent}
\end{figure}

The above expression is defined in real space. However, if one wants to
follow the evolution of the system in phase space, it is interesting to
introduce the Wigner function~\cite{wigner,AA,SH,HL} defined by the
expression (for a one-dimensional system)
\begin{eqnarray}
\label{defwigner}
W\left(q,p,t \right) &\equiv & \frac{1}{2\pi}\int {\rm d}u\, \Psi
^*\left(q-\frac{u}{2}, t\right){\rm e}^{-ipu}\Psi
\left(q+\frac{u}{2},t\right)\, .
\end{eqnarray}
A system behaves classically if the Wigner function is positive-definite
since, in this case, it can be interpreted as a classical
distribution. In addition, if the Wigner function is localized in phase
space over a small region corresponding to the classical position and
momentum, then the corresponding quantum predictions become
indistinguishable from their classical counter-parts and we can indeed
state that the system has ``classicalized''. For the
wave-function~(\ref{wfcoherent}), the Wigner function can be expressed
as
\begin{equation}
\label{wignercoh}
W\left(q,p,t \right)=\frac{1}{\pi }{\rm e}^{-k\left[q-q_{\rm
cl}(t)\right]^2} {\rm e}^{-\frac{1}{k}\left[p -p_{\rm cl}(t)\right]^2}\,
.
\end{equation}
It is represented in Fig.~\ref{wignercoherent}. We notice that the
Wigner function is always positive and, therefore, according to the
above considerations, the system can be considered as
classical. Moreover, $W(p,q,t)$ is peaked over a small region in phase
space and the wave packet follows exactly the classical trajectory (an
ellipse), as is also clear from Eq.~(\ref{wignercoh}). This confirms our
interpretation of the coherent state as the most classical state.

\par

To conclude this sub-Section, let us notice that the coherent state
$\vert \alpha \rangle $ can be obtained by applying the following
unitary operator on the vacuum state
\begin{equation}
\vert \alpha \rangle ={\rm e}^{\alpha a^{\dagger }-\alpha ^*a}\vert
0\rangle\, .
\end{equation}
This equation should be compared to Eqs.~(\ref{defBT})
and~(\ref{defs}). We see that the argument of the exponential is linear
in the creation and annihilation operators while it was quadratic in the
case of the squeezed state.

\subsection{Wigner Function of the Cosmological Perturbations}
\label{sub-sec:wignercosmo}

\begin{figure}[t]
\centering \includegraphics[height=5cm, width=5cm]{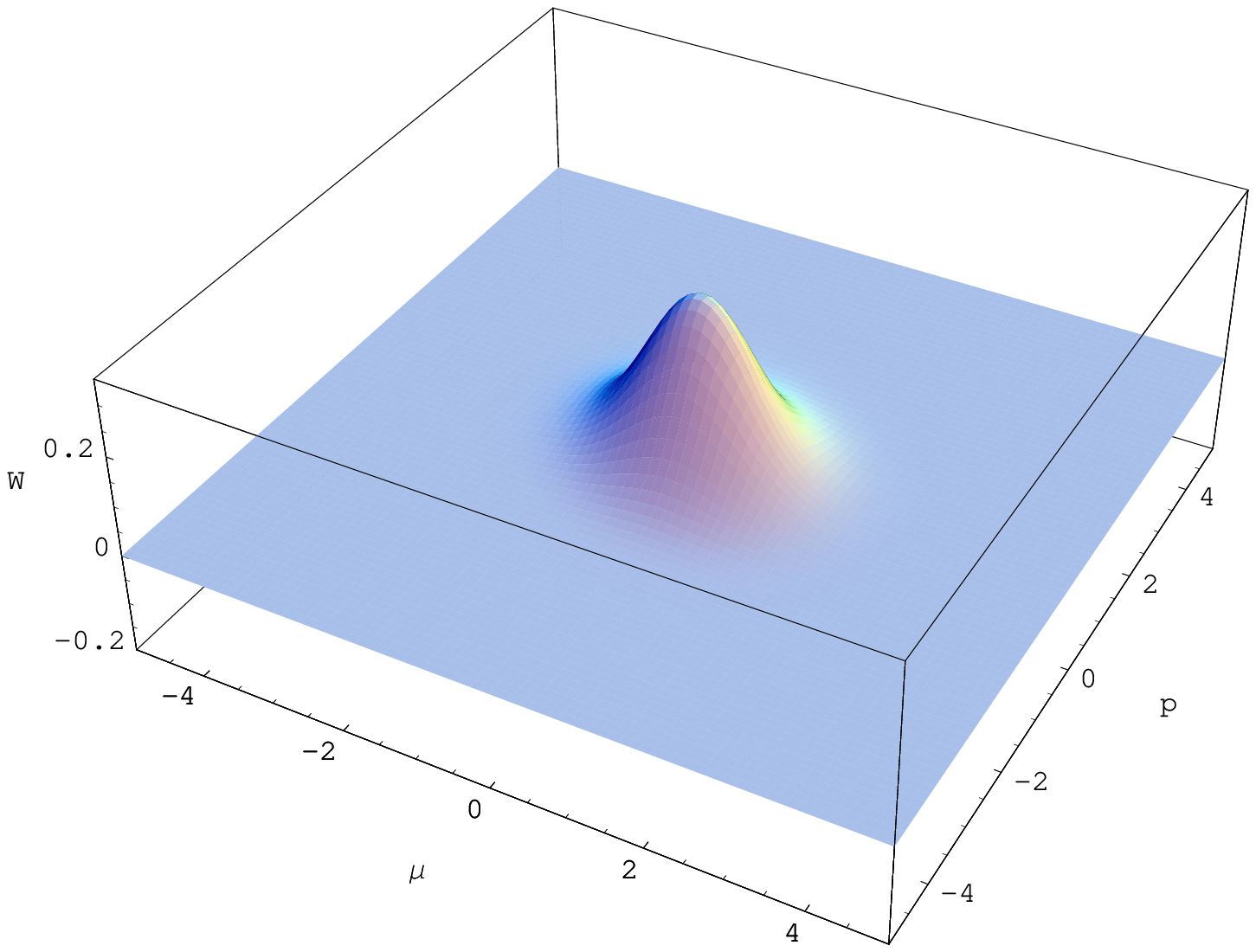}
\centering \includegraphics[height=5cm, width=5cm]{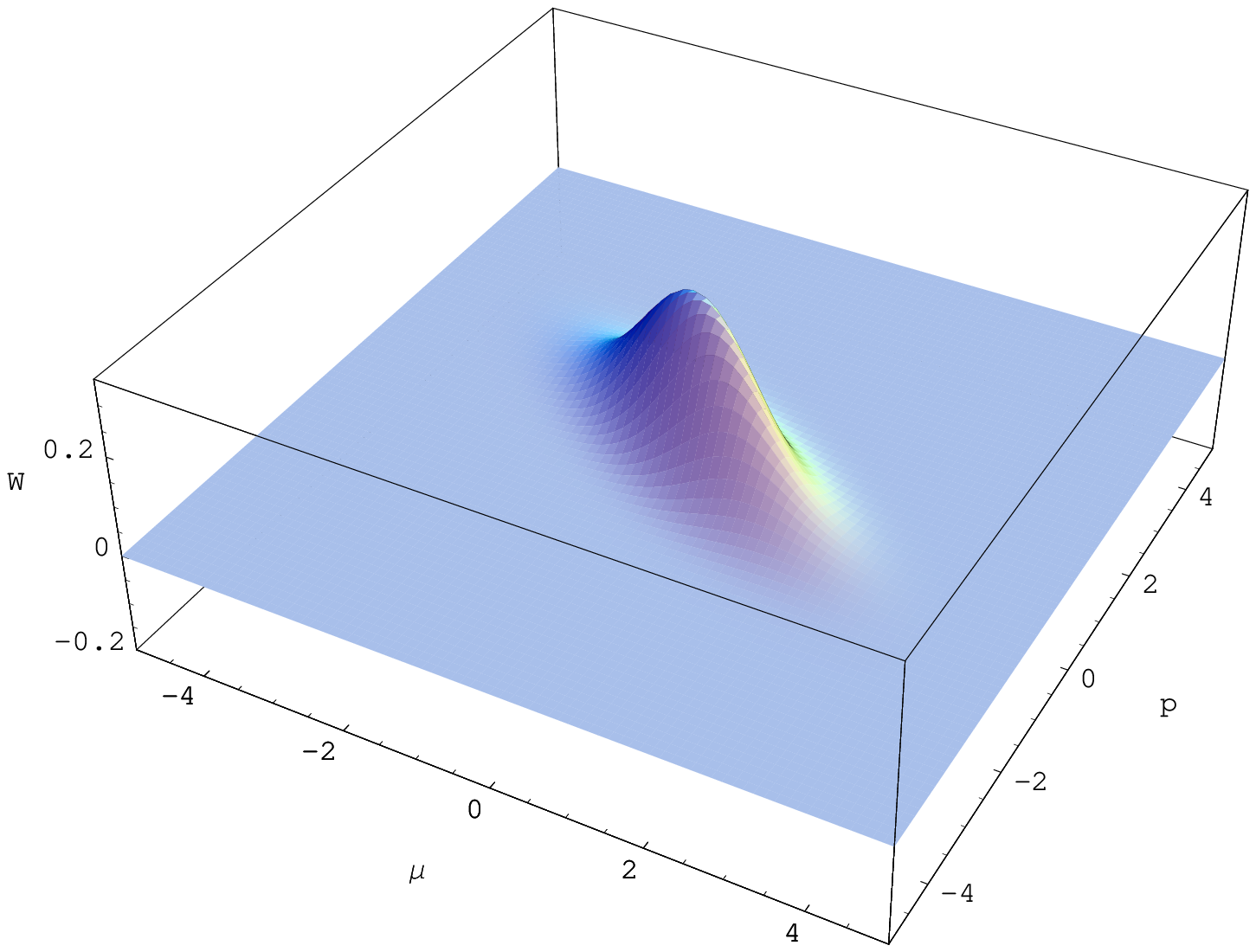}
\centering \includegraphics[height=5cm, width=5cm]{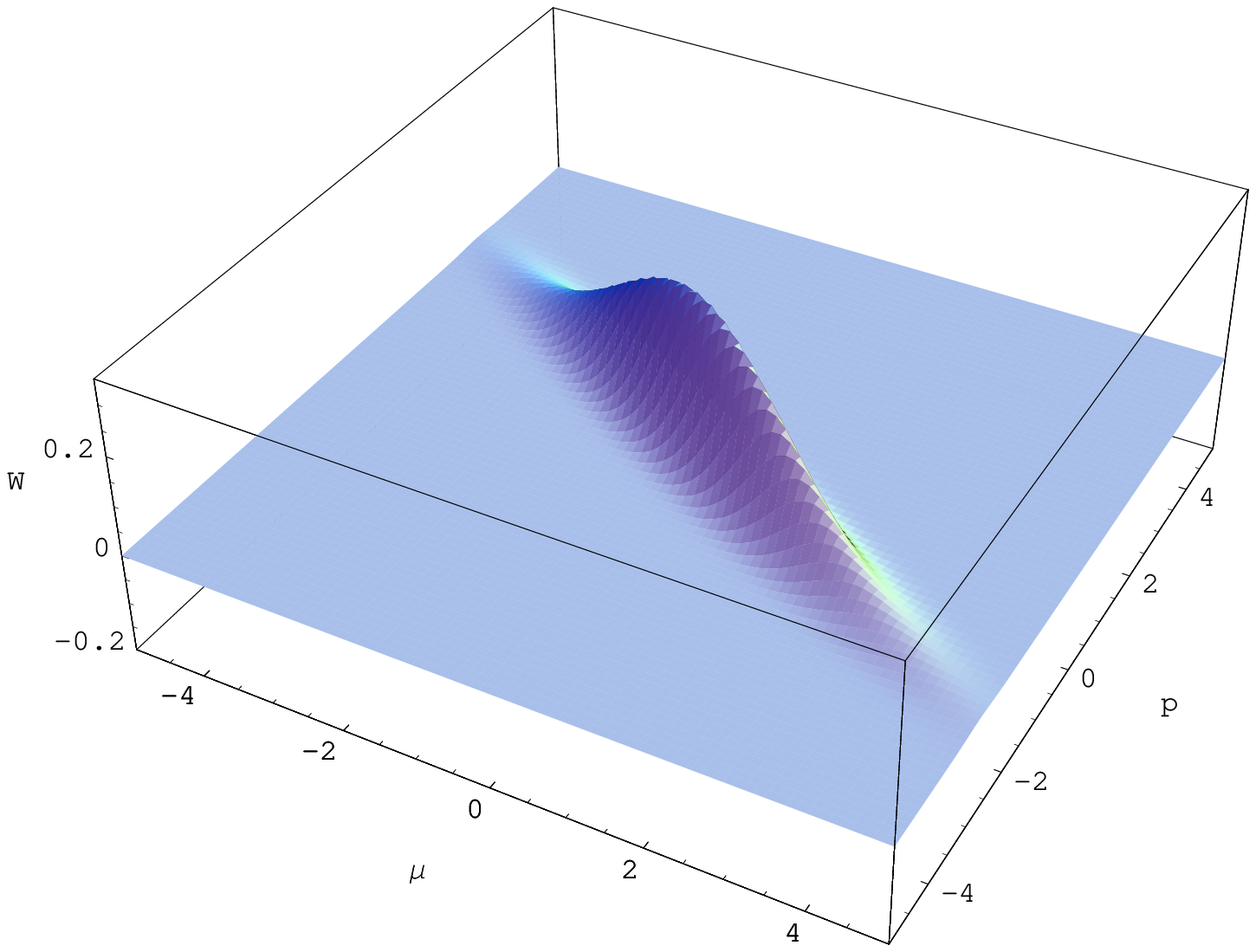}
\centering \includegraphics[height=5cm, width=5cm]{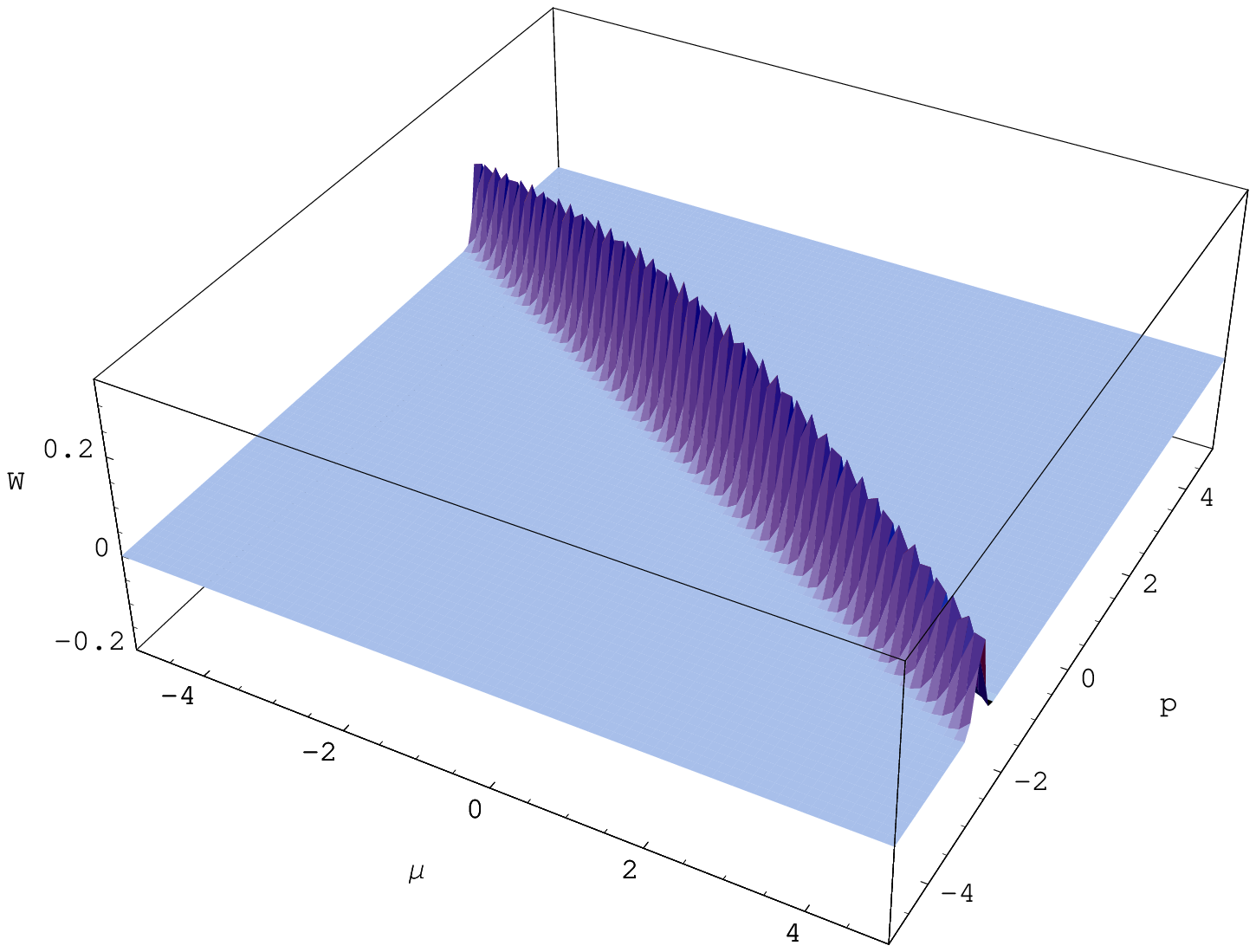}
\caption{Wigner function of cosmological perturbations obtained from
Eq.~(\ref{wignerfinal}) (for a one-dimensional system). The squeezing
parameter $r$ is chosen to be $r=0.1$, $r=0.5$, $r=1$ and $r=2$ for the
left upper, right upper, left lower and right lower panels respectively
(it is not the same time ordering as in Fig.~\ref{wignercoherent}
because, in the present case, the motion is not periodic). The other
squeezing parameters are taken to be $\phi =\pi/6$ and $\theta =0$. As
can be noticed in this figure, the Wigner function remains
positive. Since the squeezing parameter increases with time, the
different panels correspond in fact to the Wigner function at different
times. At initial time, the quantum state is the vacuum and, therefore,
the Wigner function is that of a coherent state, compare the left upper
panel with Fig.~\ref{wignercoherent}. Then, the Wigner function develops
the ``Dirac function behavior'' discussed in the text that clearly
appears on this plot.}  \label{wignerpert}
\end{figure}

In order to study whether the (super-Hubble) cosmological perturbations
have ``classicalized'', we now use the technical tool of the Wigner
function introduced before. The first application to cosmological
perturbations was made in Refs.~\cite{GP,GS2}. For a two-dimensional
system (here, we have in mind $\mu _{\vec{k}}^{_{\mathrm R}}$ and $\mu
_{\vec{k}}^{_{\mathrm I}}$ for a fixed mode $\vec{k}$), the
generalization of Eq.~(\ref{defwigner}) is straightforward and reads
\begin{eqnarray}
W\left(\mu _{\vec{k}}^{_{\mathrm R}}, \mu _{\vec{k}}^{_{\mathrm I}},
p_{\vec{k}}^{_{\mathrm R}}, p_{\vec{k}}^{_{\mathrm I}}\right) &\equiv
& \frac{1}{\left(2\pi\right)^2}\int \int {\rm d}u{\rm d}v \, \Psi
^*\left(\mu _{\vec{k}}^{_{\mathrm R}}-\frac{u}{2}, \mu
_{\vec{k}}^{_{\mathrm I}}-\frac{v}{2}\right){\rm
e}^{-ip_{\vec{k}}^{_{\mathrm R}}u-ip_{\vec{k}}^{_{\mathrm I}}v}
\nonumber \\ & & \times \Psi \left(\mu _{\vec{k}}^{_{\mathrm
R}}+\frac{u}{2}, \mu _{\vec{k}}^{_{\mathrm I}}+\frac{v}{2}\right)\, ,
\end{eqnarray}
where the wave-function is given by the
expressions~(\ref{wavefunctionpert}). Since we have to deal with
Gaussian integrations only, the above Wigner function can be calculated
exactly. One obtains
\begin{eqnarray}
\label{wignerfinal}
W\left(\mu _{\vec{k}}^{_{\mathrm R}}, \mu _{\vec{k}}^{_{\mathrm I}},
p_{\vec{k}}^{_{\mathrm R}}, p_{\vec{k}}^{_{\mathrm I}}\right) &= &
\Psi \Psi^* \frac{1}{2\pi \Re \Omega _{\vec{k}}} \exp\left[\frac{1}{2
\Re \Omega _{\vec{k}}}\left(p_{\vec{k}}^{_{\mathrm R}}+2\Im \Omega
_{\vec{k}} \mu _{\vec{k}}^{_{\mathrm R}}\right)^2\right]\nonumber \\ &
& \times \exp\left[\frac{1}{2 \Re \Omega
_{\vec{k}}}\left(p_{\vec{k}}^{_{\mathrm I}}+2\Im \Omega _{\vec{k}} \mu
_{\vec{k}}^{_{\mathrm I}}\right)^2\right]\, .
\end{eqnarray}
It is represented in Fig.~\ref{wignerpert}. The first remark is that the
Wigner function is positive (as expected since we deal with a Gaussian
state) and, therefore, can be interpreted as a classical
distribution. However, as shown in Fig.~\ref{wignerpert}, and contrary
to the case of a coherent state, $W$ is not peaked over a small region
of phase space. We are interested in the behavior of the Wigner function
for modes of astrophysical interest today. These modes have spent time
outside the Hubble radius during inflation and, as a consequence, their
squeezing parameter $r$ is big. Therefore, it is convenient to express
$\Omega _{\vec{k}}$ in terms of the squeezing parameters
\begin{equation}
\Omega _{\vec{k}}=\frac{k}{2}\frac{\cosh r -{\rm e}^{-2i\phi }\sinh
r}{ \cosh r +{\rm e}^{-2i\phi }\sinh r}\, ,
\end{equation}
and to take the strong squeezing limit, $r\rightarrow +\infty $. One has
\begin{equation}
\Re \Omega _{\vec{k}}\rightarrow 0\, ,\quad \Im \Omega
_{\vec{k}}\rightarrow \frac{k}{2}\frac{\sin \phi}{\cos \phi}\, .
\end{equation}
This implies
\begin{eqnarray}
\label{wignerdelta}
W\left(\mu _{\vec{k}}^{_{\mathrm R}}, \mu _{\vec{k}}^{_{\mathrm I}},
p_{\vec{k}}^{_{\mathrm R}}, p_{\vec{k}}^{_{\mathrm I}}\right)
&\rightarrow & \Psi \Psi^* \delta \left(\frac{k}{2}\frac{\sin
\phi}{\cos \phi}\mu _{\vec{k}}^{_{\mathrm R}}+p_{\vec{k}}^{_{\mathrm
R}}\right) \delta \left(\frac{k}{2}\frac{\sin \phi}{\cos \phi}\mu
_{\vec{k}}^{_{\mathrm I}}+p_{\vec{k}}^{_{\mathrm I}}\right)\, ,
\end{eqnarray}
where $\delta $ denotes the Dirac function. The above limit is clearly
visible in Fig.~\ref{wignerpert} for $r>1$ (lower panels).

\par

Therefore, in the large squeezing limit, the Wigner function is
elongated along a very thin ellipse in phase space. At first sight, this
means that the system is not classical since one can not single out a
small cell around some classical values $(\mu _{\rm cl},p_{\rm cl})$
that would follow a classical trajectory as it was discussed before. On
the other hand, as already mentioned above, the Wigner function remains
positive. This means that the interference term which makes the system
quantum in the sense that the amplitudes rather than the probabilities
should be summed up have become negligible. Therefore, in this sense,
the system is classical or, more precisely, is in fact equivalent to a
classical stochastic process with a Gaussian distribution (given by the
term $\Psi \Psi ^*$). We see that the nature of this classical limit is
quite different to what happens in the case of a coherent state: we
cannot predict a definite correlation between position and momentum but
we can describe the system in terms of a classical random variable. In
practice, this is what is done by astrophysicists: in particular, the
quantity $a_{\ell m}$ in Eq.~(\ref{dt/t}) is always treated as a
Gaussian random variable and any detailed quantum-mechanical
considerations avoided.

\par

As argued in Ref.~\cite{PS}, the system has become classical (in the
sense explained before) without any need to take into account its
interaction with the environment. This is ``decoherence without
decoherence'' as stressed in the above-refered article. More on this
subtle issue can be found in Refs.~\cite{PS,KP}. Of course, the question
of whether the wave-function of the perturbation has collapsed or not
(and the question of whether this question is meaningful in the present
context and/or dependent on the interpretation of quantum mechanics that
one chooses to consider) is even more delicate~\cite{PSS} and we will
not touch upon this issue here.

\subsection{Wigner function of a Free Particle}

\begin{figure}[t]
\centering \includegraphics[height=5cm, width=5cm]{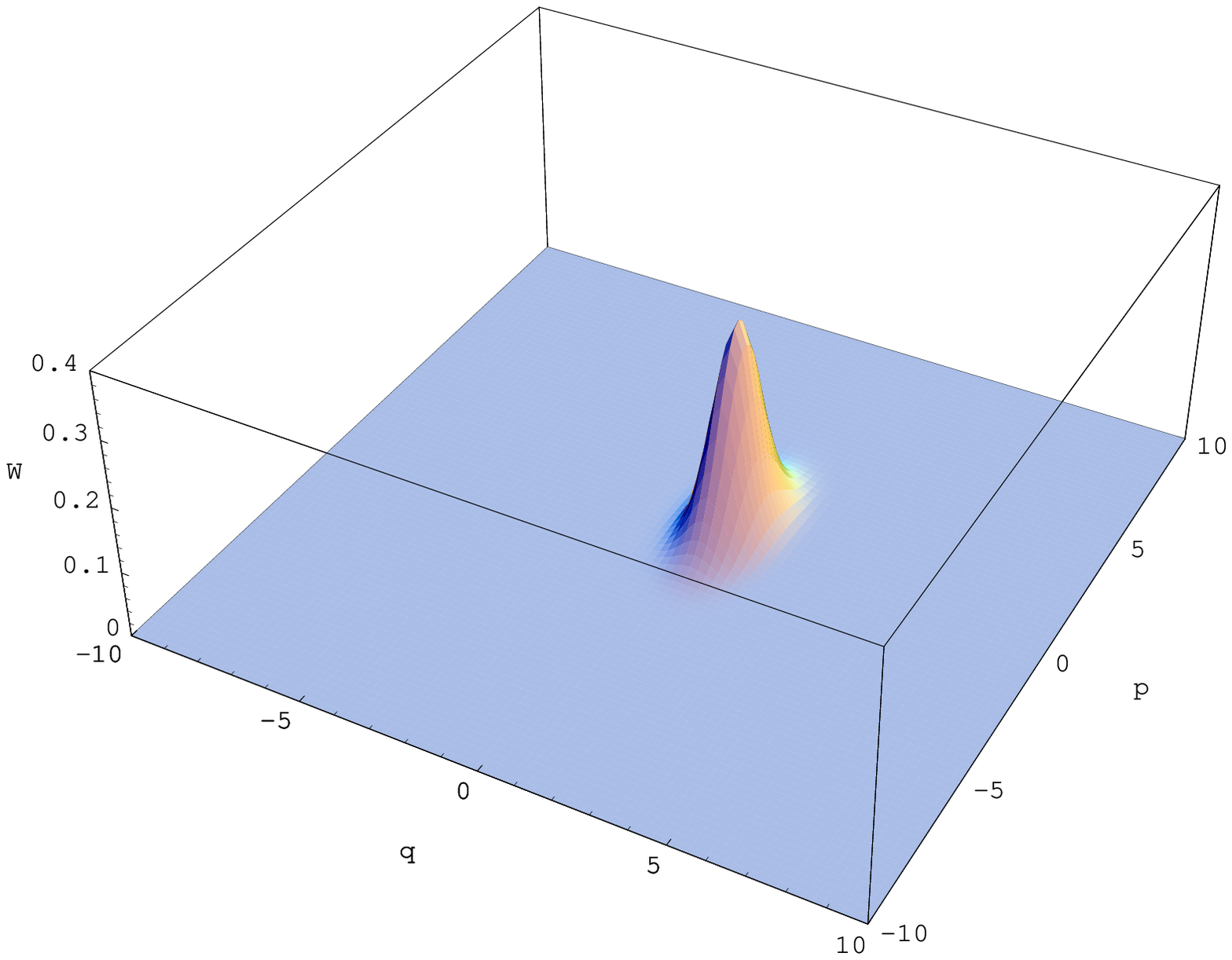}
\centering \includegraphics[height=5cm, width=5cm]{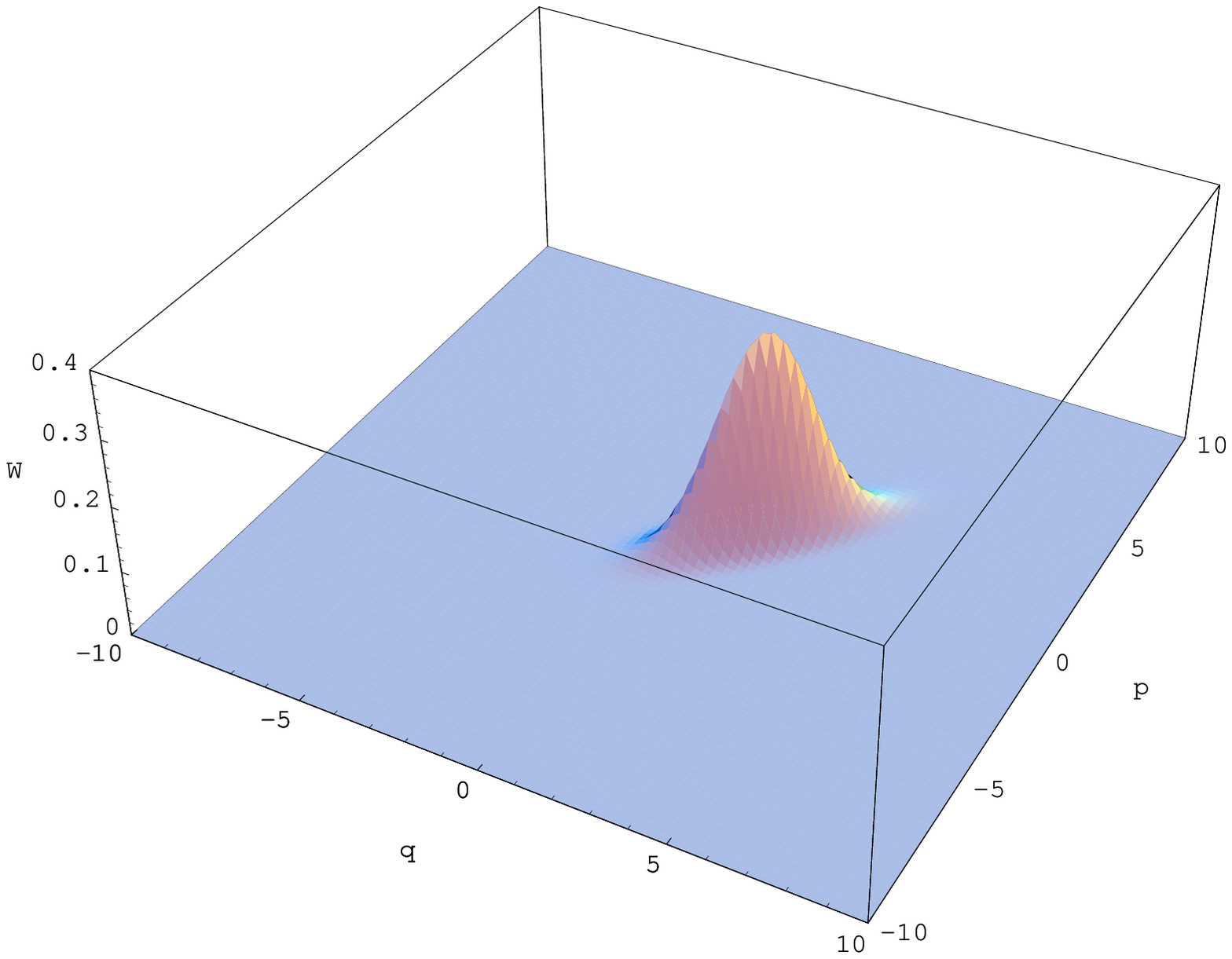}
\centering \includegraphics[height=5cm, width=5cm]{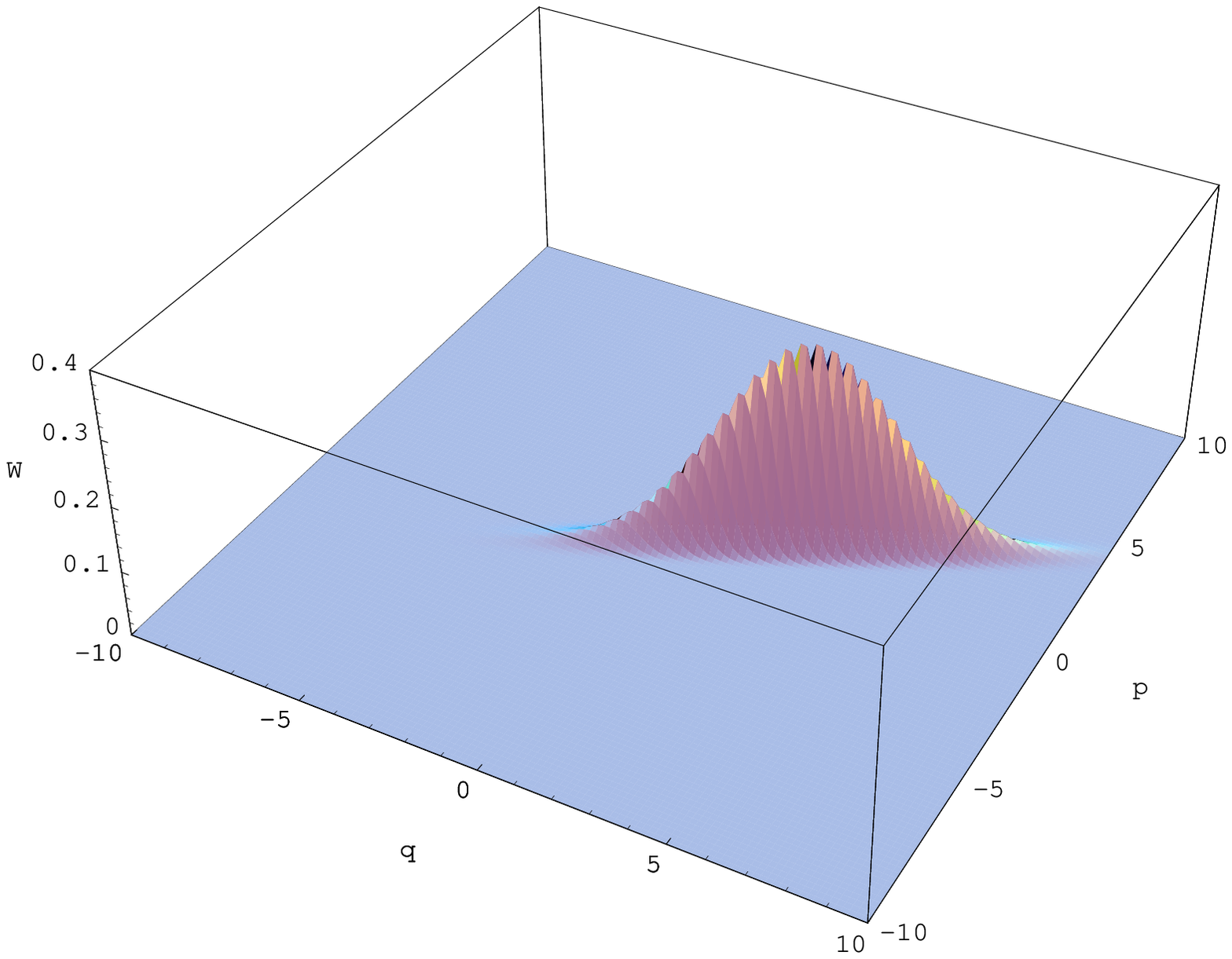}
\centering \includegraphics[height=5cm, width=5cm]{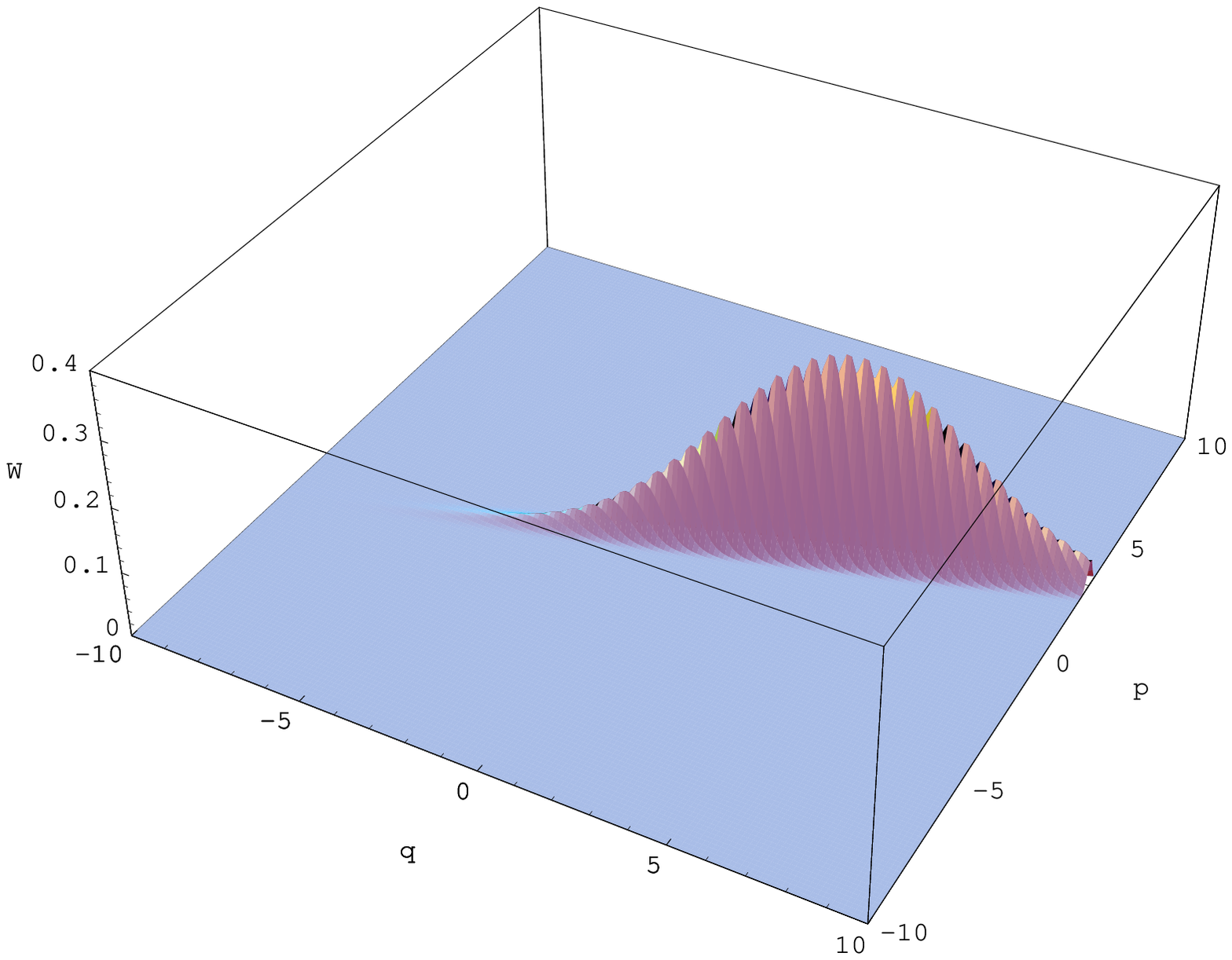}
\caption{Wigner function~(\ref{wignerfree}) of a free particle. The
parameters chosen are $k_0=1$, $q_0=1$ and $a=1$ which is in fact
equivalent to considering the dimensionless quantities $ap$, $q/a$ and
$t/a^2$. The left upper panel corresponds to $t/a^2=0$ while the right
upper, left lower and right lower panels represent the Wigner function
at times $t/a^2=0.8,2,3$ respectively (here, the time ordering is
similar to that in Fig.~\ref{wignerpert} and, therefore, different from
that in Fig.~\ref{wignercoherent}). Initially, the wave packet is
well-localized in phase space and, as time goes on, the spreading of the
wave packet becomes apparent.}  \label{fig:wignerfree}
\end{figure}

We now consider the case of a free particle since it shares common
points with the case of cosmological perturbations as first noticed in
Ref.~\cite{KP}. At the beginning of this section, we mentioned that, in
the particular case of the harmonic oscillator $V(q)\propto q^2$, the
quantum mean value of the position and momentum operators always follow
the classical trajectory whatever the quantum state $\vert \Psi \rangle$
in which the system is placed. There is obviously another situation
where this is also the case: the free massive particle where
$V(q)=0$. The wave-function is given by (we take $m=1$)
\begin{eqnarray}
\Psi(q,t)&=&\left(\frac{2a^2}{\pi
}\right)^{1/4}\frac{1}{\left(a^4+4t^2\right)^{1/4}} {\rm e}^{-i\tan
^{-1}\left(2t/a^2\right)/2+ik_0(q-q_0)-ik_0^2t/2} \nonumber \\ & &
\times {\rm e}^{-\left(q-q_0-k_0t\right)^2/\left(a^2+2it\right)}\, ,
\end{eqnarray}
where we have assumed to the wave packet is centered at $q=q_0$ at
$t=t_0$. The parameter $a$ represents the width of the wave packet while
$k_0$ parameterizes its velocity. The means of $\hat{p}$ and $\hat{q}$
can be expressed as
\begin{equation}
\langle \hat{p}\rangle =k_0=p_{\rm cl}\, ,\quad \langle \hat{q}\rangle
=q_0+k_0t=q_{\rm cl}\, .
\end{equation}
Therefore, as announced, they follow exactly the classical
trajectory. But, as argued in the sub-Section devoted to coherent
states, one must also compute the dispersions. Straightforward
calculations lead to
\begin{eqnarray}
\left\langle \hat{p}^2\right\rangle &=&k_0^2+\frac{1}{a^2}\, ,\quad
\left\langle \hat{q}^2\right\rangle =
\frac{1}{4a^2}\left(a^4+4t^2\right)+\left(q_0+k_0t\right)^2\, ,
\end{eqnarray}
from which one deduces that
\begin{equation}
\Delta \hat{p} =\frac{1}{a}\, , \quad \Delta \hat{q}
=\frac{a}{2}\sqrt{1+4\frac{t^2}{a^4}}\, .
\end{equation}
At $t=0$, one has $\Delta \hat{q}\Delta \hat{p}=1/2$ and the Heisenberg
bound is saturated: at initial time, the wave packet is minimal. But
then, and contrary to the case of the harmonic oscillator, the
dispersion on the position is increasing with time (while the dispersion
on the momentum remains constant). The wave packet does not keep its
shape unchanged while moving as it was the case for a potential
$V(q)\propto q^2$. This is the well-known phenomenon dubbed ``spreading
of the wave packet''. However, it exists a quasi-classical
interpretation of this situation. Indeed, when $t\gg a^2$, one has
\begin{equation}
\Delta \hat{q}\sim \frac{t}{a}=\Delta \hat{p}\,t= \Delta v_{\rm cl}t\, ,
\end{equation}
which reproduces the classical motion. On the contrary, for small times,
$\Delta \hat{q}$ must take values very different from the classical ones
in order to satisfy the Heisenberg inequality. Therefore, in the regime
$t\gg a^2$, the system is classical but in a sense slightly different
from the one encountered in the harmonic oscillator case.

\par

Let us now calculate the Wigner function of the free particle. One
obtains
\begin{eqnarray}
\label{wignerfree} W(p,q,t)&=& \frac{1}{\pi
}\exp\left[-\frac{2a^2}{a^4+4t^2} \left(q-q_{\rm
cl}\right)^2\right]\nonumber \\ & \times &\exp\left\{
-\frac{a^4+4t^2}{2a^2}\left[p-p_{\rm cl}-\frac{4t}{a^4+4t^2}
\left(q-q_{\rm cl}\right)\right]^2\right\}\, .
\end{eqnarray}
This equation is similar to Eq.~(\ref{wignerfinal}). The Wigner function
is the product of one exponential factor whose argument is proportional
to `` $\left(q-q_{\rm cl}^2\right)$'' [hidden in the term $\Psi^*\Psi$
in Eq.~(\ref{wignerfinal})] and of another exponential term whose
argument has the form ``$\left[p-p_{\rm cl}-f(t)\left(q-q_{\rm
cl}\right)\right]^2$, where $f(t)$ is a function of time
only. Therefore, the classical limit of cosmological perturbations can
also be understood in terms of the (quasi-) classical limit of a free
particle, as discussed in the previous paragraph.

\par

The Wigner function~(\ref{wignerfree}) is represented in
Fig.~\ref{fig:wignerfree}. This plot confirms the interpretation
presented above. First of all, the Wigner function remains positive
which indicates that a classical interpretation is meaningful. At
initial time, the Wigner function is well-localized because the wave
packet is minimal. Then, at time goes on, the spreading of the wave
packet causes the spreading of the Wigner function in phase
space. Clearly, this case bears some ressemblence with that of
cosmological perturbations, compare Figs.~\ref{wignerpert}
and~\ref{fig:wignerfree}. Therefore, inflationary fluctuations on large
scales become classical in the same sense that a free particle is
classical far from the origin. A much more detailed description of this
analogy can be found in Ref.~\cite{KP}.

\section{Conclusions}
\label{sec:conclusions}

In this review, we have presented a pedagogical introduction to the
theory of inflationary cosmological perturbations of quantum-mechanical
origin, focusing mainly on its fundamental aspects. We have shown that
the mechanism responsible for the production of the initial fluctuations
in the early universe is in fact similar to a well-known effect in
quantum field theory, namely the Schwinger effect. It is indeed the
``interaction'' of the quantum perturbed metric $\delta g_{\mu \nu}$
(and of the perturbed inflaton field $\delta \varphi $) with the
background gravitational field which is responsible for the
amplification of the initial vacuum fluctuations in the same way that
pair creation can occur in an external electrical field. Because the
gravitational field in the early universe can be strong (in Planck
units), this mechanism leads to observable effects, in particular to CMB
temperature fluctuations. Therefore, these fluctuations originate from a
remarkable interplay between general relativity and quantum mechanics.

\par

There is also another aspect associated with the inflationary mechanism
discussed above that could be relevant to probe fundamental physics. In
a typical model of inflation, the total number of e-folds is such that
the scales of astrophysical interest today were initially not only
smaller than the Hubble radius but also smaller than the Planck (or
string) scale. As a consequence, the WKB initial conditions discussed
before are maybe modified by quantum gravity (stringy) effects and,
therefore, this opens up the possibility to probe these effects through
observations of the CMB~\cite{BM1,MR2}. Of course, an open issue is the
fact that it is difficult to calculate how quantum gravity/string theory
will affect the initial conditions. Nevertheless, it is possible to draw
some generic conclusions. Firstly, the standard initial condition
consists in choosing only one WKB branch. Hence, any modification
amounts to considering that the second branch is present. As a
consequence, super-imposed oscillations in the power spectrum
unavoidably appear, the amplitude and the frequency of these
oscillations being unfortunately model dependent~\cite{BM1} (for the
observational status of these oscillations, see
Refs.~\cite{MR2,MR}). Secondly, the presence of the second WKB branch
means, in some sense, the presence of particles in the initial state
and, therefore, there is potentially a back-reaction
problem. Generically, the larger the amplitude of the super-imposed
oscillations is, the more severe the back-reaction issue. On the other
hand, predicting the effect of the energy density of the perturbations
is difficult and it is not clear whether this will spoil inflation or,
for instance, just renormalize the cosmological
constant~\cite{BM2,danielsson}. These problems are still open questions
but it is interesting to notice that the inflationary scenario is rich
enough to provide yet another means to learn about fundamental physics.

\section*{Acknowledgements}

I would like to thank P.~Brax, H.~Fried, M.~Lemoine, L.~Lorenz, P.~Peter
and C.~Ringeval for useful discussions and careful reading of the
manuscript.

%
%
%

%
%



\printindex
\end{document}